\documentclass[11pt]{amsart}
\usepackage[pdftex]{graphicx}
\usepackage{amsmath,amsthm,amssymb}
\usepackage[margin=1in]{geometry}
\usepackage{xcolor}
\usepackage{float}
\usepackage{txfonts}
\usepackage[labeled,resetlabels]{multibib}
\usepackage{fancyhdr}
\usepackage{siunitx}
\usepackage{caption}
\usepackage{subcaption}
\newcites{S}{Supplementary References}

\newcommand{\vx}{\mathbf{x}}
\newcommand{\vV}{\mathbf{V}}
\newcommand{\vX}{\mathbf{X}}
\newcommand{\cH}{\mathcal{H}} 
\newcommand{\bu}{{\mathbf u}}
\newcommand{\bn}{\mathbf n}
\newcommand{\bx}{{\mathbf x}}
\newcommand{\br}{{\mathbf r}}
\newcommand{\V}{\mathbb{V}} 
\newcommand{\vK}{\mathbf{K}}
\newcommand{\vY}{\mathbf{Y}}
\newcommand{\cL}{\mathcal{L}}
\newcommand{\Gm}{\Gamma}
\newcommand{\bnu}{\boldsymbol{\nu}}
\newcommand{\btau}{\boldsymbol{\tau}}
\newcommand{\ld}{\langle}
\newcommand{\dt}{\delta t}
\newcommand{\rd}{\rangle}
\newcommand{\inner}[2]{\left\ld #1, #2 \right\rd}
\newcommand{\wof}{W_{\mathrm{\scriptstyle OF}}}
\newcommand{\wh}{\mathcal{W}_{\begin{scriptstyle}
			{\mathrm{Hex}}
		\end{scriptstyle}}}
\newcommand{\ds}{\displaystyle}

\newtheorem{lemma}{Lemma}
\newtheorem{claim}{Claim}

\makeatletter
\let\@cleartopmattertags\relax
\newcommand\articleend
 {\enddoc@text
  \let\authors\@empty
  \let\contribs\@empty
  \let\xcontribs\@empty
  \let\toccontribs\@empty
  \let\addresses\@empty
  \let\thankses\@empty
  \newpage
 }
\let\@wraptoccontribs\wraptoccontribs
\makeatother

\makeatletter
\renewcommand\subsection{\@startsection{subsection}{2}%
  \z@{.5\linespacing\@plus.7\linespacing}{.5\linespacing}%
  {\normalfont\bfseries}}
\renewcommand\subsubsection{\@startsection{subsubsection}{3}%
  \z@{.5\linespacing\@plus.7\linespacing}{-.5em}%
  {\normalfont\scshape}}
\makeatother

\begin{document}
\pagestyle{plain}
\setlength{\footskip}{30pt}

\title[]{Toroidal nuclei of columnar lyotropic chromonic liquid crystals coexisting with isotropic phase}
\author[R.~Koizumi]{Runa Koizumi}
\address{Advanced Materials and Liquid Crystal Institute, Materials Science Graduate Program, Kent State University, Kent, OH 44242, USA}
\email{rkoizumi@kent.edu}
\author[D.~Golovaty]{Dmitry Golovaty}
\address{Department of Mathematics, The University of Akron, Akron, OH 44325-4002}
\email{dmitry@uakron.edu}
\author[A. ~Alqarni]{Ali Alqarni}
\address{Advanced Materials and Liquid Crystal Institute, Department of Physics, Kent State University, Kent, OH 44242, USA and Department of Physics, University of Bisha, Bisha, 67714, Saudi Arabia}
\email{aalqarn1@kent.edu}
\author[S.W.~Walker]{Shawn W. Walker}
\address{Department of Mathematics, Louisiana State University, Baton Rouge, LA 70803-4918}
\email{walker@lsu.edu}
\author[Y.A.~Nastishin]{Yuriy A. Nastishin}
\address{Advanced Materials and Liquid Crystal Institute, Kent State University, Kent, OH 44242, USA and Hetman Petro Sahaidachnyi National Army Academy, 32 Heroes of Maidan street, Lviv, 79012, Ukraine}
\email{nastyshyn\_yuriy@yahoo.com}
\author[M.C.~ Calderer]{M. Carme Calderer}
\address{School of Mathematics, University of Minnesota, Minneapolis, MN 55455, USA}
\email{calde014@umn.edu}
\author[O.D.~Lavrentovich]{Oleg D. Lavrentovich}
\address{Advanced Materials and Liquid Crystal Institute, Materials Science Graduate Program, Kent State University, Kent, OH 44242, USA and Department of Physics, Kent State University, Kent, Ohio 44242, USA}
\email{olavrent@kent.edu}

\begin{abstract}
Nuclei of ordered materials emerging from the isotropic state usually show a shape topologically equivalent to a sphere; the well-known examples are crystals and nematic liquid crystal droplets. In this work, we explore experimentally and theoretically the nuclei of columnar lyotropic chromonic liquid crystal coexisting with the isotropic phase that are toroidal in shape.  The geometry of toroids depends strongly on the molecular concentrations and presence of a crowding agent, polyethylene glycol.  High concentrations result in thick toroids with small central holes, while low concentrations yield thin toroids with wide holes. The multitude of the observed shapes is explained by the balance of bending elasticity and anisotropic interfacial tension.
\end{abstract}
\maketitle

\section{Introduction}

Surface tension defines the shapes of finite-size condensed matter. Tiny droplets of water in air are spherical to minimize their surface area, while solid crystals have facets due to orientational dependence of surface tension. Bulk interactions are irrelevant here: too weak to resist surface tension in the first example or too strong to permit internal curvatures in the second example. Liquid crystals show a more delicate balance between the bulk and surface energies yielding rich morphology of droplet shapes, such as smectic "batonnets" \cite{friedel1922etats}, nematic spindle-like tactoids \cite{bernal1941x}, branched \cite{wei2019molecular,peddireddy2021self} and dividing droplets \cite{lavrentovich1984division}. Although liquid crystal droplets are typically topologically equivalent to a sphere, here we explore toroidal droplets that form when a columnar liquid crystal with two-dimensional positional ordering coexists with its own isotropic melt \cite{tortora2010self}. 

The studied material is a lyotropic chromonic liquid crystal (LCLC) formed by plank-like molecules of disodium chromoglycate (DSCG) with hydrophobic polyaromatic cores and hydrophilic peripheries \cite{Oleg1,Oleg2,Oleg3,Oleg4,Oleg5}. When dispersed in water, the molecules form cylindrical aggregates by stacking face-to-face, Fig.~\ref{figS1}. Chromonic self-assembly is common for a broad family of materials, including nucleotides, dyes, food colorants, proteins, and pharmaceuticals, such as the anti-asthmatic and antiallergy drug disodium cromoglycate (DSCG). At sufficiently high concentrations, the aggregates align parallel to each other forming a nematic (N) phase. At still higher concentrations, the aggregates arrange into a hexagonal lattice producing a columnar (Col) phase, Fig.~\ref{figS1} \cite{Oleg2,Oleg3}. 

Phase transitions in LCLCs are controlled by both temperature and concentration. The isotropic (I)-Col phase transition exhibits a broad coexistence region in which Col nuclei are shaped as solid toroids \cite{tortora2010self} or spool-like structures in which the central hole shrinks into a singular line \cite{Oleg13}. These shapes resemble condensed nanoscale toroids of DNA strands in viral capsids \cite{Oleg15,Oleg16,Oleg17,Oleg18,Oleg19,Oleg20} albeit at much larger length scales of tens of micrometers \cite{Oleg13,tortora2010self} accessible to optical microscopy. The molecules within chromonic aggregates are bound by weak noncovalent forces so that chromonic aggregates can intersect, reconnect and exchange ends, avoiding entanglements known for DNA strands. 

In this work, we demonstrate that the toroidal shape of Col nuclei in the biphasic Col+I region of water dispersions of DSCG depends strongly on the concentrations $c$ of DSCG and  $C$ of a condensing agent polyethylene glycol (PEG).  PEG partitions into the I phase and helps to condense the Col phase \cite{tortora2010self}.  The increase of $c$ and $C$ yields larger toroids with narrow central holes and pronounced facets clearly revealed when the nuclei in a shape of handles are attached to the bounding glass plates. The observed shapes are explained by the balance of bend elasticity, defined by the bend modulus $K_3$, and the interfacial Col-I energy with a surface tension ${\sigma }_{\parallel }$. A dimensionless parameter $\beta =K_3/\left({\sigma }_{\parallel }V^{{1}/{3}}\right)$ measures the ratio of bending to interfacial energy and determines the shape of the cross-section of the toroidal aggregate. Orientational dependence of the interfacial tension on translations \textbf{u} within the hexagonal lattice produces faceted toroids.  The equilibrium toroid is thin with a wide central hole when $\beta \gg 1$; for $\beta \ll $1, the central hole shrinks dramatically, and the toroids resemble a faceted sphere. We first present the experimental data and then the mathematical model.
\begin{figure}[H]
\centering
    \includegraphics[width=4in]
                    {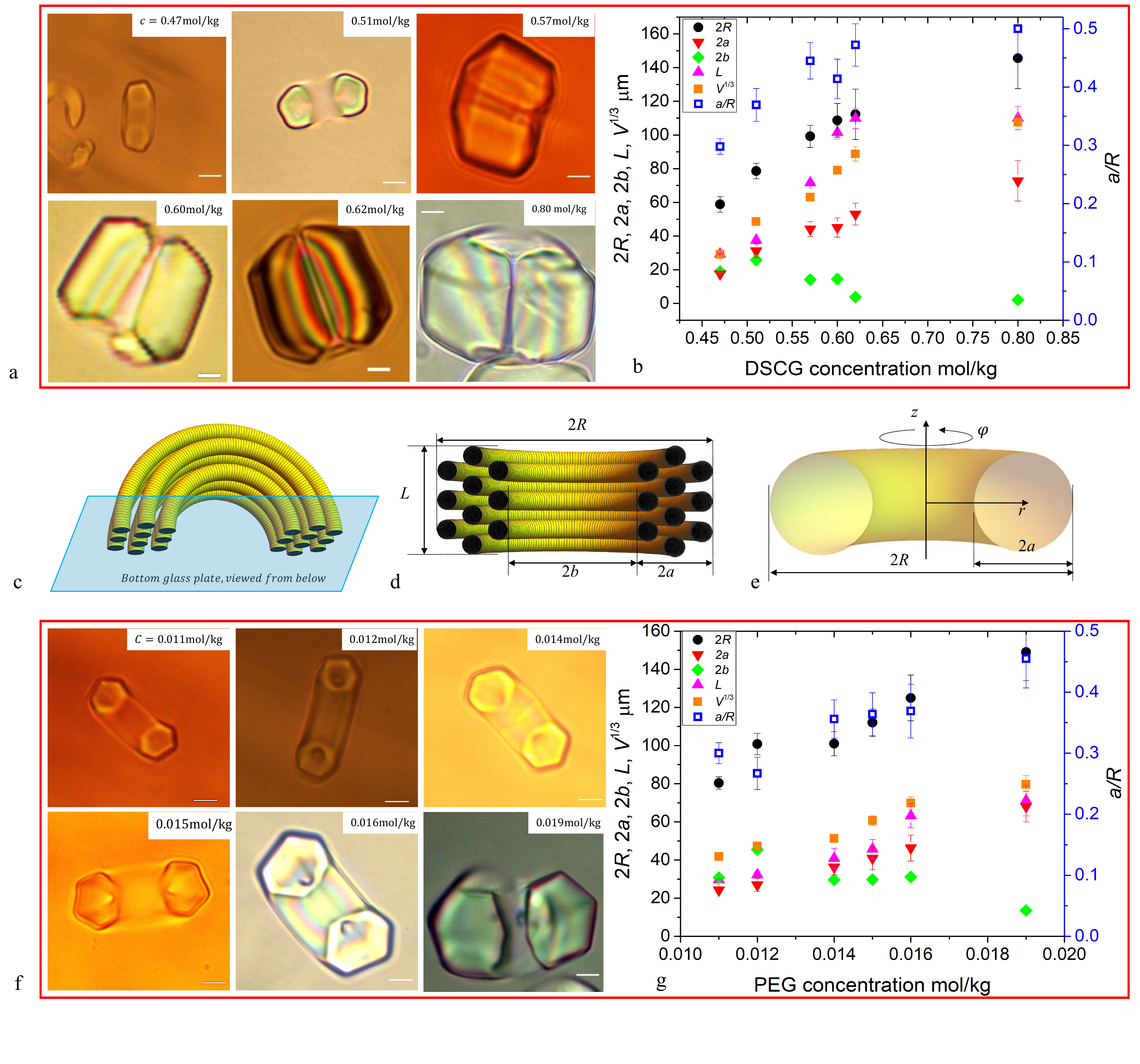} 
    \caption{(a) Optical microscopy textures of half-toroids at different concentrations of DSCG: \textit{c}=0.47, 0.51, 0.57, 0.60, 0.62, 0.80 mol/kg, scale bars $20\mu\mathrm{m}$. (b) The ratio $a/R$ (right axis, blue open square), maximum width $2R$, ``thickness'' of the toroidal cross-section $2a$, maximum opening $2b$ of the central hole, and the measure of volume $V^{\frac{1}{3}}$ as functions of $c$. Here $T=45\mathrm{{}^\circ C}$. (c) Scheme of the Col half-torus attached as a handle to the bottom glass substrate of the flat capillary; (d) Shape characteristics measured in the experiment; (e) Geometry of the toroid used in the analysis of the shape dependency on the material parameters. (f) Microscope images of half-toroids formed at different concentrations $C$ of PEG added to DSCG of a fixed concentration $c=0.34$ mol/kg: $C=0.011$, $0.012$, $0.014$, $0.015$, $0.016$, $0.019\ $mol/kg. Scale bars $20\mu\mathrm{m}$. (g) The ratio $a/R$ (right axis, blue open square), maximum width $2R$, ``thickness'' of the toroidal cross-section $2a$, maximum opening $2b$ of the central hole, and the measure of volume $V^{\frac{1}{3}},$ all plotted as a function of $C$.  Here $T=42\mathrm{{}^\circ C}$.}
  \label{olegfig4}
\end{figure}

\section{Experiment.}

\subsection{Toroids in aqueous DSCG and DSCG+PEG solutions}

The Col nuclei appear upon cooling from the I phase as thin flexible filaments that bend into toroids, Fig.~\ref{figS2}, to prevent the contact of open ends with the I phase, see Supplementary Material. Their shape depends on the concentration $c$ of DSCG, Fig.~\ref{olegfig4}a,b and $C$ of PEG, Fig.~\ref{olegfig4}f,g.  The nuclei are actually half-toroids with faceted cross-sections that are mostly attached to the bottom glass plate because of the homeotropic alignment of the director and gravity, Fig.~\ref{olegfig4}c. In Fig.~\ref{olegfig4}b,g we plot the concentration dependencies of the maximum extension $2R$ of half-toroids along the normal to the axis of bend, minimum width of the central opening $2b$, maximum width $2a$ of the solid part of the toroid, maximum extension $L$  of the toroid along the axis of bend, and the volume $V$ of half-torus, plotted as $V^{\frac{1}{3}}$.  All parameters, except for the hole width $2b$, increase with $c$ and $C$.  In pure DSCG, the hole width $2b$ decreases from $\approx 20\mu\mathrm{m}$ at $c=0.47$ mol/kg to $\le 1\mu\mathrm{m}$ at $c=0.8\ $mol/kg, Fig.~\ref{olegfig4}b. The ratio $a/R$ increases from $\mathrm{\sim}$0.3 to the maximum possible value 0.5, transforming the central hole into a singular +1 disclination coinciding with the axis of the toroid, Fig.~\ref{olegfig4}a, $c=0.62$ and $0.8\ $mol/kg. In the Col phase, this disclination is topologically stable \cite{Oleg29} and features a submicron core of a radius $a_c$.  

In condensates with a fixed at $c=0.34$ mol/kg, addition of PEG at concentration $C>0.011$ mol/kg causes phase separation into the Col and I phases, as confirmed by X-ray measurements \cite{Oleg37}. The Col inclusions are toroids with cross-sections resembling hexagons, Fig.~\ref{olegfig4}f. A distinct feature of DSCG+PEG system is that the central hole never shrinks into a disclination, as its smallest width $2b\approx 10\mu\mathrm{m}$ is relatively large, Fig.~\ref{olegfig4}g.

\begin{figure}[H]
\centering
    \includegraphics[width=3.5in]
                    {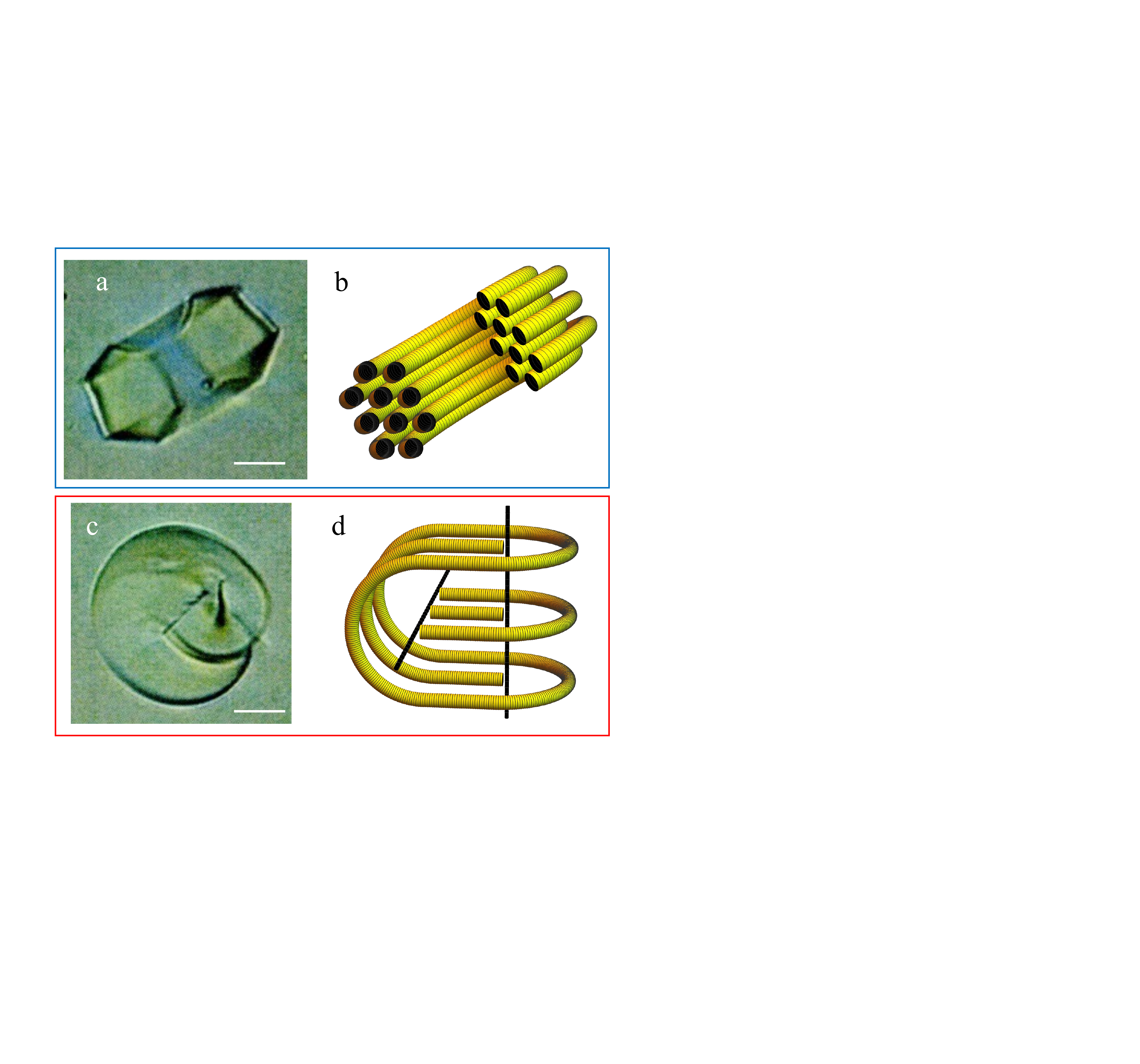} 
    \caption{(a) Optical microscopy texture and (b) inner structure of a freely suspended facetted toroid; (c) Optical microscopy texture and (d) inner structure of a compact domain with two mutually perpendicular disclinations of strength $1/2$.  Scale bars $20\ \mu$m. Here $T=50^\circ$C; DSCG, $c=0.55$ mol/kg.}
  \label{olegfig2M}
\end{figure}

It is important to stress that the homeotropic anchoring of Col aggregates at the glass substrate is not the reason for the existence of the facetted toroids, as these could be observed being freely suspended in very thick slabs, Fig.~\ref{olegfig2M}a,b. Among the freely-suspended shapes one can also observe compact domains, Fig.~\ref{olegfig2M}c, with two mutually perpendicular disclinations of strength $1/2$, Fig.~\ref{olegfig2M}d.  As suggested by Bouligand \cite{Oleg7}, the existence of two $1/2$ mutually perpendicular disclinations proves the hexagonal order (six-fold symmetry) of the Col phase; these crossed configurations were previously observed by Oswald in the bulk of a thermotropic Col phase \cite{Oleg9}.  The six-fold symmetry is an important factor shaping the Col nuclei in the studied system, as detailed in the numerical analysis below. Since their orientation is fixed, half-toroids are easier to analyze than their full counterparts; this is why we present the shape parameters in Fig.~\ref{olegfig4} for half-toroids.   Note finally that the homeotropic anchoring of LCLCs achieved in our study for the Col phase is not unique: a homeotropic alignment for the nematic phase of DSCG at a glass treated with N, N-dimethyl-N-octadecyl-3-aminopropyl trimethoxysilyl chloride (DMOAP) was reported by Nazarenko et. al. \cite{22} and by Zhou et. al. \cite{zhou2017dynamic}, while Mushenheim et. al. \cite{8302} reported the effect for graphene-coated glass plates.

\subsection{Elasticity-surface tension balance for thin toroids}

The toroidal shapes result from a balance of surface tension and bulk elasticity. To calculate the elastic energy of a toroid, we model it as a circular torus of a minor radius $a$  and a major radius $r$, associated with the geometrical parameters in Fig.~\ref{olegfig4}c as $r=R-a$ , Fig.~\ref{olegfig4}e. The elastic energy of half-torus is $F_e={\pi }^2K_3\left(r-\sqrt{r^2-a^2}\right)$, while the Col-I interfacial energy is $F_{s\parallel }=2{\pi }^2{\sigma }_{\parallel }ar$, where ${\sigma }_{\parallel }$ is the surface tension coefficient for tangential alignment of Col aggregates at the interface, assumed to be independent on the orientation of the hexagonal lattice. To enable analytical results, we consider thin toroids, $a\ll r$, in which case the surface energy at the Col-glass interface is insignificant as compared to $F_{s\parallel }$.

For  $a\ll r$ , the energy of the toroid simplifies to $F\approx {\pi }^2K_3r\xi /2+2{\pi }^2{\sigma }_{\parallel }r^2\sqrt{\xi }$, where  $\xi =\frac{V}{{\pi }^2r^3}\ll 1$. Minimizing $F$ with respect to $r$, for a fixed volume $V={\pi }^2a^2r=const$, one finds that $r={\left(\frac{V}{{\pi }^2}\right)}^{\frac{1}{5}}{\lambda }^{\frac{2}{5}}_{ec}$, where we introduce the elastocapillary length ${\lambda }_{ec}=\frac{K_3}{{\sigma }_{\parallel }}$. The result leads to$\ a={\left(\frac{V}{{\pi }^2}\right)}^{\frac{2}{5}}{\lambda }^{-\frac{1}{5}}_{ec}$ and
\begin{equation} \label{GrindEQ__1_} 
\frac{a}{r}=\frac{1}{{\pi }^{{2}/{5}}{\beta }^{{3}/{5}}},      
\end{equation} 
where
\begin{equation}
\label{eq:beta}
\beta =\frac{{\lambda }_{ec}}{V^{{1}/{3}}}=\frac{K_3}{{\sigma }_{\parallel }V^{{1}/{3}}}
\end{equation}
is a dimensionless ratio of the bulk bend energy to the surface energy normalized by the characteristic size of the nuclei. 

\begin{figure}[H]
\centering
   \includegraphics[width=4in,height=1.8in]
                    {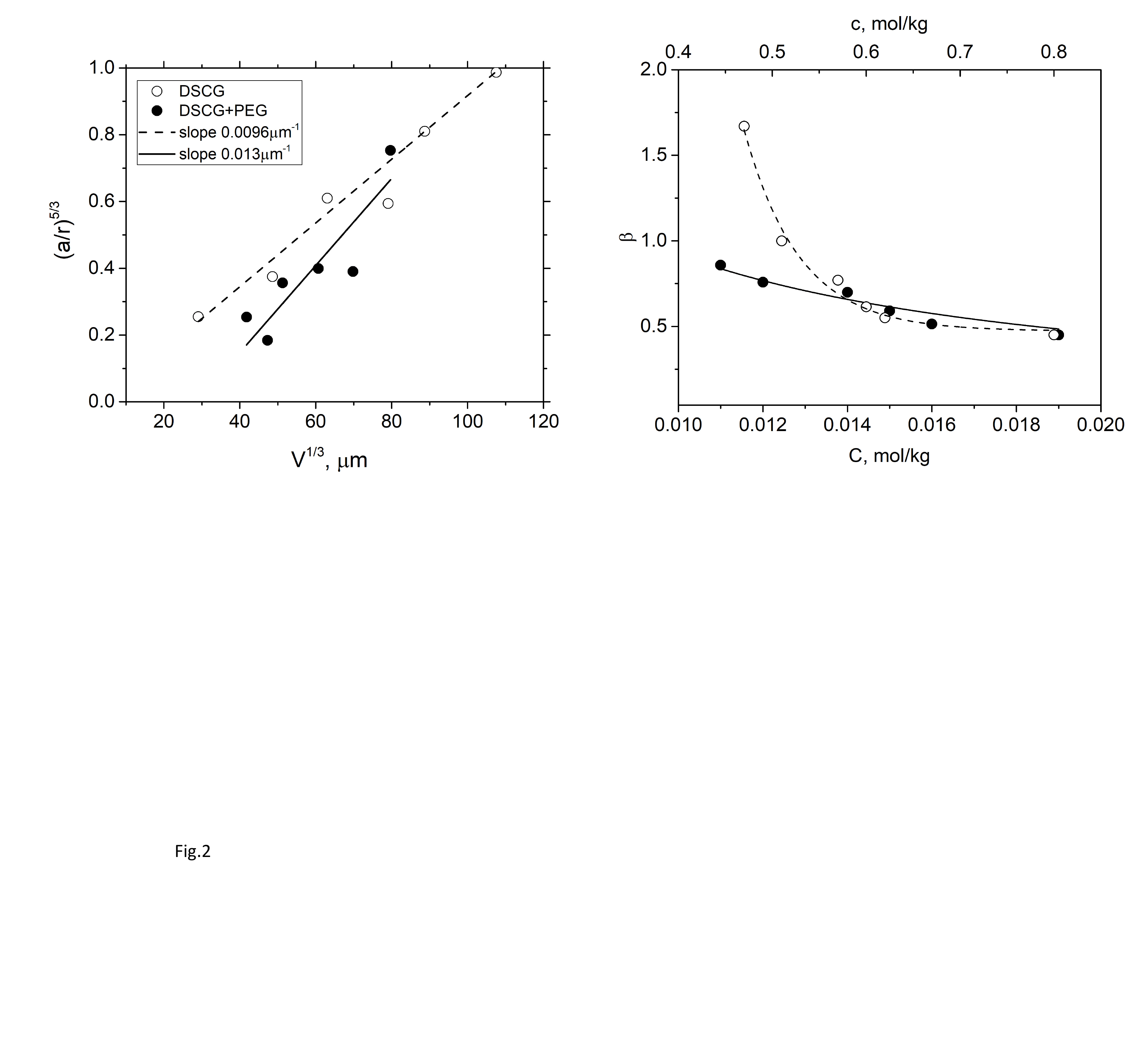} 
    \caption{Geometrical parameters of toroids in DSCG and DSCG+PEG mixtures. (a) Aspect ratio measure of toroids ${\left({a}/{r}\right)}^{\frac{5}{3}}$ is proportional to their characteristic size $V^{\frac{1}{3}}$; (b) the dimensionless parameter $\beta =K_3/\left({\sigma }_{\parallel }V^{{1}/{3}}\right)$ decreases when the concentrations $c$ and $C$ increase; the curves are guide to an eye, although in the case of DECG+PEG, the curve is close to $\beta \propto 1\mathrm{/}C$.}
  \label{olegfig6}
\end{figure}

Eq. \eqref{GrindEQ__1_} confirms the qualitative trends observed experimentally in Figs.~\ref{olegfig4}a,f, despite the limitation $a\ll r,\ R$ imposed on the model. The concentration dependencies of the geometrical parameters in Figs.~\ref{olegfig4}b,g show that $a/r$ increases with $V$, in agreement with Eq. \eqref{GrindEQ__1_}. The dependence of ${\left({a}/{r}\right)}^{\frac{5}{3}}$ vs. $V^{\frac{1}{3}}$, Fig.~\ref{olegfig6}a, is practically linear, which suggests that within the explored range of concentrations, the elastocapillary length ${\lambda }_{ec}=K_3/{\sigma }_{\parallel }$ does not change much, being on the order of tens of micrometers. For pure DSCG, least-square fitting yields ${\lambda }_{ec}=\left(49\pm 6\right)\ \mathrm{\muup }\mathrm{m}$, while for DSCG+PEG mixtures, ${\lambda }_{ec}=\left(35\pm 10\right)\ \mathrm{\muup }\mathrm{m}$. We estimated $\sigma_{||}\approx10^{-6}\,$J/m\textsuperscript{2} for Col-I interface of a pure DSCG at $c=0.47$ mol/kg by a spinning droplet technique \cite{Vonnegut}, see Supplementary Material. With $\sigma_{||}\approx10^{-6}\,$J/m\textsuperscript{2}, the fitted values of $\lambda_{ec}$ suggest that $K_3$ in the C phase is within the range of $20$-$60$ pN, which is reasonable since the highest measured value of $K_3$ in the N phase of DSCG is $50$ pN \cite{Oleg1.39}.

The dimensionless parameter $\beta$ defined in Eq. \eqref{eq:beta} and determined by the fitted $\lambda_{ec}$ and experimental values of $V$, decreases as $c$ and $C$ increase; in the case of the DSCG+PEG mixture, $\beta \propto 1/C$, while the dependency is steeper with respect to $c$, Fig.~\ref{olegfig6}b.  According to Eq. (1), a smaller $\beta $ means that $a/r$ increases, which is intuitively clear. A smaller ${\lambda }_{ec}$, $K_3$ and $\beta$ imply that the surface energy cost is high, thus the central hole shrinks to make the toroid more round and to reduce the interfacial area; the corresponding $a/r$ is large. Larger ${\lambda }_{ec}$ and $\beta$ mean a higher $K_3$; the higher elastic cost of bend is relieved by expanding the central hole, i.e., by reducing $a/r$.  

To summarize this section, the experiments uncover a rich morphology of Col toroids coexisting with the I phase. The shapes of toroids depend strongly on the concentration  $c$ of DSCG and concentration $C$ of the crowding agent PEG and shows the following trends. 

1) The cross-sections of the toroids are faceted because of hexagonal packing of the chromonic aggregates and anisotropy of the Col-I interfacial tension. The cross-section is not strictly hexagonal, with facets showing different lengths.  

2) The volume $V$ and the ratio of the minor radius $a$ to the major radius $r$ of toroids increase strongly with $c$ and $C$.  The central hole is large at low $c$ and $C$, but shrinks towards a singular line with circular bend of columns at high concentrations. 

3) A model of thin toroids that (i) assumes the interfacial tension to be independent of the orientation of the hexagonal lattice and (ii) neglects the contact with glass substrates, predicts that the ratio $a/r$ increases when $V$ increases or the elastocapillary length ${\lambda }_{ec}=K_3/{\sigma }_{\parallel }$ decreases. In liquid crystals the elastic energy of director gradients $K_3V^{1/3}\ $ scales with the linear size $V^{1/3}$ of the system. Since the interfacial tension energy scales with the surface area, ${\sigma }_{\parallel }V^{2/3}$, the elastic energy prevails for small nuclei at low concentrations, producing skinny tori with a large central core. The interfacial surface tension prevails for large nuclei and high concentrations, yielding shapes close to spherical with a shrunk central hole; these could not be described by a simple model with $a\ll r$.  

In the next section, we introduce a mathematical model of toroids that for any $a/r$ accounts for anisotropy of the I-Col interfacial tension, faceted shapes, and surface tension at the glass substrate.   

\section{Modeling and Simulations}
\label{Section 4}

\subsection{A model of toroidal columnar nuclei}
\begin{figure}[H]
     \centering
     \begin{subfigure}[c]{0.5\textwidth}
         \centering
         \includegraphics[width=\textwidth]{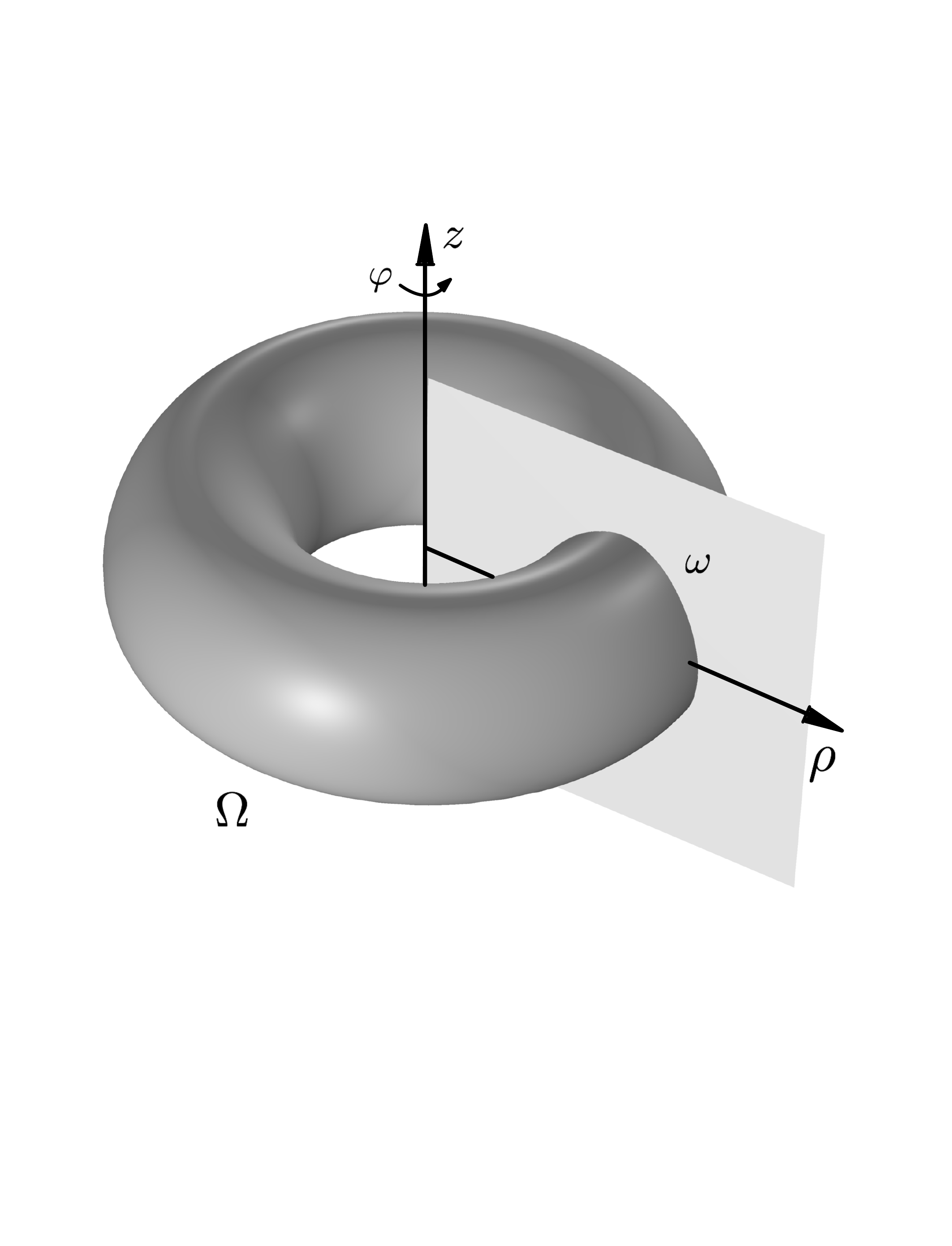}
         \caption{}
     \end{subfigure}
     \qquad
     \begin{subfigure}[c]{0.3\textwidth}
         \centering
         \includegraphics[width=\textwidth]{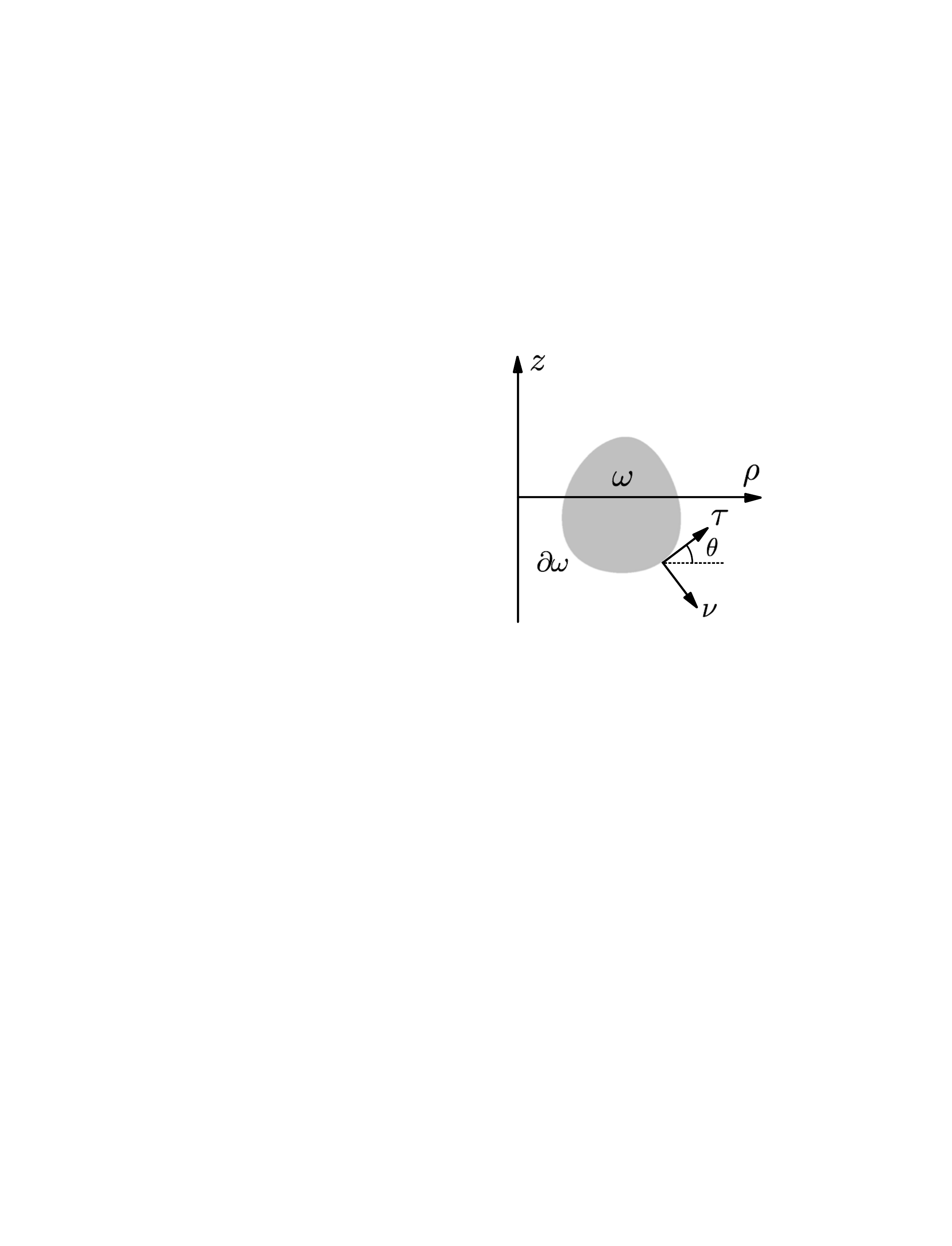}
         \caption{}
     \end{subfigure}
     \caption{Geometry of the problem: (a) the toroidal Col nucleus $\Omega$; (b) cross-section of $\Omega$ by a vertical plane gives a two-dimensional domain $\omega$ with the boundary $\partial\omega$.}
        \label{fig:setup}
\end{figure}

Suppose that $(\rho,\phi,z)$ are cylindrical coordinates in $\mathbb R^3$ and let $$\omega\subset\mathbb R^2_+:=\left\{(\rho,z)\subset \mathbb R^2:\rho>0\right\}$$ be a simply-connected domain in the right half of the $\rho z$-plane. Let the Col nucleus $\Omega$ be axially symmetric (Fig.~\ref{fig:setup}) and obtained by rotating $\omega$ around the $z$-axis, \[\Omega:=\left\{(\rho,\phi,z)\subset\mathbb R^3:(\rho,z)\in\omega,\ \phi\in[0,2\pi)\right\}.\]

The Col nucleus is composed of circularly bent "columns" centered on and lying in planes perpendicular to the $z$-axis. Each cross-section of the nucleus by a plane that contains the $z$-axis reveals a triangular lattice of points corresponding to the cross-sections of the columns; we assume that this lattice is fixed with one of the corresponding two-dimensional lattice vectors being parallel to the $z$-axis. The deformation of the chromonic columns is therefore limited to bending. The bending energy of a given column is proportional to the square of the column curvature so that
\begin{equation}
\label{eq:bending}
E_b:=\frac{K_3}{2}\int_\Omega \rho^{-2}dV=\pi K_3\int_\omega \rho^{-1}dA.
\end{equation}
where $dV=\rho\,d\rho\,dz\,d\phi$ and $dA=d\rho\,dz$.

The Col-I interfacial tension depends on orientation of the columnar lattice vectors with respect to the surface normal. Because of the rotational invariance, the surface energy density can only depend on the relative angle between one of the lattice vectors and the surface normal. The outward normal to $\partial\Omega$ at a given point coincides with a normal at the same point to an appropriately rotated $\partial\omega$. It follows that the surface energy density is a function $\sigma:\mathbb S^2\to\mathbb R$ of $\bnu$, where $\bnu$ is normal to both $\partial\Omega$ and $\partial\omega$. The surface energy of the columnar chromonic nucleus is then given by
\begin{equation}
\label{eq:surface}
E_s:=\int_{\partial\Omega} \sigma(\bnu)dS=2\pi \int_{\partial\omega} \sigma(\bnu)\rho\,ds.
\end{equation}

We seek the optimal shape of Col nuclei that minimizes the total energy functional 
\begin{equation}
\label{eq:esum}
E[\omega]:=E_b+E_s=\pi K_3\int_\omega \rho^{-1}dA+2\pi \int_{\partial\omega} \sigma(\bnu)\rho\,ds,
\end{equation}
among $\omega\subset\mathbb R^2_+$ that satisfy a fixed volume constraint $\mathrm{Vol}(\Omega)=V>0$ or
\begin{equation}
\label{eq:volc}
V=2\pi\int_\omega \rho\,dA.
\end{equation}
We nondimensionalize the problem, 
\[\tilde \rho=\frac{\rho}{V^{1/3}},\ \tilde z=\frac{z}{V^{1/3}},\ \tilde s=\frac{s}{V^{1/3}},\ \tilde \sigma=\frac{\sigma}{\sigma_{||}},\ \tilde E=\frac{E}{\pi\sigma_{||}V^{2/3}},\]
where $\sigma_{||}>0$ is a reference surface energy density. Then, dropping tildes and using the same symbols for the rescaled domain $\omega$, we have
\begin{equation}
\label{eq:etot}
E[\omega]=\beta\int_\omega \rho^{-1}dA+2 \int_{\partial\omega} \sigma(\bnu)\rho\,ds-4\lambda\int_\omega \rho\,dA,
\end{equation}
where the nondimensional $\beta$ describes the relative contribution of the bulk bending and surface energies and $\lambda$ is the Lagrange multiplier corresponding to the constraint \eqref{eq:volc}.

Now suppose that the boundary curve
\[\partial\omega=\left\{\br(s)=(\rho(s),z(s))\in\mathbb R_+^2:0\leq s<L,\ \br(L)=\br(0),\ \br^\prime(L)=\br^\prime(0)\right\}\]
is positively oriented and parametrized with respect to the arc length $s$, where $L>0.$ Suppose further that $\br$ is smooth everywhere except for some $s=c$ where $\theta$ experiences a jump $[\theta]_{x=c}:=\theta_+(c)-\theta_-(c)=\lim_{x\to c^+}\theta(x)-\lim_{x\to c^-}\theta(x)$. Here we define the orthonormal frame $(\btau(s), \bnu(s))$ as
\begin{gather}
\btau(s)=\br^\prime(s)=(\cos{\theta(s)},\sin{\theta(s)}),\label{eq:tau1}\\
\bnu(s)=\br^\prime_\perp(s)=(\sin{\theta(s)},-\cos{\theta(s)}), \label{eq:nu1}
\end{gather}
where $\theta(s)$ is the angle between the tangent to the curve and the positive direction of the $\rho$-axis for all $s\in[0,L]$ and ${\mathbf a}_\perp=(a_2,-a_1)$ for every ${\mathbf a}=(a_1,a_2)$. Consider a general smooth variation $\delta\br$ of $\br$, then the same procedure as presented in Supplementary Material for smooth curves result in the following weak form of the Euler-Lagange equation
 \begin{equation}
\label{eq:weak}
\int_0^L\left\{\sigma(\bnu)\,{\mathbf e}_\rho\cdot\delta\br+\rho\left[\sigma(\bnu)\btau-\left(\nabla\sigma(\bnu)\cdot\btau\right)\bnu\right]\cdot\delta\br^\prime\right\}\,ds+\beta \int_0^L\rho^{-1}\bnu\cdot\delta\br\,ds-2\lambda \int_0^L\rho\bnu\cdot\delta\br\,ds=0.
\end{equation}
Standard arguments involving integration by parts, the strong form of the Euler-Lagrange equation, and continuity of the variation $\delta\br$ give the appropriate jump condition
\begin{equation}
\label{eq:jump0}
\left[\sigma(\bnu)\btau-\left(\nabla\sigma(\bnu)\cdot\btau\right)\bnu\right]_{s=c}=0,
\end{equation}
or
\begin{equation}
\label{eq:jump}
\left[\gamma\btau-\gamma_\theta\bnu\right]_{s=c}=0,
\end{equation}
when the condition is written in terms of \[\gamma(\theta):=\sigma\left(\sin{\theta},-\cos{\theta}\right).\] Up to the definition of the orthonormal frame and the weight $\rho$, this corresponds to a standard condition of continuity of the capillary force
\begin{equation}
\label{eq:capillary}
{\mathbf C}(\theta)=\rho\left(\gamma(\theta)\btau(\theta)-\gamma_\theta(\theta)\bnu(\theta)\right)
\end{equation}
across a corner \cite{GA89}. 

To gain some insight into \eqref{eq:jump} we follow the exposition in \cite{GA89} reviewed in Supplementary Material. Let ${\mathbf g}(\theta)=\gamma(\theta)^{-1}\bnu(\theta)$ be a Frank potential associated with the surface energy density $\gamma$. The polar plot of ${\mathbf g}$ is the so-called Frank diagram \cite{frank1963metal}. The {\it capillary force} 
\[{\mathbf g}_\theta(\theta)=\frac{1}{\gamma^2(\theta)}{\mathbf C}(\theta),\]
is tangent to the Frank diagram. As demonstrated in \cite{GA89}, a range of angles $[\theta_-,\theta_+]$ corresponds to a corner if and only if the points ${\mathbf g}(\theta_-)$ and ${\mathbf g}(\theta_+)$ on the Frank diagram share a common tangent line (a Maxwell line). 

The full problem for the unknown $(\rho,\theta,z,L)$ satisfied by the energy-minimizing curve is given by the system of ODEs
\begin{equation}
\label{eq:system1}
\left\{
\begin{array}{l}
 \rho\left(\gamma_{\theta\theta}+\gamma\right)\theta^\prime+\gamma\sin{\theta}+\gamma_\theta\cos{\theta}+\frac{\beta }{\rho}-2\lambda\rho=0,    \\
\rho^\prime=\cos{\theta},   \\
z^\prime=\sin{\theta},   
\end{array}
\right.
\end{equation}
subject to the conditions
\begin{equation}
\label{eq:cond1}
\theta(L)=\theta(0)+2\pi,\quad \theta^\prime(L)=\theta^\prime(0),\quad \int_0^L\cos{\theta}\,ds=\int_0^L\sin{\theta}\,ds=0, \quad 2\pi\int_\omega \rho\,dA=1,
\end{equation}
as described in Supplementary Material.

In addition to considering full toroidal configurations, we also explore half-toroids attached to a glass substrate by considering the modified energy over $\omega,$ where
\begin{equation}\label{eq:total_energy_h}
	E_h [\omega] = \frac{\pi K_3}{2} \int_{\omega} \rho^{-1} dA + 2\sigma_g\int_{\omega} dA +\pi\int_{\partial \omega} \sigma (\bnu) \rho\, ds,
\end{equation}
among all $\omega$ for which the corresponding toroid $\Omega$ has a prescribed volume. Here $\sigma_g$ is the surface energy density of a chromonics/glass interface. With a slight abuse of notation, the nondimensional expression for $E_h$ is given by
\begin{equation}\label{eq:total_energy_h}
	E_h [\omega] = \frac{\beta}{2} \int_{\omega} \rho^{-1} dA + \frac{2\chi}{\pi} \int_{\omega} dA +\int_{\partial \omega} \sigma (\bnu) \rho\, ds,
\end{equation}
with the nondimensional parameter $\chi$ defined as
\[\chi=\frac{\sigma_g}{\sigma_{||}}.\]
The system of Euler-Lagrange ODEs for \eqref{eq:total_energy_h} is similar to \eqref{eq:system1}.

Next, we discuss behavior of solutions of \eqref{eq:system1}-\eqref{eq:cond1}. We use simulations to find stationary points of \eqref{eq:etot} via a gradient flow for this energy. In order to be able to simulate curves with corners, we use the standard regularization technique by adding a curvature-penalizing term to \eqref{eq:etot}. The numerical scheme is described in the {\it Materials and Methods} section.

\subsection{Numerical results}
All simulations in this section were done using FELICITY \cite{Walker_SJSC2018}. We first simulate the shapes of toroids for the anisotropic surface tension
\[\sigma(\bnu)=\sigma_{||}+\sigma^{\,a}_{||}\sin^2(3\theta),\]
where the relationship between $\bnu$ and $\theta$ is given by \eqref{eq:nu1}. This expression can be rewritten in a nondimensional form
\begin{equation}
\label{eq:anis}
\gamma(\theta)=1+\gamma_1\sin^2(3\theta),
\end{equation}
where $\gamma_1=\sigma^{\,a}_{||}/\sigma_{||}.$  

\begin{figure}[H]
\centering
    \includegraphics[width=5in]{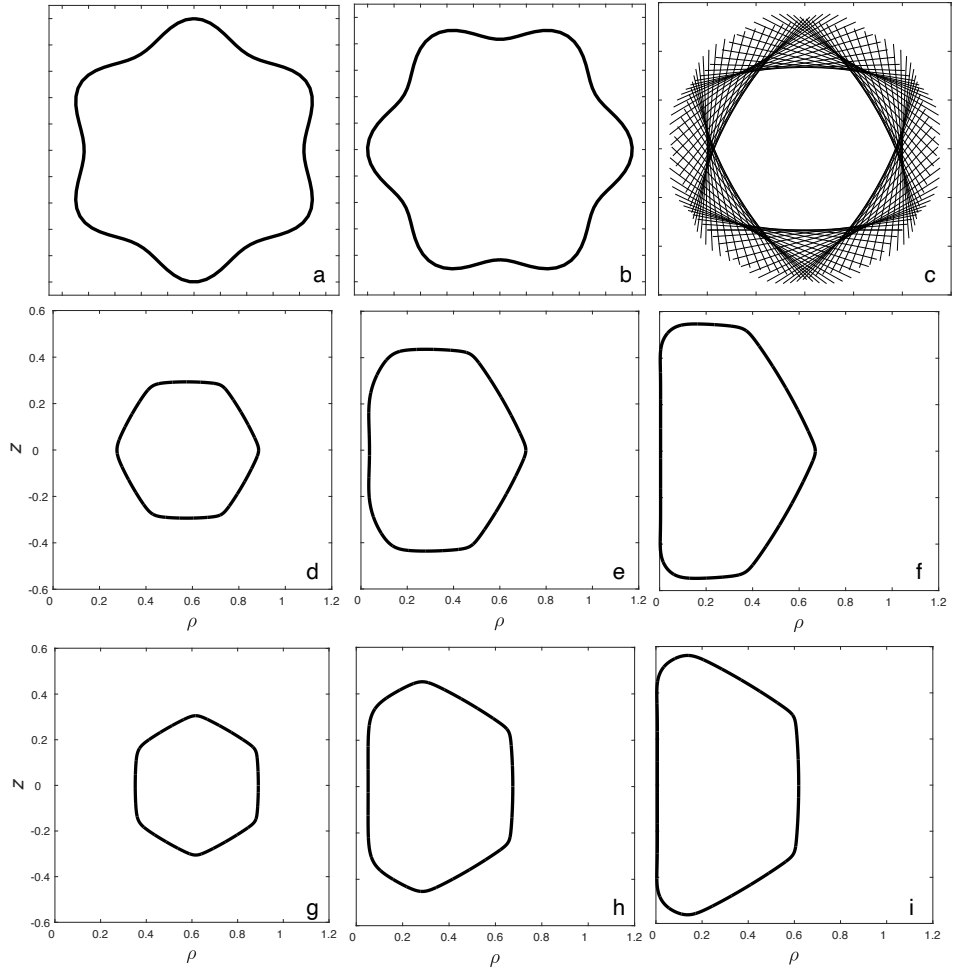}
    \caption{Numerical simulations of the toroidal shapes for the anisotropic interfacial energy $\gamma(\theta)=1+\gamma_1\sin^2(3\theta)$ with $\gamma_1=0.2$ (a-f) and $\gamma_1=-0.2$ (g-i); (a) Frank diagram; (b) Wulff plot; (c) Wulff construction; (d-f) cross-sections of toroids minimizing the sum of the elastic and interfacial energy for $\gamma_1=0.2$ where (d) $\beta=1$, (e) $\beta=0.1$, and (f) $\beta=0.01$; (g-i) the same as in (d-f), but $\gamma_1=-0.2$. In these simulations the regularization parameter $\varepsilon=$ 5e-4 (See Section~\ref{sec:num}).}
  \label{fig6}
\end{figure}

Figs.~\ref{fig6}a,b,c show the Frank diagram, Wulff plot, and Wulff construction, respectively, corresponding to \eqref{eq:anis} with $\gamma_1=0.2$ in the absence of bend elasticity. Note that the Frank diagram is not convex and there are six Maxwell lines that indicate that the equilibrium shape must have six facets and six corners. The Wulff construction in Fig.~\ref{fig6}c produces shape that is close to the one obtained by accounting for both the surface tension and bend elasticity, when $\beta$ is relatively large. For large $\beta$, the major radius $r$ of the torus is large while its minor radius $a$ is small in order to accommodate significant bending energy and the volume constraint. Because $a/r\ll1$, the variation of $\rho$ across the cross-section is smaller than $\rho$ itself so that $\rho$ is essentially constant. The shape of the cross-section in Fig.~\ref{fig6}d is close to that obtained via the Wulff construction in Fig.~\ref{fig6}c, however, the corners in the hexagon are rounded due to regularization employed in the gradient flow simulations. On the other hand, decreasing $K_3$ and $\beta$ or increasing $\sigma_{||}$ and volume, shrinks a wide "donut hole" into a narrow central core (Fig. \ref{fig6}e,f). We also observe that the corners and facets facing the $z$-axis become more rounded with decreasing $\beta$ and the part of the curve closest to the $z$-axis transform into a large facet.

Changing the sign of $\gamma_1$ by setting $\gamma_1=-0.2$ in \eqref{eq:anis} corresponds to rotating the plots in Fig.~\ref{fig6}a,b,c by $\pi/3$, so that a facet faces the $z$-axis instead of a corner. Upon decreasing $\beta$ and $K_3$ or increasing $S\sigma_{||}$, in Fig.~\ref{fig6}g,h,i this facet expands and approaches progressively closer to the $z$-axis.

\begin{figure}[H]
\centering
    \includegraphics[width=3in]{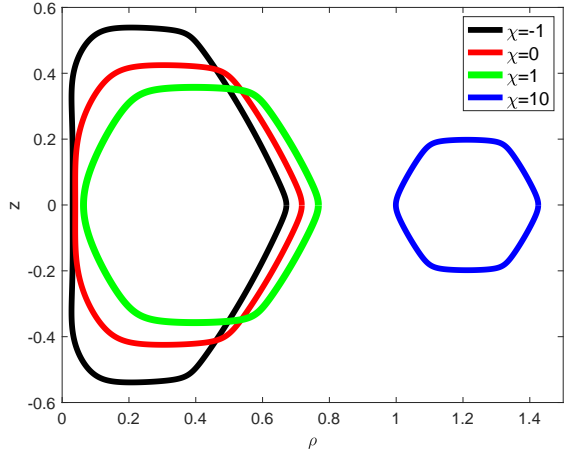}
    \caption{Cross-sections of the energy-minimizing half-toroids for the anisotropic surface energy $\gamma(\theta)=1+\gamma_1\sin^2(3\theta)$ with $\gamma_1=0.2,$ $\beta=0.06$ and $\varepsilon=10^{-3}$.}
  \label{fig13}
\end{figure}

Decreasing $|\gamma_1|$ produces produces a convex Frank diagram and less faceted shapes, Figs.~\ref{fig6.1}-\ref{fig9.1}. 
Fig.~\ref{fig13} demonstrates how the shape of {\it half-toroids} is influenced by the surface tension of the chromonic/substrate interface. Here we use gradient flow for the energy \eqref{eqn:total_energy_h} to obtain equilibrium shapes. The interfacial I-Col energy for the director being normal to the I-Col interface is estimated (see Supplementary Material) to be larger than $10^{-4}\,$N/m. Thus the dimensional glass-Col anchoring energy $\sigma_{||}\,\chi$ must be lower than this number; a good starting estimate is $10^{-6}\,$N/m, i.e., $|\,\chi|$ is on the order of 1. Qualitative argument in the Supplementary Material shows that $\chi$ might be either positive or negative. As $\chi$ increases, the principal radius of the toroid increases, while it becomes more "skinny", Fig. \ref{fig13}.

\section{Discussion and Conclusions}
The numerical model above captures well the faceted shapes of toroids. The sign of $\gamma_1$ defines qualitatively different shapes: $\gamma_1<0$ yields an outermost facet parallel to the axis of the nucleus, Fig.~\ref{fig6}g,h,i, while $\gamma_1>0$ corresponds to two outer facets separated by a corner, Fig.~\ref{fig6}d,e,f.  In the experiment, pure DSCG shows a prevalence to form  outermost facets parallel to the axis of the nucleus, Fig.~\ref{olegfig4}a, while DSCG+PEG shows facets separated by a corner, Fig.~\ref{olegfig4}f.  The prevalence is not absolute, as in each case, the preferred shapes are observed with a probability of about 70\%. It shows that both  $\gamma_1<0$ and $\gamma_1>0$ correspond to the minima of the surface tension with only a small difference in the depth. We selected $\gamma_1=-0.2$ for pure DSCG and $\gamma_1=0.2$ for DSCG+PEG as the best match with the facets observed experimentally. 

\begin{figure}[H]
\centering
    \includegraphics[width=5in]{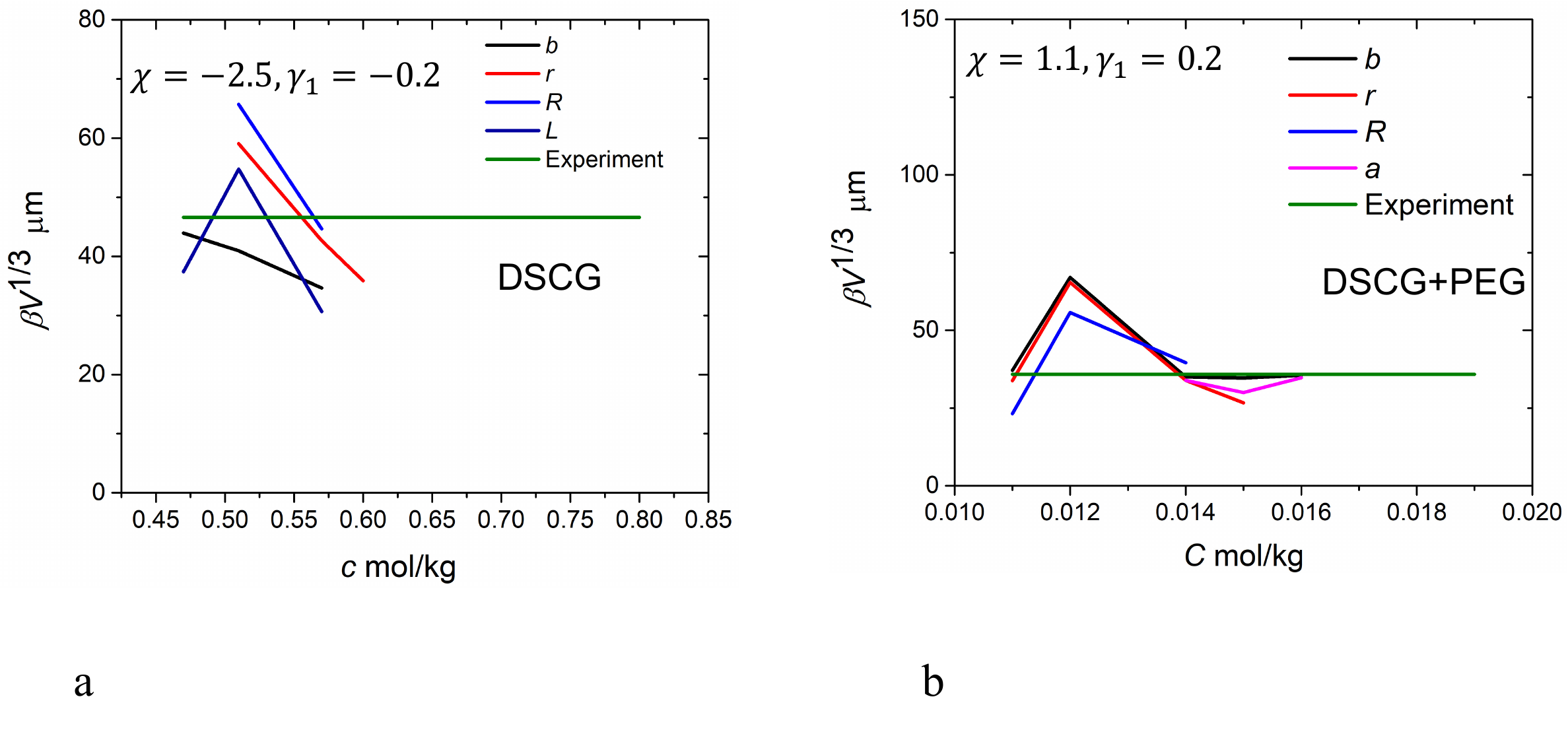}
    \caption{Comparison of simulated $\beta V^{1/3}$ to the experimentally obtained values for (a) pure DSCG and (b) DSCG+PEG. The best match between the simulations and the experimental results is achieved when $\chi=-2.5$ and $\gamma_1=-0.2$ for pure DSCG and $\chi=1.1$ and $\gamma_1=0.2$ for DSCG+PEG.}
  \label{fig6dis}
\end{figure}

Numerical simulations find the dimensionless shape parameters $a/V^{1/3}$, $r/V^{1/3}$, $R/V^{1/3}$, $b/V^{1/3}$, and $L/V^{1/3}$ as functions of $\beta$ for fixed $\gamma_1$ and $\chi$ and the concentrations $c$ and $C$. The numerical dependencies are compared to the experimental data to extract the matching $\beta$.  These values are used to calculate the elastocapillary length $\lambda_{ec}=K_3/\sigma_{||}=\beta V^{1/3}$ and to plot it as a function of $c$, Fig.~\ref{fig6dis}a, and $C$, Fig.~\ref{fig6dis}. Different shape parameters produce somewhat different values of $\beta$ and $\lambda_{ec}$. In some cases, $\beta$ could not be determined because the simulated shape parameters are outside of the range of positive-definite $\beta$. For this reason, $a/V^{1/3}$ data is not included in Fig.~\ref{fig6dis}a. The scatter of data is natural since the experimental textures show a range of shapes even at the fixed $c$ and $C$. The closest agreement of the experimental and numerical dependencies $\lambda_{ec}(c,C)$ is chosen to estimate $\gamma_1$ and $\chi$.
For DSCG, the closest correspondence is found for $\chi=-2.5$ and $\gamma_1=-0.2$ as shown in Fig.~\ref{fig6dis}.  A departure from these values causes a disagreement between the simulations and the experiment; an example for $\chi=1.0$ and $\gamma_1=-0.2$ is illustrated in Supplementary Material, Fig.~\ref{figS21n}a. For DSCG+PEG, the simulation matches the experiments best (Fig.~\ref{fig6dis}b) when $\chi=1.1$ and $\gamma_1=0.2$. When $\chi$ departs from this value, the discrepancy grows; see Supplementary Material, Fig.~\ref{figS21n}b for $\chi=-4.5$ and $\gamma_1=0.2$. The values $\chi=-2.5$ and $1.1$ are within the expected range, as discussed in the Supplement \ref{sec:supest}, see Supplementary Material, Fig.~\ref{figS2n}.
In conclusion, we described the unusual faceted toroidal shapes of the columnar nuclei coexisting with the isotropic phase of a lyotropic chromonic liquid crystal. The shapes are reminiscent of the DNA condensates but occur at a much larger scale, suitable for optical microscopy observation. Experiments show that the toroidal shapes depend strongly on the concentration of the chromonic molecules in the aqueous solutions. Theoretical and numerical analysis demonstrates that the faceted shapes result from the anisotropy of the interfacial tension, associated with different orientations of the hexagonal lattice at the interface. The facets are not equal in length, as the ones closer to the center of toroids tend to be elongated more than the facets farther away from the center. Numerical simulations demonstrate that the bending elasticity of the columns is the primary cause of this behavior. The balance of bending energy with the elastic constant $K_3$ and anisotropic interfacial energy $\sigma_{||}$, expressed as the elastocapillary length $\lambda_{ec}=K_3/\sigma_{||}$ controls the width of the central opening of the toroids, which in the experiments varies from tens of micrometers to submicrometer. Larger openings and skinny toroids, observed at low concentrations of the chromonic molecules and crowding agent, are facilitated by a smaller interfacial tension $\sigma_{||}$, smaller volume of the nuclei and a larger bend constant $K_3$, while the opposite behavior with toroids approaching faceted spheres, is observed at high concentrations. Although the experiments were performed for half-toroids attached to the glass plates of the cells, the general trends are expected to be the same for full toroids.  

\section{Materials and Methods} 

\subsection{Experiments}

Disodium cromoglycate (DSCG), Fig.~\ref{figS1}, with a purity of 98\% was purchased from Alfa Aesar and used without further purification. De-ionized water with resistivity $\ge 18.0\ \mathrm{M}\mathrm{\Omega }\mathrm{cm}$ was used to prepare the aqueous solutions of DSCG. At $\mathrm{25{}^\circ C}$, a homogeneous nematic (N) phase appears when the concentration of DSCG, $c$, is in the range of $0.3\le c\le 0.65\ \mathrm{mol/}\mathrm{kg}$ ($13.5\ \le c\le 25\ \mathrm{wt\%}$). In the range $0.65\le c\le 0.70\ \mathrm{mol/kg}\ \ (25\le c\le 26.5\ \mathrm{wt\%\ })$, the nematic + columnar (N+Col) coexistence region appears, and at $c>0.70\ \mathrm{mol/kg}$, the solution is in the homogeneous columnar (Col) phase \cite{Oleg23}. At T=45$\mathrm{{}^\circ C}$ , the solution remains in the I phase when $c\le $0.40 mol/kg ($\le $17wt\%); above this concentration, the Col and I phases coexist.  The liquid crystalline regions exhibit a higher DSCG concentration compared to the overall concentration in the entire sample  \cite{tortora2010self,Oleg24,Oleg25,Oleg37}. Fig.~\ref{figS1} schematizes the molecular arrangements within the aggregates and their mutual alignments in the I, N, and Col phases. 

Addition of the crowding agent PEG to aqueous solutions of DSCG expands the temperature range of the biphasic regions and condenses Col nuclei, Fig.~\ref{figS2}.  For example, an isotropic DSCG solution with $c=0.34\ \mathrm{mol/kg}$ (15 wt\%) at $\mathrm{45}\mathrm{{}^\circ }\mathrm{C}$ transforms into a biphasic I+Col state when PEG is added at the concentrations $C=\left(0.03-0.07\right)\ \mathrm{mol/kg}$  \cite{tortora2010self,Oleg37}. We use PEG with a molecular weight of 3.35 kg/mol (Sigma Aldrich). The gyration diameter of the PEG molecules $2r_g\approx 4.4\ \mathrm{nm}$ is larger than the inter-columnar separation $\approx (1.9-3.2)\ \mathrm{nm}$ in the columnar phase of DSCG \cite{Oleg3,tortora2010self}. The PEG molecules are thus expelled from the liquid crystalline regions and partition into the isotropic regions \cite{tortora2010self}.  

The facets of toroids are best observed when the axis of rotational symmetry are in the plane of the view.  A tangential alignment of the director $\widehat{\boldsymbol{\mathrm{n}}}$ at bounding plates does not allow one to observe the faceted structures clearly. Because of this, we use clean glass rectangular capillaries (purchased from VitroCom) that yield a homeotropic alignment after cleaning in an ultrasonic bath for 15 minutes at 60$\mathrm{{}^\circ C}$ followed by an isopropanol rinse and drying in an oven for 10 min at 80$\mathrm{{}^\circ C}$. The capillaries are 0.2 mm thick and 4.0 mm wide. After filling with the LC, the two ends of the capillary are sealed with an epoxy glue to prevent evaporation of water. The samples are placed inside a hot stage (Linkam Model PE94) and observed using an optical polarizing microscope (Nikon Optiphot 2 POL) in the transmission mode with parallel polarizers. The materials are cooled down from the homogeneous I phase to the biphasic I+Col region at a rate of $0.1\mathrm{{}^\circ C/min}$ and then the temperature of the sample is fixed, either $\mathrm{45{}^\circ C}$  (DSCG samples) or at $\mathrm{42{}^\circ C}$ (DSCG+PEG). The Col nuclei form half-toroidal handles, Figs.~\ref{olegfig4},\ref{figS2}, attached to the bottom glass plate of the capillary as shown in Fig.~\ref{olegfig4}c since their mass density is higher than that of the I phase \cite{tortora2010self}. The Col nuclei were explored 20 min after the cooling stopped and the temperature fixed, to ensure that they have reached their stationary state.

\subsection{Numerical method}
\label{sec:num}
\subsubsection{Computing energy minimizers via gradient flow for the regularized energy}\label{sec:compute_energy_min} 

We seek to minimize the energy $E [\omega] = 2 \pi J [\omega]$ over $\omega$ where
\begin{equation}\label{eqn:total_energy}
	J [\omega] = \frac{\beta}{2} \int_{\omega} \rho^{-1} dA + \int_{\partial \omega} \sigma (\bnu) \rho\, ds + \frac{\varepsilon}{2} \int_{\partial \omega} \kappa^2\,ds,
\end{equation}
subject to the constraint that the volume is fixed; note that $\varepsilon \geq 0$ is a regularization parameter.  The associated Lagrangian is
\begin{equation}\label{eqn:lagrangian}
	\cL [\omega,\lambda] = J [\omega] + \lambda \left( C_v - \int_{\omega} \rho dA \right),
\end{equation}
where $C_v > 0$ is the desired volume (without the factor of $2 \pi$).

Note that one can rewrite \eqref{eqn:lagrangian} entirely in terms of $\Gamma := \partial \omega$.  By Gauss' divergence theorem (in the plane), we have
\begin{equation}\label{eqn:reduce_integrals_to_boundary}
	\int_{\omega} \rho^{-1}\, dA = \int_{\Gamma} (\bnu \cdot {\mathbf e}_1) \log \rho \, ds, \qquad \int_{\omega} \rho\, dA = \frac{1}{2} \int_{\Gamma} (\bnu \cdot {\mathbf e}_1) \,\rho^2 \, ds.
\end{equation}
Therefore, $J[\omega] \equiv J[\Gamma]$ and $\cL [\omega,\lambda] \equiv \cL [\Gamma,\lambda]$.

For the energy of half-toroids introduced in \eqref{eq:total_energy_h}, the associated relaxed energy can be written as 
\begin{equation}\label{eqn:total_energy_h}
	J_h [\omega] = \frac{\beta}{2} \int_{\omega} \rho^{-1} dA + \frac{2\chi}{\pi} \int_{\omega} dA +\int_{\partial \omega} \sigma (\bnu) \rho\, ds + \frac{\varepsilon}{2} \int_{\partial \omega} \kappa^2\,ds,
\end{equation}
with the Lagrangian given by 
\begin{equation}\label{eqn:lagrangian_h}
	\cL_h [\omega,\lambda] = J_h [\omega] + \lambda \left( C_v - \int_{\omega} \rho dA \right),
\end{equation}
where $C_v > 0$ is the desired volume (without the factor of $2 \pi$).
One can still rewrite \eqref{eqn:lagrangian_h} entirely in terms of $\Gamma := \partial \omega$ because
\begin{equation}\label{eqn:reduce_integrals_to_boundary_h}
	\int_{\omega} dA = \int_{\Gamma} (\bnu \cdot {\mathbf e}_1) \rho \, ds.
\end{equation}
Therefore, $J_h[\omega] \equiv J[\Gamma]$ and $\cL_h [\omega,\lambda] \equiv \cL_h [\Gamma,\lambda]$.

\subsubsection{Gradient Flow}\label{sec:grad_flow}

Since the first-order conditions for a critical point of \eqref{eqn:lagrangian} are non-linear, we use a gradient flow strategy to find the minimizer.   Suppose $\vX$ is a parameterization of $\Gamma$ that depends on a pseudo-time variable $t$.  In other words, $\Gamma(t)$ is time-varying and is parameterized (instantaneously) by $\vX(\cdot,t)$.  Hence, our goal is to create a ``velocity'' $\vV = \partial_{t} \vX$ so that the energy $J[\Gamma(t)]$ is monotonically decreasing and $\partial_{t} \int_{\omega(t)} \rho dA = 0$ (volume is preserved).

We achieve this energy decrease by a gradient flow, i.e. we define $\vV$ to be minus the ``shape gradient'' of \eqref{eqn:lagrangian} with a (time-varying) Lagrange multiplier that enforces volume conservation.  The next section describes this more specifically.

\subsubsection{Weak Formulation}\label{sec:weak_form}

We assume that, for each $t$, $\vV(t)$ is defined on $\Gamma(t)$ and lies in a Hilbert space $\cH = \cH(t)$. Let $\inner{\vV}{\vY}$ be an inner product on $\cH$ for any functions $\vV, \vY \in \cH$.  With this, we define the gradient flow weakly, i.e. $\vV(t)$ solves (for each $t$)
\begin{equation}\label{eqn:weak_form_grad_flow}
\begin{split}
	\inner{\vV}{\vY} = - \delta_{\Gm} \cL[\Gm(t),\lambda(t); \vY],
\end{split}
\end{equation}
for all admissible ``shape'' perturbations $\vY$.  Here, $\delta_{\Gm}$ denotes the shape derivative, which is explained in \cite{Delfour_Book2011,Walker_book2015}.  Combining \eqref{eqn:weak_form_grad_flow} with the boundary motion equation:
\begin{equation}\label{eqn:bdy_motion}
	\frac{d}{dt} \vX = \vV(\vx,t), \quad \text{for all } \vX \in \Gamma(t),
\end{equation}
completely defines the evolution of the boundary $\Gamma(t)$.  For example, if $J[\Gamma] = \int_{\Gamma} 1$, then the evolution would simply be mean curvature flow: $\vV = -\kappa \bnu$.

\subsubsection{Fully Discrete Approximation}\label{sec:discrete}

In order to have a tractable problem, we discretize \eqref{eqn:weak_form_grad_flow} with a variant of a method found in \cite{Barrett_JCP2007,Barrett_SJSC2008,Barrett_NM2008}.  We first discretize the curve $\Gamma$ by a polygonal curve.  Hence, $\vX$ is a vector-valued, continuous piecewise linear finite element function \cite{Alberty_NA1998,BrennerScott_book2008}.
In addition, we use a backward-Euler method for approximating \eqref{eqn:bdy_motion} with a fixed time step $\dt$.  Therefore, given the current guess for the polygonal curve $\vX^{m}$ (at time index $m$), we introduce the continuous piecewise linear finite element space $\V^m$ defined on $\Gamma^{m}$.  Thus, $\vX^{m} \in [\V^m]^2$, which is a vector valued finite element space.

The polygonal curve at the next time index is obtained by solving the following system of equations, i.e. find $(\kappa_{\sigma}^{m+1},\kappa^{m+1},\vX^{m+1},\lambda^{m+1})$ such that:
\begin{equation}\label{eqn:FEM_system_grad_flow}
\begin{split}
	\inner{\frac{\vX^{m+1} - \vX^{m}}{\dt}}{\bnu^m_h \eta}_{m}^{h} + \inner{\kappa_{\sigma}^{m+1}}{\eta}_{m}^{h} & + \varepsilon \inner{\partial_s \kappa^{m+1}}{\partial_s \eta}_{m} \\
	&\quad  - \lambda^{m+1} \inner{\rho}{\eta}_{m}^{h} = \frac{\varepsilon}{2} \inner{(\kappa^{m})^3}{\eta}_{m}^{h}, ~ \forall \eta \in \V^m,
\end{split}
\end{equation}
\begin{multline}\label{eqn:FEM_system_ansiotropic_curv}
	\inner{\kappa_{\sigma}^{m+1}}{\bnu^m_h \cdot \vY}_{m}^{h}  - \inner{\rho \sigma (\bnu^m) \partial_s \vX^{m+1}}{\partial_s \vY}_{m} - \frac{\beta}{2} \inner{\rho^{-2} (\vX^{m+1} \cdot {\mathbf e}_1)}{\vY \cdot \bnu^m}_m  \\
- \frac{2\chi}{\pi} \inner{1}{\vY \cdot \bnu^m}_m=- \inner{\rho (\bnu \cdot \partial_s \vY) \btau \cdot \sigma' (\bnu^m)}{1}_{m} +  \inner{\sigma(\bnu^m) (\vY \cdot {\mathbf e}_1)}{\eta}_{m}^{h}, ~ \forall \vY \in [\V^m]^2,
\end{multline}
\begin{equation}\label{eqn:FEM_system_std_curv}
\begin{split}
\inner{\kappa^{m+1}}{\bnu^m_h \cdot \vY}_{m}^{h} - \inner{\partial_s \vX^{m+1}}{\partial_s \vY}_{m} =0, ~ \forall \vY \in [\V^m]^2,
\end{split}
\end{equation}
\begin{equation}\label{eqn:FEM_system_vol_constraint}
\begin{split}
	-\inner{\frac{\vX^{m+1} - \vX^{m}}{\dt}}{\bnu^m_h \rho}_{m}^{h} = \frac{1}{\dt} \left[ \int_{\omega^m} \rho dA - C_v \right],
\end{split}
\end{equation}
where $\inner{\cdot}{\cdot}_{m}$ denotes the inner product on $\Gm^m$, $\inner{\cdot}{\cdot}_{m}^{h}$ is a mass lumped inner product on $\Gm^m$, and $\bnu^{m}_{h}$ is a ``discrete'' vertex normal vector of $\Gamma^m$. The system \eqref{eqn:FEM_system_grad_flow}-\eqref{eqn:FEM_system_vol_constraint} is linear and \emph{semi-implicit} which allows for taking large time steps (see \cite{Barrett_JCP2007,Barrett_SJSC2008,Barrett_NM2008} with the caveat about anisotropic surface energy and the regularization parameter).

The first equation, \eqref{eqn:FEM_system_grad_flow}, is an approximation of the gradient descent equation \eqref{eqn:weak_form_grad_flow}.
Equation \eqref{eqn:FEM_system_std_curv} is essentially a ``weak'' definition of the standard curvature of $\Gamma^{m+1}$ (viewed as a planar curve), whereas \eqref{eqn:FEM_system_ansiotropic_curv} is a weak definition of the ``anisotropic, radially weighted'' curvature.  The last equation accounts for the volume constraint, whose right-hand-side corrects for any deviations from the desired volume $C_v$.

After solving the system, we obtain $\vX^{m+1}$ which is defined on $\Gamma^{m}$.  We obtain the new curve $\Gamma^{m+1}$ by simply taking the nodal values of $\vX^{m+1}$ as the new vertex positions of the polygonal curve.

Therefore, given an initial curve $\Gamma^{0}$, we obtain $\Gamma_{m}$ by iteratively solving the above system $m$ times.  This requires an initial guess for $\kappa^{0}$ because it appears in the right-hand-side of \eqref{eqn:FEM_system_grad_flow}.   This is accomplished by using a similar equation to \eqref{eqn:FEM_system_std_curv}.  More specifically, we find $\vK^{0} \in [\V^{0}]^2$ such that
\begin{equation}\label{eqn:init_curv_solve}
\begin{split}
	\inner{\vK^{0}}{\vY}_{0}^{h} = \inner{\partial_s \vX^{0}}{\partial_s \vY}_{0}, ~ \forall \vY \in [\V^0]^2,
\end{split}
\end{equation}
and then computing $\kappa^{0} (\vx_i) := \bnu^m_h (\vx_i) \cdot \vK^{0} (\vx_i)$ at each vertex $\vx_i$ in $\Gamma^{0}$.

\section{Acknowledgments} This work was supported by NSF grants DMS-2106675 (ODL), DMS-2106551 (DG), DMS-1555222-CAREER and DMS-2111474 (SWW), DMS-1435372, DMS-1729589 and DMS-1816740 (MCC). The authors would like to thank Dr. Thomas A. Everett from the Enhanced Oil Recovery Lab at Purdue University for his help using the spinning drop tensiometer. MCC also aknowledges the hospitality of the Isaac Newton Institute at Cambridge University, UK and the support from the Simons Foundation.

\bibliographystyle{ieeetr}
\bibliography{chrom,dmref17final,Oleg_Refs,MasterBibTeX.bib}

\articleend
\newpage
\setcounter{section}{0}
\renewcommand{\thesection}{S\arabic{section}} 
\setcounter{equation}{0}
\renewcommand{\theequation}{S\arabic{equation}} 
\setcounter{figure}{0}
\renewcommand{\thefigure}{S\arabic{figure}} 
\setcounter{page}{1}
\renewcommand{\thepage}{S\arabic{page}} 
\begin{center}
{\Large{\bf Supplementary Material}}
\end{center}
\bigskip

\title{Toroidal nuclei of columnar lyotropic chromonic liquid crystals coexisting with isotropic phase}
\author[R.~Koizumi]{Runa Koizumi}
\address{Advanced Materials and Liquid Crystal Institute, Materials Science Graduate Program, Kent State University, Kent, OH 44242, USA}
\email{rkoizumi@kent.edu}
\author[D.~Golovaty]{Dmitry Golovaty}
\address{Department of Mathematics, The University of Akron, Akron, OH 44325-4002}
\email{dmitry@uakron.edu}
\author[A. ~Alqarni]{Ali Alqarni}
\address{Advanced Materials and Liquid Crystal Institute, Department of Physics, Kent State University, Kent, OH 44242, USA and Department of Physics, University of Bisha, Bisha, 67714, Saudi Arabia}
\email{aalqarn1@kent.edu}
\author[S.W.~Walker]{Shawn W. Walker}
\address{Department of Mathematics, Louisiana State University, Baton Rouge, LA 70803-4918}
\email{walker@lsu.edu}
\author[Y.A.~Nastishin]{Yuriy A. Nastishin}
\address{Advanced Materials and Liquid Crystal Institute, Kent State University, Kent, OH 44242, USA and Hetman Petro Sahaidachnyi National Army Academy, 32 Heroes of Maidan street, Lviv, 79012, Ukraine}
\email{nastyshyn\_yuriy@yahoo.com}
\author[M.C.~ Calderer]{M. Carme Calderer}
\address{School of Mathematics, University of Minnesota, Minneapolis, MN 55455, USA}
\email{calde014@umn.edu}
\author[O.D.~Lavrentovich]{Oleg D. Lavrentovich}
\address{Advanced Materials and Liquid Crystal Institute, Materials Science Graduate Program, Kent State University, Kent, OH 44242, USA and Department of Physics, Kent State University, Kent, Ohio 44242, USA}
\email{olavrent@kent.edu}

\maketitle
\section{Experiment}
\smallskip\noindent\textbf{Interfacial tension.} In the past, the surface tension of the C-I interface has been measured for surfactant-based lyotropic phase by the so-called grain boundary groove method \citeS{Oleg1.41}, borrowed from metallurgy \citeS{Oleg1.42}. 
\begin{figure}[H]
\centering
    \includegraphics[width=3in]
                    {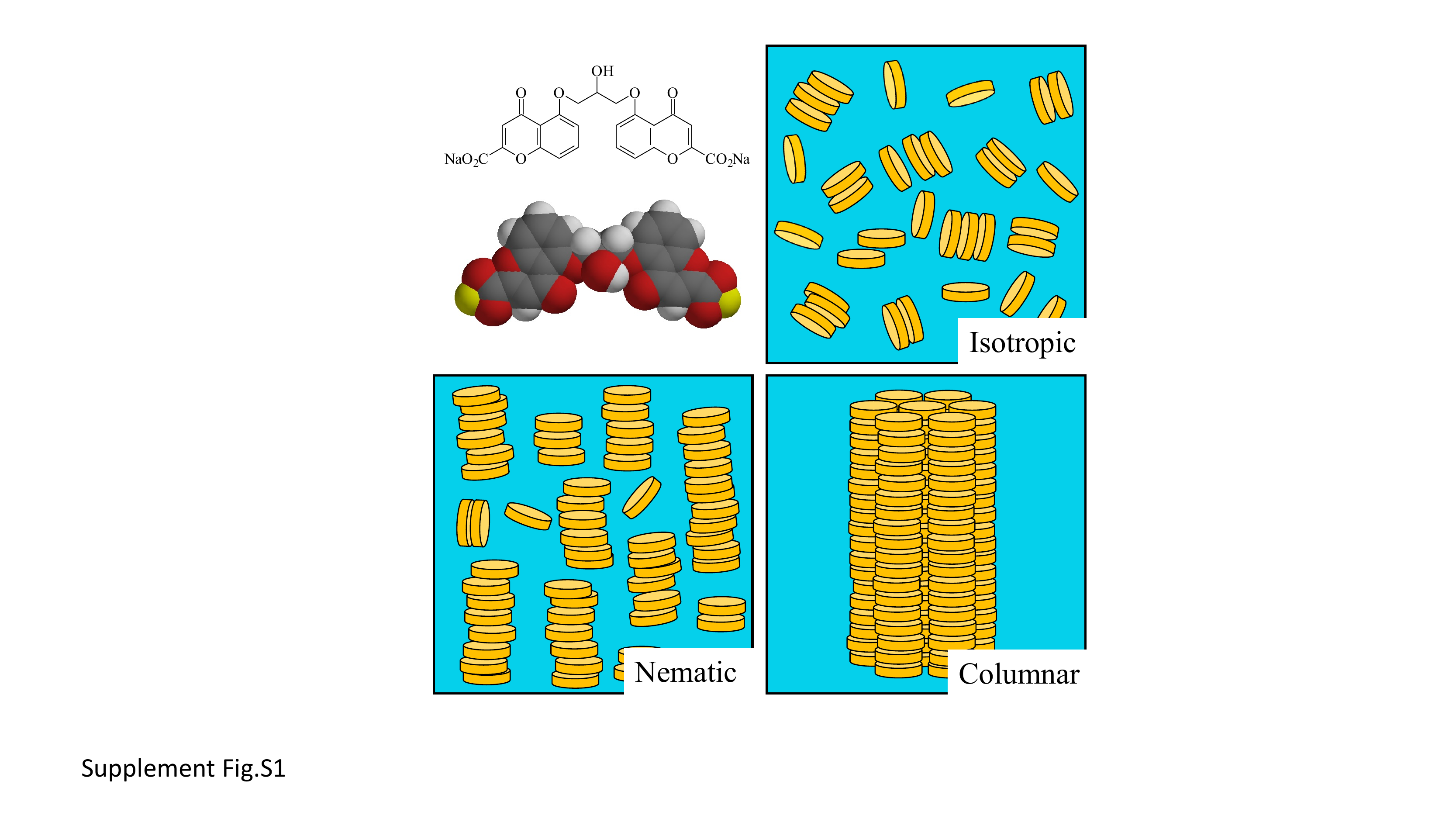} 
    \caption{Chemical and space-filling structure of DSCG molecule and molecular arrangements in the isotropic, nematic, and in a filament of a columnar phase. The director  $\widehat{\boldsymbol{\mathrm{n}}}$  is along the average alignment direction of the columnar aggregates.}
  \label{figS1}
\end{figure}
In this approach, one creates a Col-I interface by placing a sample in a temperature gradient and explores the depth $h$ of an indentation caused by the presence of a grain boundary between two domains of the Col phase. If the interface is strictly perpendicular to the temperature gradient, i.e., not curved in the plane normal to the glass plates, the measurement of $h$ yields the surface tension coefficient, $\sigma_{||}\propto h^2$. Although we do observe an increase of $h$ with $c$ and $C$, the data could not be used to determine $\sigma_{||}$ since the Col-I interface makes an angle of about $62^\circ$-$73^\circ$ with the glass, as the Col phase wets the substrate better than the I phase. Instead, we use the technique of a spinning drop to determine $\sigma_{||}$ at the Col-I interface by a spinning droplet technique, which is based on the fact that centrifugal forces acting on a droplet (I phase) placed in a more dense second fluid (Col phase) would elongate. The elongation is limited by the interfacial tension between the two fluids, which allows one to measure the interfacial tension, see \citeS{Oleg1.41,Oleg1.43,Oleg1.44,Oleg1.45,Oleg1.46} and references therein. The denser fluid (Col phase) fills a round capillary; the droplet of the lighter fluid (I phase) is injected into the central part of the tube. The tube rotates around the symmetry axis with a rotational speed $\omega$. Measuring the thinning diameter $\delta$ of the droplet, the interfacial tension is calculated as $\sigma_{||}=\frac{\delta^3\omega^2\Delta\rho}{4}$, where $\Delta\rho$ is the density difference between the two fluids. We centrifuged the $c=0.47\,$mol/kg solution of DSCG at $4000$ rpm at $T=45^\circ\,$C in order to separate the Col and the I phases from each other and measured their densities using a DE45 Density Meter (Mettler Toledo) as $\rho_C=1.12\times10^3\,$kg/m\textsuperscript{3} and $\rho_I=1.05\times10^3\,$kg/m\textsuperscript{3}, respectively; therefore, $\Delta\rho=0.07\times10^3\,$kg/m\textsuperscript{3}. The interfacial tension is determined by the spinning drop tensiometer DataPhysics Instruments, SVT 20. To enhance contrast between the Col and I phases, the I solution is doped with $0.05\,\mathrm{wt}\%$ Methylene Blue dye. The measurements of $\delta$ at $\omega=9000\,$rpm yield $\sigma\approx10^{-6}\,$J/m\textsuperscript{2}. Because of high light scattering at the Col phase, more accurate measurements could not be possible. 

\smallskip\noindent\textbf{Bend elastic constant.} The independent measurements of $K_3$ for the Col islands are difficult. Elastic moduli of LCLCs have been measured only for the homogeneous N phase. These measurements suggest that $K_3\propto\lambda_pc$, where $\lambda_p$ is the persistence length which could become smaller at higher concentrations $c$ or $C$ since because of the electrostatic effects such as the decrease of the Debye screening length \cite{Oleg1.39},\citeS{Oleg1.47,Oleg1.48,Oleg1.49}. Therefore, the product $\lambda_pc$ and thus $K_3$ might not increase much with $c$ and $C$. The estimate of $K_3=50$-$60\,$ pN based on the fitted value of the elastocapillary length $\lambda_{ec}$ is slightly larger than the maximum value $K_3=50\,$ pN measured in the homogeneous N phase \cite{Oleg1.39}.

\begin{figure}[H]
\centering
    \includegraphics[width=4in]
                    {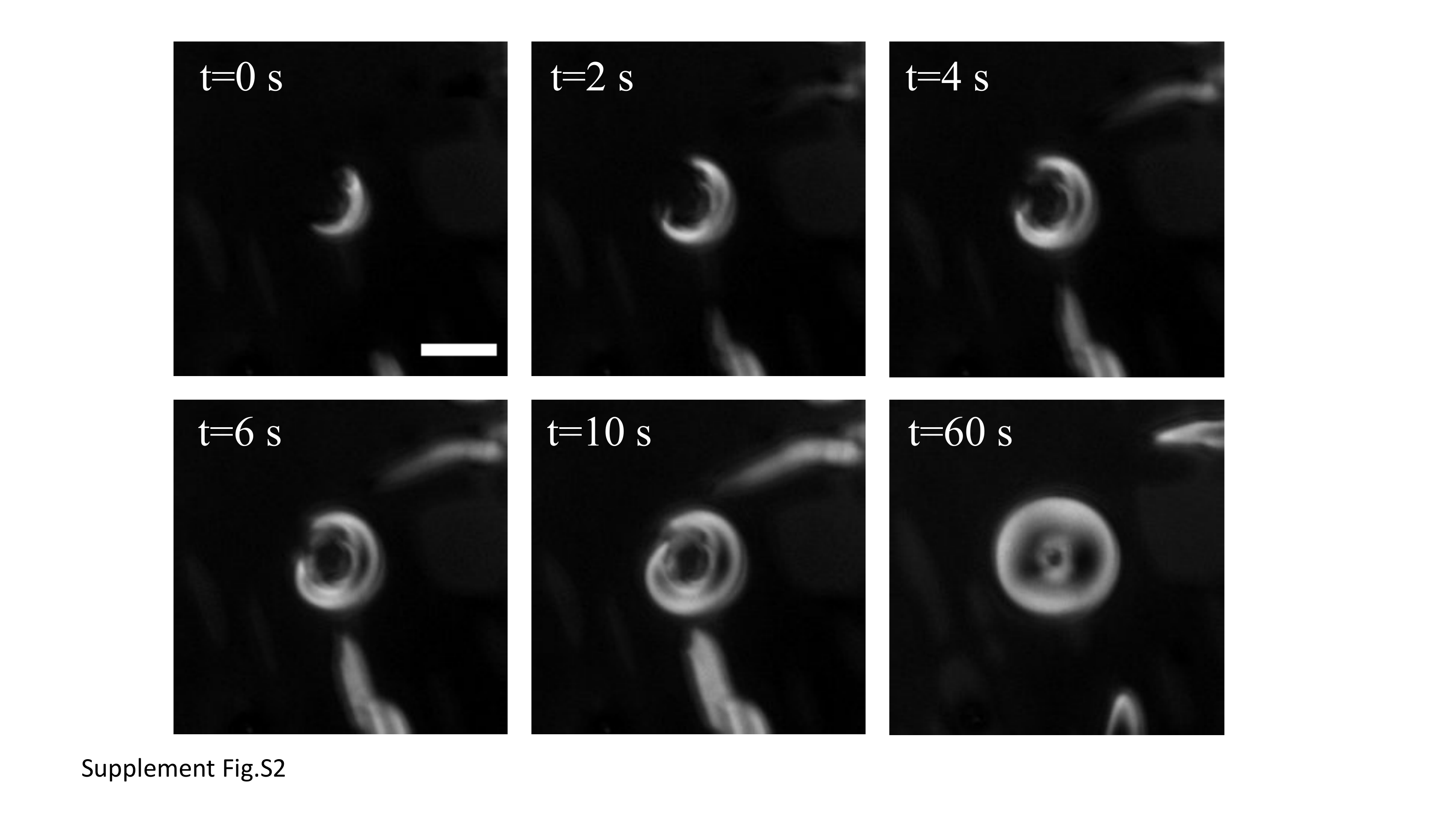} 
    \caption{Dynamics of the Col toroid formation from bent filaments in the mixture of DSCG, $c=0.34$ mol/kg (15 wt\%), and PEG, $C=0.016$ mol/kg (5 wt\%); $t=0\ $s refers to the time when the cooling rom the isotropic phases was stopped, and the temperature was fixed at $T=42\mathrm{{}^\circ C}$. Scale bar $50\mu\mathrm{m}$.}
  \label{figS2}
\end{figure}

\smallskip\noindent\textbf{Closure of toroids.  }The toroidal shapes result from a delicate balance of the anisotropic surface tension and bulk elasticity. The preferred alignment of the director at the Col-I interface is tangential, with a surface tension coefficient ${\sigma }_{||}$. This preference is evidenced by parallel alignment of chromonic aggregates at the Col-I interfaces, Figs.~\ref{olegfig4},\ref{figS2}. Transient perpendicular alignment of ${\boldsymbol{\mathrm{n}}}$ at the Col-I interface is observed only when the growing filament are not yet curled into toroids, Fig.~\ref{figS2}. The perpendicular alignment is accompanied by a higher surface tension coefficient ${\sigma }_{\bot }>{\sigma }_{||}$ . The two open ends of the filaments in Fig.~\ref{figS2} at $t<$10 s that are exposed to water carry a surface energy proportional to their surface area $F_{s\bot }=2\pi a^2{\sigma }_{\bot }$, where we approximate the filament as a circular cylinder of a radius $a.$  The two open ends would tend to come together in order to minimize the exposure of the hydrophobic aromatic cores to water and maximize the exposure of the hydrophilic groups of the DSCG molecules, as clearly seen in Fig.~\ref{figS2} for $t=6$0 s. The toroidal shape is preferred over a filament if the bending energy is smaller than the surface energy $F_{s\bot }=2\pi a^2{\sigma }_{\bot }$ that the two open ends carry. 

To calculate the elastic energy of a toroid in Fig.~\ref{figS2}, $t=60$ s, we model it as a circular torus of a minor radius $a$  and a major radius $r$, associated with the geometrical parameters in Fig.~\ref{olegfig4}d as $r=R-a$ , Fig.~\ref{olegfig4}e. The elastic energy density of bend \cite{Oleg29} is  $f_e=\frac{1}{2}K_3{\left({\boldsymbol{\mathrm{n}}}\times \mathrm{curl}{\boldsymbol{\mathrm{n}}}\right)}^2$,  where $K_3$ is the bend modulus. We introduce a system of coordinates $\left(\rho ,\theta ,\phi \right)$ related to the Cartesian coordinates and the geometric parameters defined above as $$\left\{x,y,z\right\}=\left\{\left(R-a-\rho \mathrm{cos}\theta \right)\mathrm{cos}\phi ,\left(R-a-\rho \mathrm{cos}\theta \right)\mathrm{sin}\phi ,\ \rho \mathrm{sin}\theta \right\},$$ Fig.~\ref{olegfig6}(a). In these coordinates, ${\boldsymbol{\mathrm{n}}}\boldsymbol{=}\left(0,0,1\right)$ and ${\left({\boldsymbol{\mathrm{n}}}\times \mathrm{curl}{\boldsymbol{\mathrm{n}}}\ \ \right)}^2\ =1/{(R-a-\rho \mathrm{cos}\theta \ )}^2,$ the scale factors are $g_{rr}=1;\ g_{\theta \theta }={\rho }^2;\ g_{\phi \phi }={\left(r-\rho \mathrm{cos}\theta \right)}^2$; the volume element is $dV=\rho {(r-\rho \mathrm{cos}\theta \ )}^2$. Integration over $0\le \rho \le a;\ 0\le \theta ,\phi \le 2\pi $ yields the bend energy of the Col torus as $F_e=2{\pi }^2K_3\left(r-\sqrt{r^2-a^2}\right)$.  The surface energy of the toroid with tangential director at the Col-I interface is $F_{s\parallel }=4{\pi }^2{\sigma }_{\parallel }ar$, where ${\sigma }_{\parallel }$ is the surface tension coefficient for tangential anchoring at the Col-I interface. The sum $F_e+F_{s\parallel }$ should be compared to the surface energy $F_{s\bot }+F_{s\parallel }$=$2\pi a^2{\sigma }_{\bot }+4{\pi }^2{\sigma }_{\parallel }ar$ of a cylindrical filament that is a straight circular cylinder of the same cross-sectional radius $a$ and volume as the torus, which means that the length of the cylinder is $2\pi r$. In the straight cylinder, the director is not distorted, thus the elastic energy is 0.  The balance of $F_e=2{\pi }^2K_3\left(r-\sqrt{r^2-a^2}\right)$ and $F_{s\bot }=2\pi a^2{\sigma }_{\bot }$ shows that the open-ended cylinder would prefer to curl into toroid, Fig.~\ref{figS2}, whenever  ${\sigma }_{\bot }\ge \frac{\pi K_3r}{a^2}$. The highest value of $K_3$ measured in the N phase of DSCG is approximately 50 pN \citeS{Oleg39}; with $r\approx 30\ \mu m,\ a\approx 10\ \mu m$, the last condition can be recast as  ${\sigma }_{\bot }\ge 0.5\times {10}^{-4}\ J\ m^{-2}$.

\smallskip\noindent\textbf{Bend elasticity vs surface tension for complete thin toroids with circular cross-section} The ratio of the minor radius to the major radius for thin toroids that are complete and do not touch the substrates is calculated similarly to the main text, as  
\begin{equation} \label{GrindEQ__2_} 
\frac{a}{r}={\left(\frac{V}{{2\pi }^2}\right)}^{\frac{1}{5}}{\left(\frac{{\sigma }_{\parallel }}{K_3}\right)}^{\frac{3}{5}}={\left(\frac{V}{{2\pi }^2}\right)}^{\frac{1}{5}}\frac{1}{{\lambda }^{{3}/{5}}_{ec}}=\frac{1}{2^{1/5}{\pi }^{2/5}{\beta }^{3/5}}.      
\end{equation} 

\section{Estimate of the potential range of $\chi$ values.}
\label{sec:supest}
The parameter $\chi $ characterizes the relative strength of the surface tension at the glass-LC interface, written as $\chi {\sigma }_{||}$, where ${\sigma }_{||}$ is the interfacial tension coefficient at the Col-I interface for the columns align parallel to it.  The chromonic columns are perpendicular to the glass substrate since tilted alignment would result in a contact of polyaromatic hydrophobic cores of the molecules with water. This alignment and elasticity of the Col phase make the measurements of the surface tension  $\chi {\sigma }_{||}$ difficult. Below we resort to a simple analytical argument to find the range of plausible values of $\chi $.

Consider a semicircular cylinder of a length $L$ and a radius $R$ attached to the glass by its cross-section of area $2LR$. The elastic energy of bent columns that strike the glass substrate perpendicularly and thus bend over by $\pi $,  is $\frac{\pi }{2}K_3\ L\ \mathrm{ln}\ \frac{R}{r_c}$ , where $r_c$ is the radius of the central core. The surface energy of the footprint is $2\chi {\sigma }_{||}LR$, the surface energy of the Col-I interfaces is  $\pi {\sigma }_{||}LR+2{\sigma }_{||}\frac{V}{L}$, where $V=\frac{\pi LR^2}{2}$ =const is the volume of the semicylinder. We use the relationship $V=\frac{\pi LR^2}{2}$ =const to express $R$ through $V$ and$\ L$. The surface energy is then $\sqrt{\frac{2VL}{\pi }}\left(2\chi +\pi \right){\sigma }_{||}+2\frac{V}{L}{\sigma }_{||}$. 

\begin{figure}[H]
\centering
    \includegraphics[width=3in]
                    {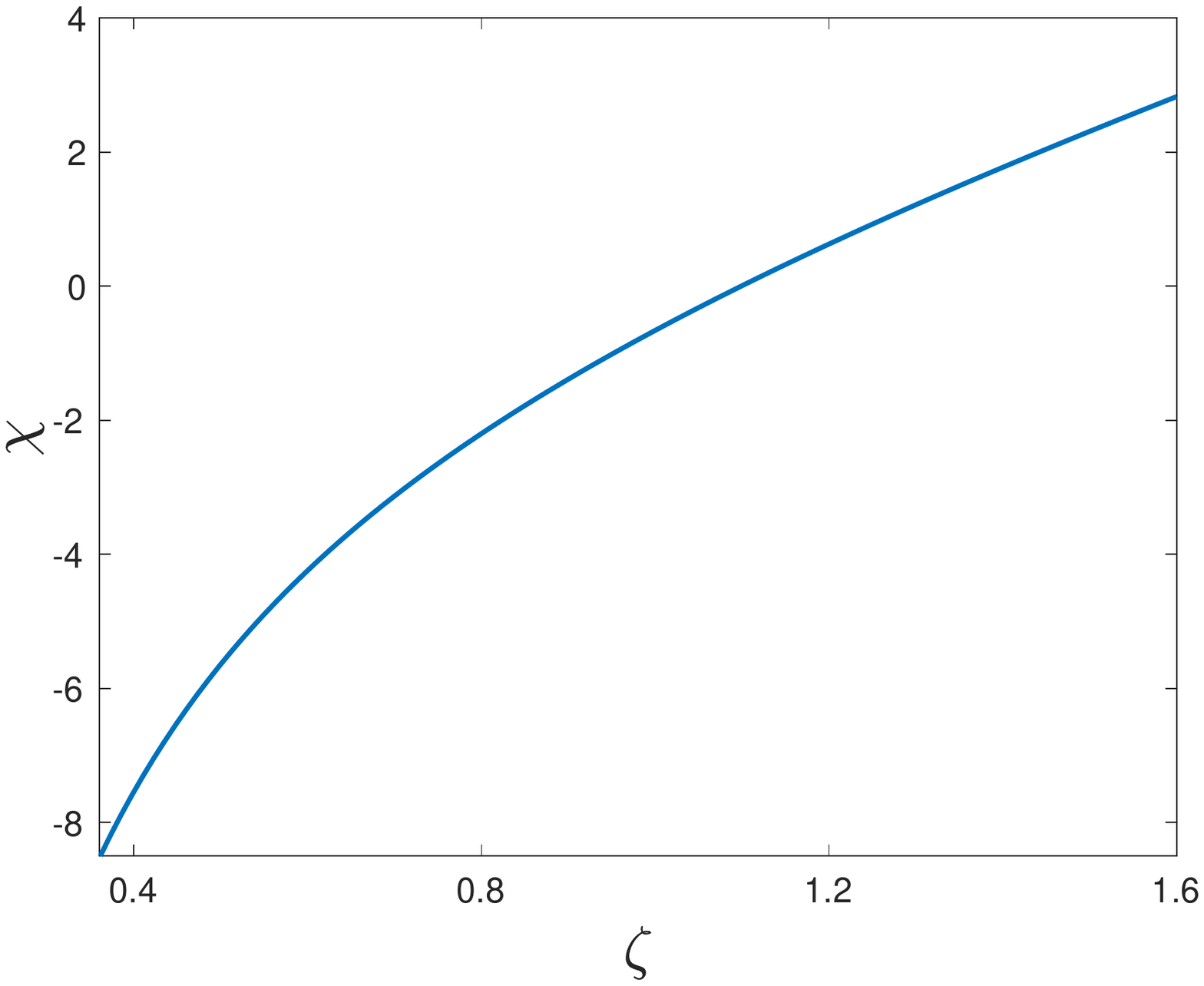} 
    \caption{Dependence $\chi \left(\zeta \right)$ calculated from \eqref{eq:oS1} for $\frac{4K_3}{{\sigma }_{||} L_{eq}}=2.4$. Note that the parameter $\chi$ could be either positive or negative for $0.5\leq\zeta\leq 1.8$.}
  \label{figS2n}
\end{figure}

Minimizing the sum of the elastic and surface energies with respect to $L$, one finds $$\frac{\pi }{2}K_3\ \mathrm{ln}\ \frac{R}{r_c}+\sqrt{\frac{V}{2\pi L}}\left(2\chi +\pi \right){\sigma }_{||}-2\frac{V}{L^2}{\sigma }_{||}=0.$$ For rough estimates, let us represent $V=\frac{\pi LR^2}{2}$ =const as $V=\frac{\pi {\xi }^2L^3}{2}$, using a substitution $R=\zeta L$, where $\zeta\sim1$ is a scaling factor. Since a logarithm is a slow function for large arguments, we treat it as a constant, $\mathrm{ln}\ \frac{R}{r_c}\approx \mathrm{ln}\ 50$ $\approx $4. Then one finds the equilibrium $L_{eq}=\frac{4\frac{K_3}{{\sigma }_{||}}\ }{{2\zeta }^2-\zeta \left(1+\frac{2\chi }{\pi }\right)}$. The last expression could be rewritten as 
\begin{equation}
\label{eq:oS1}
\chi =\frac{\pi }{2}\left(2\zeta -1-\frac{4K_3}{{\sigma }_{||}\zeta L_{eq}}\right)
\end{equation}
In the experiments, nuclei show the shapes that could be roughly characterized by the ratio $R/L$ in the range (0.5 - 1.8). For the typical $L_{eq}\approx 100\ \mathrm{\muup }\mathrm{m}$, $\frac{K_3}{{\sigma }_{||}}=60\ \mathrm{\muup }\mathrm{m}$ and $\frac{4K_3}{{\sigma }_{||} L_{eq}}=2.4$, the plot $\chi(\zeta)$ suggests that $\chi$ might be either positive or negative if the scaling factor is in the plausible range $0.5\leq\zeta\leq1.8$,  Fig.~\ref{figS2n}. This conclusion is only qualitative since the experimental shapes with hexagonal footprints are different from the rectangular footprints considered in the model above. 
\begin{figure}[H]
\centering
    \includegraphics[width=5in]
                    {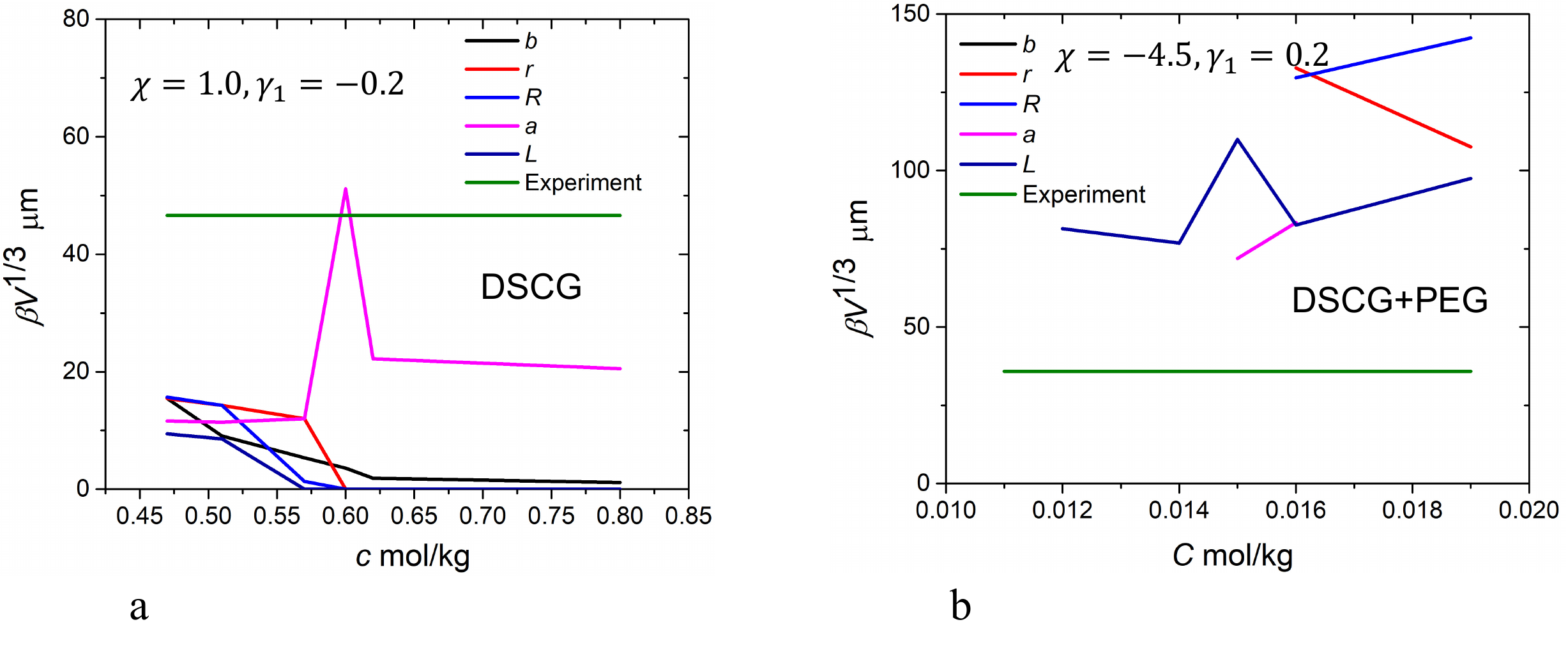} 
    \caption{Comparison of simulated $\beta V^{1/3}$ to the experimentally obtained values for (a) pure DSCG and (b) DSCG+PEG for parameters that do not match the experimental data: $\chi=1.0$ and $\gamma_1=-0.2$ for pure DSCG and $\chi=-4.5$ and $\gamma_1=0.2$ for DSCG+PEG.}
  \label{figS21n}
\end{figure}

\section{Thermomechanics of plane curves}
\label{Section 3}
Here we will follow \cite{GA89} to briefly survey the mechanics of a closed plane curve $\mathcal C$ separating two different phases of a material undergoing a phase transition. Suppose that $\bnu$ represents an outward unit normal vector to $\mathcal C$ and let the line energy density be given by the scalar function $\gamma(\bx)=\sigma(\bnu(\bx))>0.$ The function $\sigma$ reduces to a positive constant when the line energy is isotropic. Further, let $F$ represent the difference between the bulk energies of the phases.

A positively oriented smooth closed curve is a periodic map  $p\Rightarrow \br(p)=(\rho(p),z(p))$ from $\mathbb R$ into ${\mathbb R}^2$, such that $\left|\br'(p)\right|\neq 0$.  Assuming that the curve is positively oriented and parametrized with respect to the arc length $s$, we define the orthonormal frame $(\btau(s), \bnu(s))$ as
\begin{gather}
\btau(s)=\br^\prime(s)=(\cos{\theta(s)},\sin{\theta(s)}),\label{eq:tau}\\
\bnu(s)=\br^\prime_\perp(s)=(\sin{\theta(s)},-\cos{\theta(s)}), \label{eq:nu}
\end{gather}
where $\theta(s)$ is the angle between the tangent to the curve and the positive direction of the $\rho$-axis for all $s\in[0,L]$ and ${\mathbf a}_\perp=(a_2,-a_1)$ for every ${\mathbf a}=(a_1,a_2)$. The vectors of the frame are related by the Frenet formulas
\begin{equation}
\label{eq:fres}
\btau^\prime(s)=-\kappa(s)\bnu(s),\quad \bnu^\prime(s)=\kappa(s)\btau(s),
\end{equation}
where
\[
\kappa(s)= \theta^\prime(s),
\]
is the curvature of $\br(s)$.  A curve is convex if $\kappa> 0$; this allows to use $s$ and   $\theta$ as equivalent parameters of the curve.
The force sustained by the interface, the capillary force is given by
\begin{equation}
\mathbf C= \gamma(\theta)\btau + \gamma^{\prime}(\theta)\bnu.
\end{equation}
with a trivial change in notation. For our purposes, we will allow for piecewise smooth curves, that is for $\br=\{\br_1, \br_2, \dots \}$, being a finite or countable collection of continuous arcs, with possibly discontinuous tangent and normal vectors at their junction points. 

The concept of stability of the curve with the line energy density $\gamma$ at a point $\theta$ is central in determining the presence of corners and facets on the boundary.  Let $\Sigma=\int_{\partial\omega}\gamma\,ds$ denote the total interfacial energy. A necessary and sufficient condition for $\Sigma$ to have a minimum at $\theta$ is that  $\gamma(\theta)+ \gamma^{\prime\prime}(\theta)>0$. The interfacial energy is strictly stable, stable, or unstable at $\theta$ provided that
\begin{equation}
\gamma(\theta)+ \gamma^{\prime\prime}(\theta)>0, \quad  \gamma(\theta)+ \gamma^{\prime\prime}(\theta)\geq 0, \quad  \gamma(\theta)+ \gamma^{\prime\prime}(\theta)<0,
\end{equation}
respectively.   The analogous concepts of global stability or instability follow by requiring the latter to be satisfied for all $\theta\in\mathbb R$. When one considers  the problem of time evolution of interfaces, instability corresponds to the loss of parabolicity of  the governing equation, which becomes backward parabolic at the points of instability of $\gamma$.

The construction of the interface curve that minimizes the total line energy is given by Wulff's theorem, stated as follows: 

\smallskip\noindent {\it Assume that  the difference between the bulk energies of the phases $F\neq 0$.  Then 
	$$\br(\theta)=F^{-1}\big(\gamma^\prime(\theta)\btau(\theta)-\gamma(\theta)\bnu(\theta)\big)$$ defines the interface which is closed, convex and parametrized by $\theta$. Moreover, the curvature is given by} \[\kappa(\theta)=F(\gamma(\theta)+ \gamma^{\prime\prime}(\theta))^{-1}.\]

\smallskip We now turn to the description of corners and facets observed in experiments. The loss of stability of the function $\gamma(\theta)$  on some $\theta$-intervals can be resolved by allowing corners on the energy minimizing curve $\br(\theta)$. These correspond to jumps in $\theta$ 
across the unstable portions of $\gamma(\theta)$.  For a pair of distinct angles $\{\theta^{-}, \theta^{+}\} $ to determine an admissible corner, the following conditions must be satisfied: (i) $\gamma(\theta)
$ is unstable for $\theta\in(\theta^-, \theta+)$, (ii) $ |\theta^+-\theta^-|<\pi$, and  (iii) $\mathbf C(\theta^-)=\mathbf C(\theta^+)$, representing the continuity of the capillary force at the corner. 

The Frank diagram is a central tool to identify corners in the curve. It is the polar diagram of $\gamma^{-1}$, and thus the locus of the Frank potential 
$\boldsymbol \sigma= \gamma^{-1}(\theta)\bnu(\theta)$.  The capillary force is related to the Frank potential in the following way:
$\mathbf C= -{\gamma(\theta)}^2 {\boldsymbol \sigma}^\prime(\theta)$; it is tangent to the Frank diagram, pointing in the direction of decreasing $\theta$.

The convexity-stability theorem provides an important tool to identify corners:
{\it (i) The Frank diagram is convex if and only if $\gamma$ is stable. (ii) More generally, $\gamma$ is stable on the globally-convex sections of the Frank diagram.
	If $ (\theta^-, \theta^+)$ is an open interval separating two adjacent globally-convex sections,
	then $(\theta^-, \theta^+)$  is a corner and $\gamma$ is unstable somewhere in  $(\theta^-, \theta^+)$.}
Hence, the Frank diagram corresponding to  a curve with corners has nonconvex regions.

The convexification of the Frank diagram is the polar diagram of a polar function  $\Sigma(\theta)$, and hence the locus of a vector potential $\boldsymbol \Sigma= \Sigma(\theta)\bnu(\theta)$, the convexified Frank potential. On the globally stable sections of the energy, $\boldsymbol{\sigma}(\theta)= \boldsymbol{\Sigma}(\theta)$. Between such sections, 
$\boldsymbol{\Sigma}$ coincides with the Maxwell lines of the Frank diagram. Consequently, the points of tangency of a  Maxwell line with the Frank diagram identify an ordered angular pair corresponding to a corner of the curve (a proper interpretation has to be given to the case that a Maxwell line shares three or more points of tangency with the diagram). Facets correspond to Maxwell lines in the diagram.  

 \section{Models of Lyotropic Chromonic Liquid Crystals}
 \label{Section 2}

The simple model that we study assumes that a Col nucleus is composed of circular chromonic "columns" centered on and lying in planes perpendicular to the $z$-axis. Each cross-section of the nucleus by a plane that contains the $z$-axis then reveals a triangular lattice of points corresponding to the cross-sections of the columns; we assume that this lattice is fixed with one of the corresponding two-dimensional lattice vectors being parallel to the $z$-axis. The deformation of the chromonic columns is therefore limited to bending. The bending energy of a given column is proportional to the square of the column curvature as long as we work within the framework of the linear elasticity. 

In addition to bending, the total energy  also includes the anisotropic surface energy contribution on the boundary between the toroidal aggregate and the surrounding isotropic phase. The equilibrium shape of a toroidal Col nucleus is determined in competition between the surface and the bulk energy contributions. Assuming that the rate of growth of a toroidal aggregate is much slower than the rate at which the balance develops between the surface and the bulk energies, we analyze this balance assuming that the volume of a toroid is fixed.

Given that each toroidal Col nucleus is assumed to be axially symmetric, the geometry of the aggregate is fully determined by its cross-section with a plane that contains the $z-$axis. In order to recover the equilibrium shape of the planar curve that forms the boundary of each cross-section, we will utilize the machinery developed in thermomechanics of plane curves \citeS{Angenent1999}. Briefly, a curve in \citeS{Angenent1999} is assumed to represent an interface between two phases with different bulk energies. The equilibrium shape of the curve is dictated by the balance between the bulk energies and the anisotropic line tension within the interface. This is essentially the same situation as we encounter here, except that the bulk energy in our case is spatially-dependent in the columnar phase---this is a principal contribution of the present work. Note that, when a toroid is represented by its cross-section, the surface energy of its boundary reduces to the line energy of the curve bounding the cross-section.

We now postulate the free energy whose minimization  gives the optimal shapes of the aggregates in equilibrium. We first introduce the Oseen-Frank energy of nematic liquid crystals as proposed by Oseen, Zocher and Frank, 
\begin{align*}
\wof=& \frac{1}{2}K_1(\nabla\cdot\bn)^2 + \frac{1}{2}K_2(\bn\cdot\nabla\times\bn)^2+\frac{1}{2}K_3|\bn\times\nabla\times\bn|^2\\ +& \frac{1}{2}(K_2+K_4)[(\bn\cdot\nabla)\bn- (\nabla\cdot\bn)\bn)], \quad |\bn|=1,
\end{align*}
where the material constants $K_i>0, \, i=1, 2, 3$ satisfy Ericksen's inequalities \cite{Virga_book1994}. 

The energy of a columnar chromonic liquid crystal consists of the sum of the Oseen-Frank and the transverse elastic energy $\wh$, formulated in terms of the displacement vector $\bu$
\begin{equation}
E_h= \int_{\Omega} \wof(\bn(\bx), \nabla\bn(\bx)) + \wh(\nabla\bu(\bx)), \,\, \, \bn\cdot\bu=0,
\end{equation}
penalizing distortion on planes perpendicular to $\bn.$ In the expression proposed by de Gennes, \cite{4110}, 
\begin{align}
\wh=&\frac{1}{2} B\left(u_{x}+ v_{y}\right)^2 +\frac{1}{2} C\left[\left(u_{x}- v_{y}\right)^2 +\left(u_{y}+v_{x}\right)^2\right], \quad \bu=(u,v).\label{Wh}
\end{align}
Note that the total energy includes the elastic free energy of the lattice deformation as well as the bending term of the Oseen-Frank energy, that for small displacements and  director distortions,  $\delta \bn$, can also be  expressed in terms of the displacement vector $\bu$. The positive constants $B$ and $ C$ correspond to compression and shear moduli, respectively. Moreover $B\approx \frac{K_3}{d^2}$, where $d>0$ represents the typical period of the lattice. 

In order to model aggregates of the columnar phase, we assume that the splay $K_1$ and the twist $K_2$ elastic constants are much larger than the bending constant $K_3,$ effectively prohibiting both splay and twist deformations. We denote by $\Omega$ the domain of a single aggregate with volume $\text{vol} (\Omega)=V$ and with the piecewise smooth boundary $\partial\Omega$. We postulate an anisotropic surface energy density $\sigma_{||}(\bx)=\sigma(\bnu(\bx),\bn(\bx)))$,  where $\bnu$ denotes the normal to the boundary $\partial\Omega$ at $\bx$. In the current approach, we neglect the elastic energy $\wh$. The free-boundary problem is then formulated as finding the vector field and domain, $(\bn, \Omega)$, respectively, that minimize the following energy
\begin{equation}
E[\bn]=\frac{K_3}{2}\int_{\Omega} |\bn\times\nabla\times\bn|^2\,d\bx+ \int_{\partial\Omega}\sigma(\bnu,\bn)\,ds,\label{EnergyGeneral}\end{equation}
where the admissible director fields $\bn$ and sets $\Omega$ are subject to the constraints
\begin{align}
&\bn\cdot\bn=1,\quad\mathrm{div}\,\bn=0,\quad \bn\cdot\mathrm{curl}\,\bn=0\quad \text{in}\,\, \Omega, \label{eq:bulkcons}\\
&\bnu\cdot\bn=0, \quad \text{on}\,\, \partial\Omega, \label{boundary-condition0}\\
&\text{Vol}(\Omega)= V. \label{volume0}
\end{align}
Note that the minimization here is performed with respect to both $\bn$ and $\Omega.$ 

As stated, the problem \eqref{EnergyGeneral}-\eqref{volume0} does not include any features of the columnar phase, and thus it can also be used to describe nematic clusters when $K_1,K_2\gg K_3$.

\section{Derivation of governing equations for smooth curves}
Given the energy functional defined in \eqref{eq:etot}, suppose that
\[\partial\omega=\left\{\br(s)=(\rho(s),z(s))\in\mathbb R_+^2:0\leq s<L,\ \br(L)=\br(0),\ \br^\prime(L)=\br^\prime(0)\right\},\]
where $L>0$ and $\br$ is a simple, smooth, positively oriented closed curve parametrized with respect to its arclength. Recall from \eqref{eq:tau}-\eqref{eq:nu} that  $(\btau(s),\bnu(s))$ is the orthonormal frame associated with $\br(s)$ and that the curvature  of $\br(s)$ at $s\in[0,L]$ is given by $\kappa(s)=\theta^\prime(s),$  where $\theta(s)$ is the angle between the positive direction of the curve and the $\rho$-axis. 

Because we are interested in optimizing the shape of $\partial\omega$, we consider variations $\delta\br$ of $\br$ of the form $\delta\br=(\delta f)\bnu$ where $\delta f:[0,L]\to\mathbb R$ satisfy $\delta f(0)=\delta f(L)$ and $\delta f^\prime(0)=\delta f^\prime(L)$. The variations of the first and the last integrals are then given by 
\begin{equation}
\label{eq:varel}
\delta\left(2\beta \int_\omega \rho^{-1}dA\right)=2\beta \int_{\partial\omega}\rho^{-1}\delta f ds=2 \beta \int_0^L\rho(s)^{-1}\delta f(s)\,ds
\end{equation}
and
\begin{equation}
\label{eq:varvol}
\delta\left(4\lambda \int_\omega \rho\,dA\right)=4\lambda \int_{\partial\omega}\rho\,\delta f ds=4\lambda \int_0^L\rho(s)\delta f(s)\,ds,
\end{equation}
respectively. At the same time
\begin{equation}
\label{eq:varsur}
\delta\left(2 \int_{\partial\omega} \sigma(\bnu)\rho\,ds\right)=2\int_0^L\left\{\nabla\sigma(\bnu(s))\cdot\delta\bnu(s)\rho(s)+\sigma(\bnu(s))\delta\rho(s)+\sigma(\bnu(s))\rho(s)\btau(s)\cdot\delta\br^\prime(s)\right\}\,ds.
\end{equation}
Note that
\[\delta\br^\prime(s)=\delta f^\prime(s)\bnu(s)+\delta f(s)\kappa(s)\btau(s)\]
and
\[\delta\bnu(s)=-\delta f^\prime(s)\btau(s).\]
Substituting these expressions into \eqref{eq:varsur} and integrating by parts, we obtain
\begin{equation}
\label{eq:varsur1}
\delta\left(2 \int_{\partial\omega} \sigma(\bnu)\rho\,ds\right)=2\int_0^L\left\{\left[\rho(s)\nabla\sigma(\bnu(s))\cdot\btau(s)\right]_s+\sigma(\bnu(s))\left({\mathbf e}_\rho\cdot\bnu(s)+\rho(s)\kappa(s)\right)\right\}\delta f(s)\,ds,
\end{equation}
where ${\mathbf e}_\rho=(1,0)$. Now, combining \eqref{eq:varel}, \eqref{eq:varvol}, and \eqref{eq:varsur1} and omitting the $s$-dependence, we find that the Euler-Lagrange equation satisfied by a critical point of the energy \eqref{eq:etot} is
\begin{equation}
\label{eq:elode}
\left[\rho\nabla\sigma(\bnu)\cdot\btau\right]_s+\sigma(\bnu)\left({\mathbf e}_\rho\cdot\bnu+\rho\kappa\right)+\frac{\beta }{\rho}-2\lambda\rho=0.
\end{equation}

If we recall \eqref{eq:tau}-\eqref{eq:nu} and set
\[\gamma(\theta):=\sigma\left(\sin{\theta},-\cos{\theta}\right),\]
then it is immediately follows that
\[\gamma_\theta=\nabla\sigma(\bnu)\cdot\btau.\]
Using this expression and \eqref{eq:tau}-\eqref{eq:fres} in \eqref{eq:elode} gives an alternative form of the Euler-Lagrange equation
\begin{equation}
\label{eq:elode1}
\rho\kappa\left(\gamma_{\theta\theta}+\gamma\right)+\gamma\sin{\theta}+\gamma_\theta\cos{\theta}+\frac{\beta }{\rho}-2\lambda\rho=0.
\end{equation}
Combining the equations obtained in this section, we determine the full problem for the unknown $(\rho,\theta,z,L)$ satisfied by the energy-minimizing curve. The problem is given by the system of ODEs
\begin{equation}
\label{eq:system}
\left\{
\begin{array}{l}
 \rho\left(\gamma_{\theta\theta}+\gamma\right)\theta^\prime+\gamma\sin{\theta}+\gamma_\theta\cos{\theta}+\frac{\beta }{\rho}-2\lambda\rho=0,    \\
\rho^\prime=\cos{\theta},   \\
z^\prime=\sin{\theta},   
\end{array}
\right.
\end{equation}
subject to the conditions
\begin{equation}
\label{eq:cond}
\theta(L)=\theta(0)+2\pi,\quad \theta^\prime(L)=\theta^\prime(0),\quad \int_0^L\cos{\theta}\,ds=\int_0^L\sin{\theta}\,ds=0, \quad 2\pi\int_\omega \rho\,dA=1.
\end{equation}

Using the second equation in \eqref{eq:system} we observe that
\[\rho^\prime\left(\rho\left(\gamma_{\theta\theta}+\gamma\right)\theta^\prime+\gamma\sin{\theta}+\gamma_\theta\cos{\theta}\right)=\left(\rho\gamma_\theta\cos{\theta}+\rho\gamma\sin{\theta}\right)^\prime.\]
We proceed by multiplying the first equation in \eqref{eq:system} by $\rho^\prime$ and integrating with respect to $s$ to obtain
\begin{equation}
\label{eq:int}
\rho\left(\gamma_\theta\cos{\theta}+\gamma\sin{\theta}\right)+\beta \log{\rho}-\lambda\rho^2=\mathcal{D}
\end{equation}
on $[0,L]$ where $\mathcal{D}$ is a constant. The equation \eqref{eq:int} provides the relationship between the $\rho$-coordinate of a point on the curve $\partial\omega$ and the angle between $\partial\omega$ and the $\rho$-axis at the same point. An immediate consequence of \eqref{eq:int} is that flat facets on $\partial\omega$ may be present only if they are parallel to the $z$-axis, when $\rho$ on these facets remains constant. 

Although in the preceding discussion we assumed that the energy-minimizing curve is smooth, it is possible that this curve will develop corners where the curvature is not defined and the angle function $\theta$ is not differentiable. To this end, observe that the only term in \eqref{eq:etot} where the derivative of $\theta$ would appear following integration by parts is the surface energy term. We can adapt our variational argument in order to handle this situation as demonstrated in the main text of the paper. 

\section{Analysis}
From now on, we suppose that $\gamma:\mathbb R\to\mathbb R$ is a smooth, $2\pi$-periodic and strictly positive function. Borrowing once again from \cite{GA89}, we have that (i) the Frank diagram associated with $\gamma$ is convex as long as $\gamma_{\theta\theta}+\gamma>0$ for all $\theta\in[0,2\pi)$ and the energy minimizing curve $\br$ cannot have corners; (ii) if the Frank diagram is not convex, then the curve $\br$ has corners, but not cusps. The angles at which corners occur for a given $\gamma$ are fixed by the corresponding Frank diagram and are well-separated if the Maxwell lines of a given diagram have no common points (regular diagram in the terminology of \cite{GA89}).

We look for a convex, piecewise-smooth curve $\br$ that solves \eqref{eq:system}-\eqref{eq:cond}. If $\gamma_{\theta\theta}+\gamma$ is not strictly positive on $[0,2\pi)$, we impose an additional technical assumption on $\gamma$ that it achieves a local minimum at some $\theta_0\in(0,\pi)$. 
\begin{lemma}
\label{l:1}
The Lagrange multiplier $\lambda$ is strictly positive.
\end{lemma}
\begin{proof}
Since $\br$ is convex, $\theta^\prime\geq0$ away from the corners. If $\gamma_{\theta\theta}+\gamma$ is strictly positive on $[0,2\pi)$, then there are no corners and the result immediately follows by evaluating \eqref{eq:system} at $\pi/2$. Otherwise, $\theta_0$ is not inside a corner and $\gamma_{\theta\theta}(\theta_0)+\gamma(\theta_0)>0$ hence $\lambda>0$ by evaluating \eqref{eq:system} at $\theta_0$.
\end{proof}

Next, observe that the jump condition \eqref{eq:jump} implies that $\gamma_\theta\cos{\theta}+\gamma\sin{\theta}$ is continuous across the corners. Due to continuity of $\br$ we then have that \eqref{eq:int} holds even when $\br$ has corners.

With Lemma \ref{l:1} in mind, we introduce the rescaling
\[\tilde\rho=\lambda\rho,\ \tilde{z}=\lambda z,\ \tilde{s}=\lambda s,\ \tilde{L}=\lambda L,\ \tilde{\beta }=\lambda\beta \]
then, dropping tildes for notational convenience and taking advantage of \eqref{eq:int}, we can rewrite the problem \eqref{eq:system}-\eqref{eq:cond} in the form
\begin{equation}
\label{eq:simsys}
\left\{
\begin{array}{l}
 \rho\left(\gamma_{\theta\theta}+\gamma\right)\theta^\prime+\gamma\sin{\theta}+\gamma_\theta\cos{\theta}+\frac{\beta }{\rho}-2\rho=0,   \\
 \rho\left(\gamma_\theta\cos{\theta}+\gamma\sin{\theta}\right)+\beta \log{\rho}-\rho^2=\mathcal{D},    \\
z^\prime=\sin{\theta}
\end{array}
\right.
\end{equation}
for $s\in[0,L]$, subject to the conditions
\begin{equation}
\label{eq:simcond}
\theta(L)=\theta(0)+2\pi,\quad \theta^\prime(L)=\theta^\prime(0),\quad \int_0^L\sin{\theta}\,ds=0 
\end{equation}
and \eqref{eq:jump} at the corners whenever discontinuities of $\theta$ are present. Note that we can drop the condition $\int_0^L\cos{\theta}\,ds=0$ and the volume constraint in \eqref{eq:cond} because periodicity of $\rho$ is ensured by the first equation in \eqref{eq:simsys} and the volume constraint can be satisfied by an appropriate rescaling, respectively. In addition to the functions $\rho$, $z$, and $\theta$, the unknowns of the problem \eqref{eq:simsys}-\eqref{eq:simcond} also include the constants $\mathcal{D}$ and $L$.

\subsection{Analysis of the model for curves without corners.}
\label{san}

Assume that $\gamma_{\theta\theta}+\gamma$ is strictly positive on $[0,2\pi)$ so that the energy minimizing curve is smooth. As long as this curve is also convex, we have that 
\[\min_{s\in[0,L]}\rho(s)=\rho\left(-\frac{\pi}{2}\right)\ \mbox{and}\ \max_{s\in[0,L]}\rho(s)=\rho\left(\frac{\pi}{2}\right).\]
Let \[\rho_{m}:=\rho\left(-\frac{\pi}{2}\right)\ \mbox{and}\ \rho_{M}:=\rho\left(\frac{\pi}{2}\right)\] and suppose that \[h(\theta):=\gamma_\theta\cos{\theta}+\gamma\sin{\theta}.\] The following claim holds.
\begin{claim}
\label{c:1}
The function $h$ is strictly increasing from \[h_m:=h\left(-\frac{\pi}{2}\right)=-\gamma\left(-\frac{\pi}{2}\right)\] to \[h_M:=h\left(\frac{\pi}{2}\right)=\gamma\left(\frac{\pi}{2}\right)\] when $\theta\in\left(-\frac{\pi}{2},\frac{\pi}{2}\right)$, while it is strictly decreasing  from $h_M$ to $h_m$ when $\theta\in\left(\frac{\pi}{2},\frac{3\pi}{2}\right)$.
\end{claim}
\begin{proof}
The claim immediately follows from
\[\left(\gamma_\theta\cos{\theta}+\gamma\sin{\theta}\right)_\theta=\left(\gamma_{\theta\theta}+\gamma\right)\cos{\theta},\]
along with $2\pi$-periodicity of $\gamma$ and positivity of $\gamma_{\theta\theta}+\gamma$.
\end{proof}

With the help of Claim 1, we conclude that there exist unique $\theta_+\in\left[-\frac{\pi}{2},\frac{\pi}{2}\right]$ and $\theta_-\in\left[\frac{\pi}{2},\frac{3\pi}{2}\right]$ such that $h(\theta_-)=h(\theta_+)=0$. Now, fix a constant $a>0$ and set $\rho(\theta_-)=a$. Then, the second equation in \eqref{eq:simsys} can be written as
\begin{equation}
\label{eq:1}
h\rho+\beta \log{\rho}-\rho^2=\beta \log{a}-a^2,
\end{equation}
for every $\theta\in[0,2\pi]$.

\begin{claim}
\label{c:2}
Let $a_*$ be the positive solution of
\begin{equation}
\label{eq:3}
h_m\,t_*+\beta \log{t_*}-t_*^2=\beta \log{a_*}-a_*^2,
\end{equation}
where
\[t_*=\frac{\sqrt{h_m^2+8\beta }+h_m}{4}.\]
For every $a>a_*$ and $h\in\left[h_m,h_M\right]$, there is a unique solution $\varrho(h,a)>a$ of \eqref{eq:1} that continuously increases with $h$. For $a<a_*$, no such solution of \eqref{eq:1} exists.
\end{claim}
\begin{proof}
Consider two functions
\[\phi(t)=h_m\,t+\beta \log{t}-t^2\]
and
\[\psi(t)=\beta \log{t}-t^2\]
when $t>0$. Then 
\[\phi(t)<\psi(t)\]
for every $t>0$ because $h_m<0$ and \eqref{eq:1} at $\theta=-\frac{\pi}{2}$ can be written in the form
\[\phi(\rho_m)=\psi(a),\]
where $\rho_m<a$. Further, the function $\phi$ and $\psi$ both have a single maximum and no minima when $t>0$ and the maximum of $\phi$ is at 
\[t_*=\frac{\sqrt{h_m^2+8\beta }+h_m}{4}.\]
It follows that the equation $\phi(\rho_m)=\psi(a)$ can only be satisfied if $\psi(a)\leq\phi(t_*)$. When combined with the condition $\rho_m<a$, it is easy to see that this will be true when $a>a_*$ with $a_*$ defined in the statement of the claim (cf. Fig. \ref{fig0}). 
\begin{figure}[htb]
\centering
    \includegraphics[width=2.5in]
                    {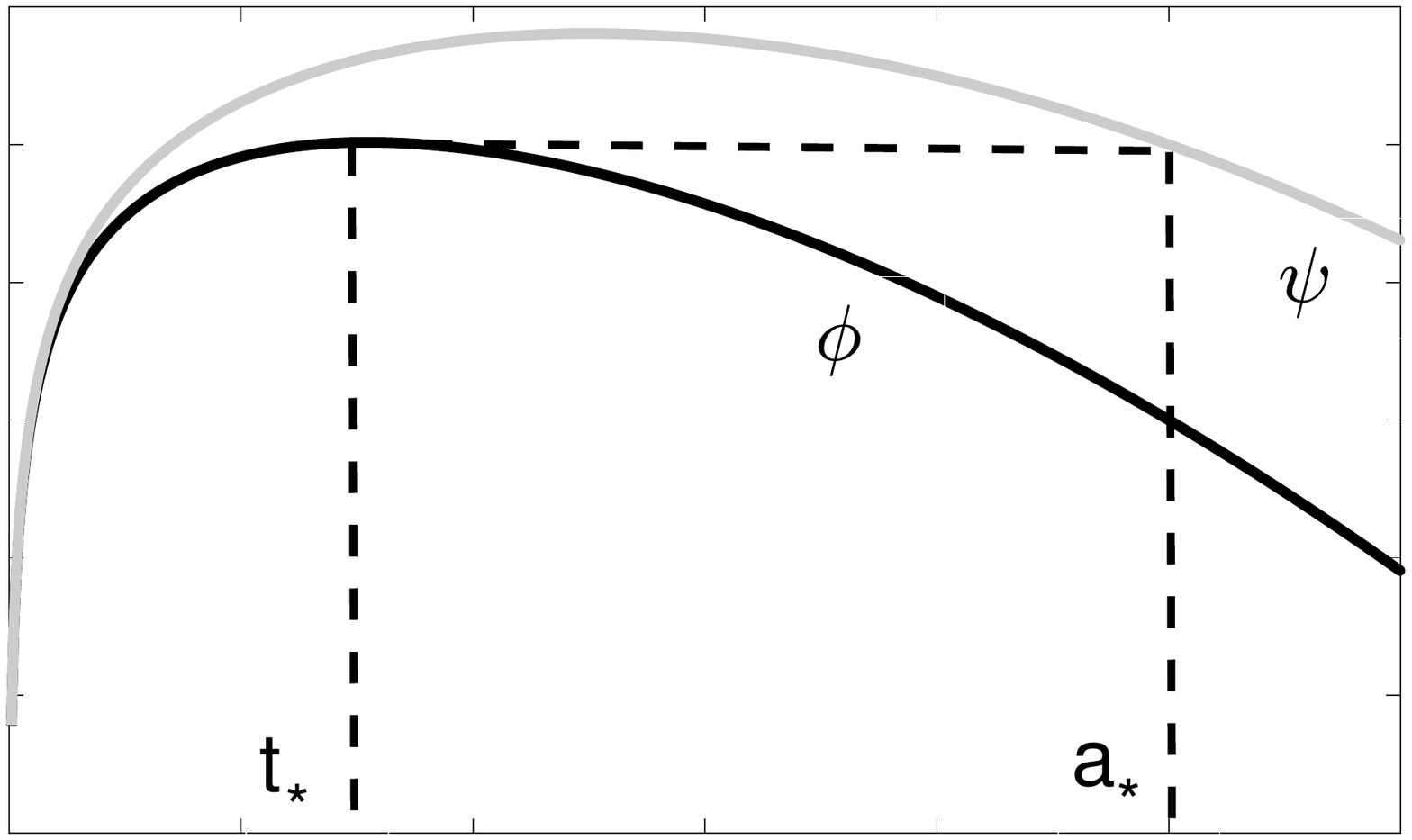}
    \caption{Choice of $t_*$ and $a_*$.}
  \label{fig0}
\end{figure}
It is also easy to see that, as long as the equation \eqref{eq:1} can be solved at $h_m$, it can also be solved for every $h>h_m$. Indeed, rewriting the equation \eqref{eq:1} again, now as
\begin{equation}
\label{eq:2}
h\rho=\rho^2-a^2-\beta \log{\rho/a},
\end{equation}
we observe that the graphs of the functions of $\rho$ on the right and left sides of \eqref{eq:1} intersect at the point $(\rho_m,\rho_m\,h_m)$. As $h$ increases from $h_m$ to $h_M$, the slope of the linear function on the left side of \eqref{eq:2} increases. The function on the right of \eqref{eq:1} is independent of $h$, it has a single minimum on $\mathbb R_+$, and it goes to infinity when $\rho\to 0^+$ or $\rho\to\infty$. We conclude that, as $h$ monotonically increases from $h_m$ to $h_M$, the equation \eqref{eq:2} has two continuous solution branches: one monotonically increasing and another monotonically decreasing with $h$. Since the solution curve $\br$ that we seek is convex, we select $\varrho(h,a)$ from the branch along which $\rho$ increases with $h$. 
\end{proof}
Note that the function $\varrho$ is independent of the surface energy density $\gamma$ and the surface energy is only used in Claim 2 to describe the range of $h$. The solution $\rho_a(\theta)$ of the second equation in \eqref{eq:simsys} with $\rho_-=a$ and parametrized with respect to $\theta$ is then given by
\[\rho_a(\theta)=\varrho(h(\theta),a).\]
Observe that exactly the same solution branch will be traversed by $\rho(h(\theta),a)$ in opposite directions as $\theta$ varies between $\left[-\frac{\pi}{2},\frac{\pi}{2}\right]$ and $\left[\frac{\pi}{2},\frac{3\pi}{2}\right]$, respectively. Further, it follows that $\rho_+=\rho_-$.

The remainder of the solution procedure can be significantly simplified if the curve is strictly convex, because it can be parametrized with respect to $\theta$ instead of $s$. With a slight abuse of notation, the reparametrization $(\rho(\theta),z(\theta))$ solves 
\begin{equation}
\label{eq:simsys1}
\left\{
\begin{array}{l}
 h\rho+\beta \log{\rho}-\rho^2=\beta \log{a}-a^2,\\
\ds z^\prime=\frac{\rho^2\left(\gamma_{\theta\theta}+\gamma\right)\sin{\theta}}{2\rho^2-h\rho-\beta},
\end{array}
\right.
\end{equation}
when $\theta\in[0,2\pi]$ along with a condition that $z(0)=z(2\pi)$.

We can now outline a possible procedure for finding a solution of \eqref{eq:simsys}--\eqref{eq:simcond} corresponding to a strictly convex curve $\br$. Choose $a>0$ and, for every $h\in\left[h\left(-\frac{\pi}{2}\right),h\left(\frac{\pi}{2}\right)\right]$, determine $\varrho(h,a)$ as outlined in Claim \ref{c:2}. The function $\rho_a(\theta)=\varrho(h(\theta),a)$ solves the second equation in \eqref{eq:simsys}. The function $z_a(\theta)$ can then be determined by integrating already known expression on the right hand side of the second equation in \eqref{eq:simsys1}. In order to satisfy the constraint $z(0)=z(2\pi)$ we must have 
\begin{equation}
\label{eq:3.11}
\int_0^{2\pi}\frac{\rho_a^2(\theta)\left(\gamma_{\theta\theta}(\theta)+\gamma(\theta)\right)\sin{\theta}}{2\rho_a^2(\theta)-h(\theta)\rho_a(\theta)-\beta }\,d\theta=0,
\end{equation}
and this can be ensured by iterating in $a$. 

As we will show below, however, the existence of a strictly convex minimizing curve is not guaranteed even in the case of an isotropic surface energy because the curve may develop vertical facets when $\beta $ is sufficiently small. Indeed, it is possible that there is no $a>a_*$ such that \eqref{eq:3.11} holds. In that case, and as long as $z(2\pi)-z(0)>0$, the following modification to the solution can be made. Set $a=a_*$ so that $\rho_m=t_*$ according to \eqref{eq:3.11}. Because $t_*$ is a critical point of the function $\phi$ in the proof of Claim \ref{c:2}, we have 
\[h_m\,\rho_m+\beta \log{\rho_m}-\rho_m^2=0,\]
then the curvature 
\[\theta^\prime\left(-\frac{\pi}{2}\right)=0,\]
according to the first equation in \eqref{eq:simsys}. The solution curve then can be closed in a smooth way by completing it with a straight vertical facet. This procedure is illustrated in the next section, where we numerically solve the algebraic equations derived above assuming that the surface energy density is isotropic.

\subsection{Numerical results for isotropic surface energy.}
\label{Section 4.3}
In this section we set $\gamma\equiv 1$ so that $h(\theta)=\sin{\theta}.$ We solved \eqref{eq:simsys} in MATLAB using the iterative process described at the end of the previous section. The results are illustrated below in Figs. \ref{fig3}-\ref{fig2}. All computations in the remainder of this paper are done in nondimensional variables and, in particular, assuming that the volume of the toroid is equal $1$.

\begin{figure}[H]
\centering
    \includegraphics[width=2in]
                    {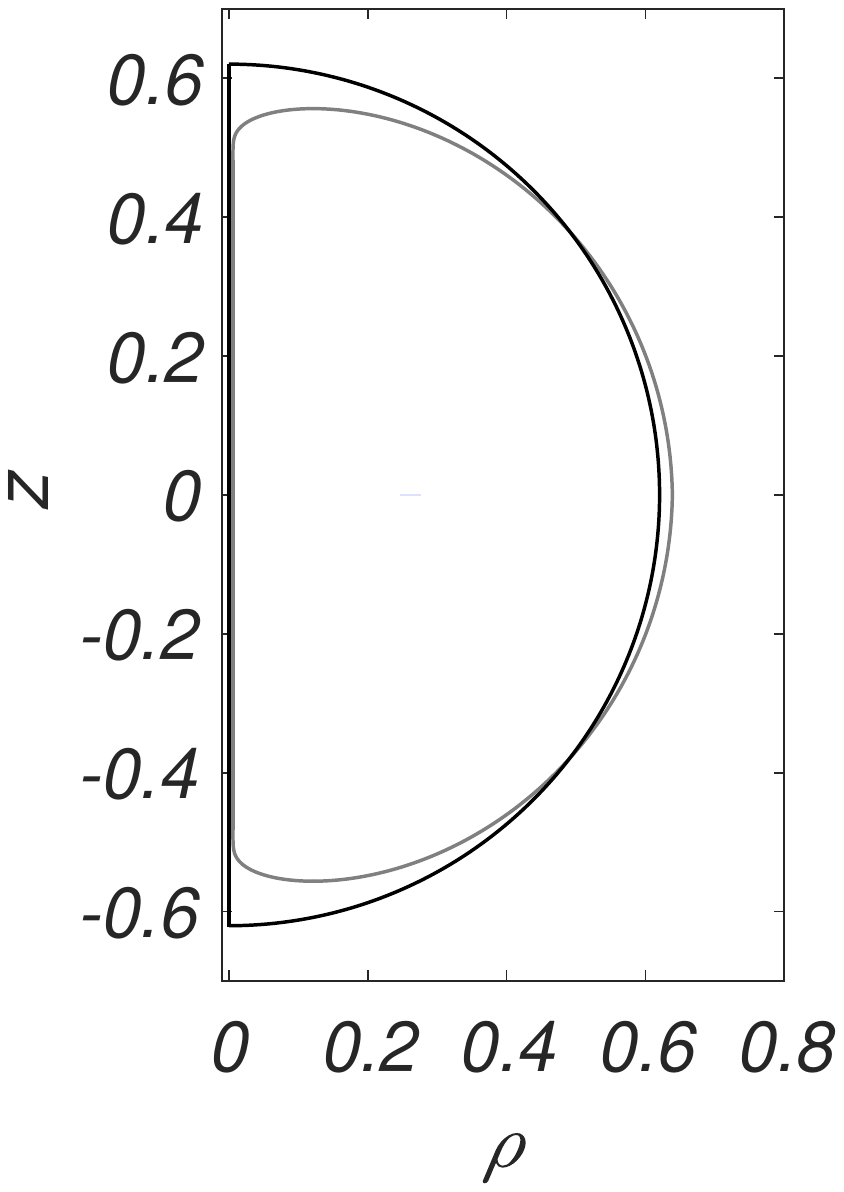}
    \caption{Comparison between the shapes of cross-sections of a toroid with $\beta=0.006$ and the limiting half-disk corresponding to the sphere of the  same volume as the toroid.}
  \label{fig3}
\end{figure}

Note that, if $\beta=0$, there is no contribution from the bending energy and the variational problem reduces to minimizing the surface area of the axially symmetric three-dimensional region subject to a volume constraint. In this case the energy-minimizing domain $\Omega$ has the shape of a sphere and, hence,  $\omega$ becomes a half of a disk. In turn, we expect that for a small $\beta>0$ the shape of a minimizing curve will be close to that of a half of a circle where the endpoints of the arc are connected by the diameter. Indeed, this is what can be observed in Fig. \ref{fig3} for $\beta=0.006$. 
\begin{figure}[H]
\centering
    \includegraphics[width=2in]
                    {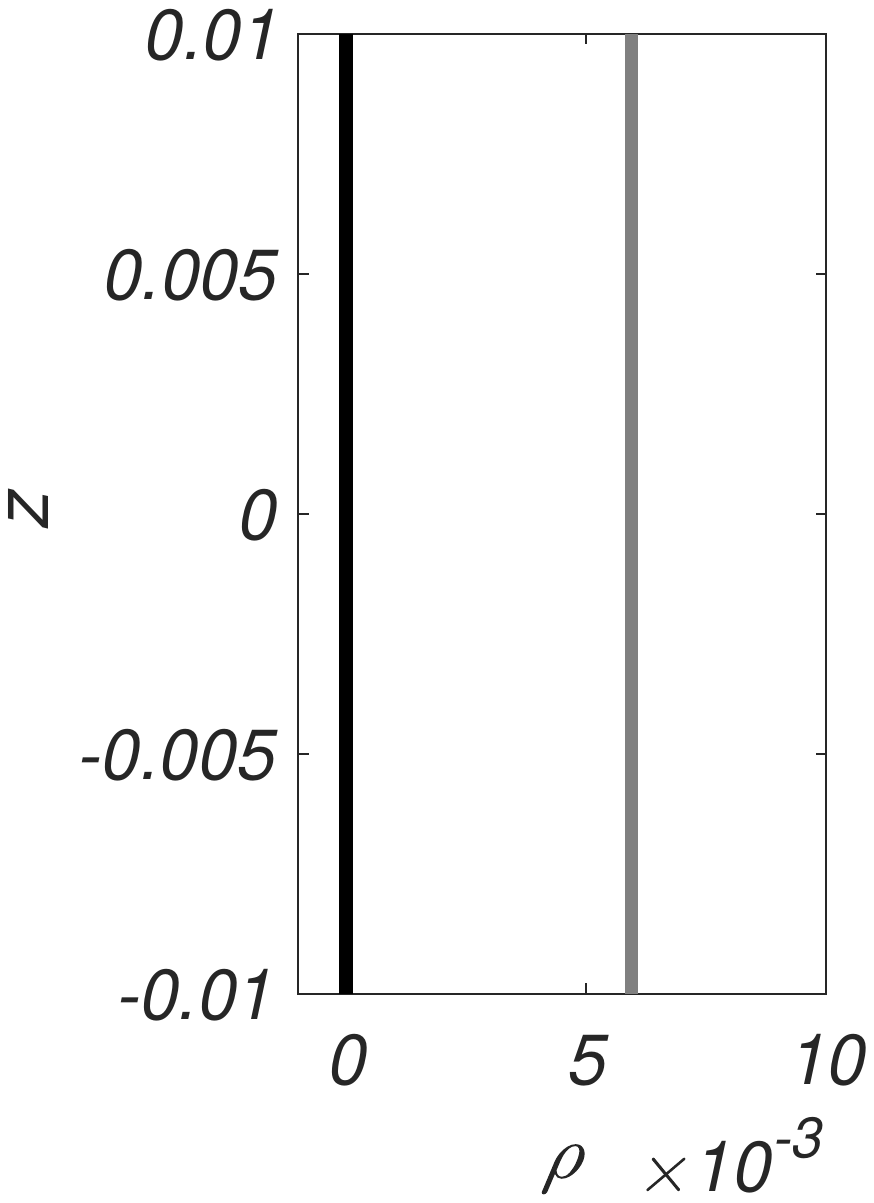}
    \caption{The same setup as in Figure \ref{fig3}, zoomed in at $z$-axis. Note that the facet (grey line) is indeed at the distance comparable to $\beta$ from the axis as indicated by \eqref{eq:dist}.}
  \label{fig4}
\end{figure}
Because the cost of bending diverges as the curve approaches $z$-axis, we also expect that the energy-minimizing curve should remain separate from that axis, with the distance decreasing as $\beta$ becomes smaller. Here an expansion of $t_*$ with respect to a small $\beta$ leads to the following estimate
\begin{equation}
\label{eq:dist}
d:=\mbox{dist}\left(\partial\omega,\left\{\rho=0\right\}\right)\sim\beta
\end{equation}
for the distance between the curve and the $z$-axis in the isotropic case. Fig. \ref{fig4}, corresponding to the configuration depicted in Fig. \ref{fig3} zoomed in near the origin, confirms this estimate.

As $\beta$ gets larger, bending becomes more expensive in comparison with the surface energy---even away from the $z$-axis---and the size of the toroidal domain should increase with $\beta$. At the same time, the diameter of a cross-section of the torus by a plane containing the $z$-axis should decrease with $\beta$ increasing while the shape of this cross-section approaches that of a disk. This behavior can be seen in Fig. \ref{fig1}. 

\begin{figure}[H]
\centering
    \includegraphics[width=2.in]
                    {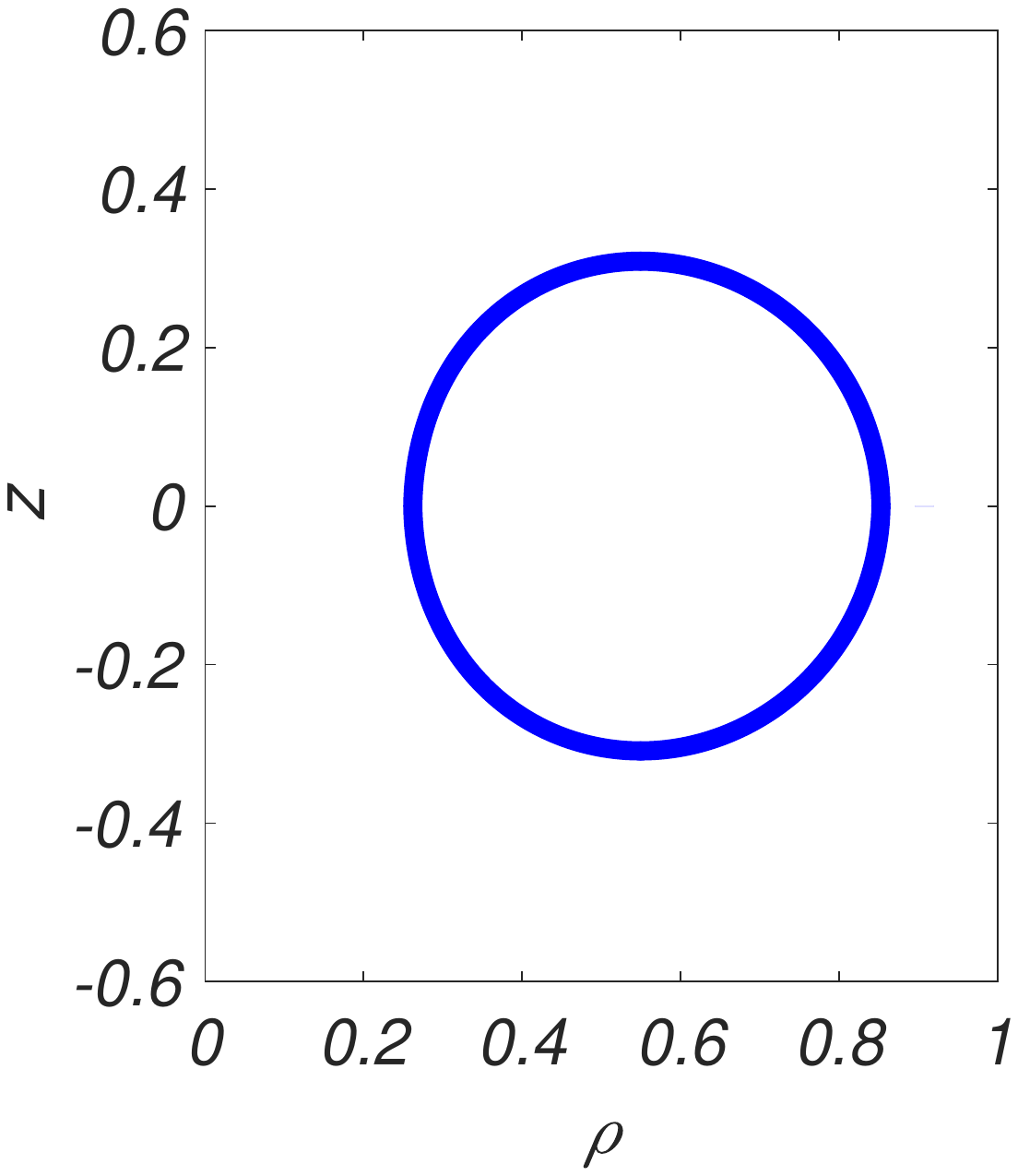}     \includegraphics[width=2.in]
                    {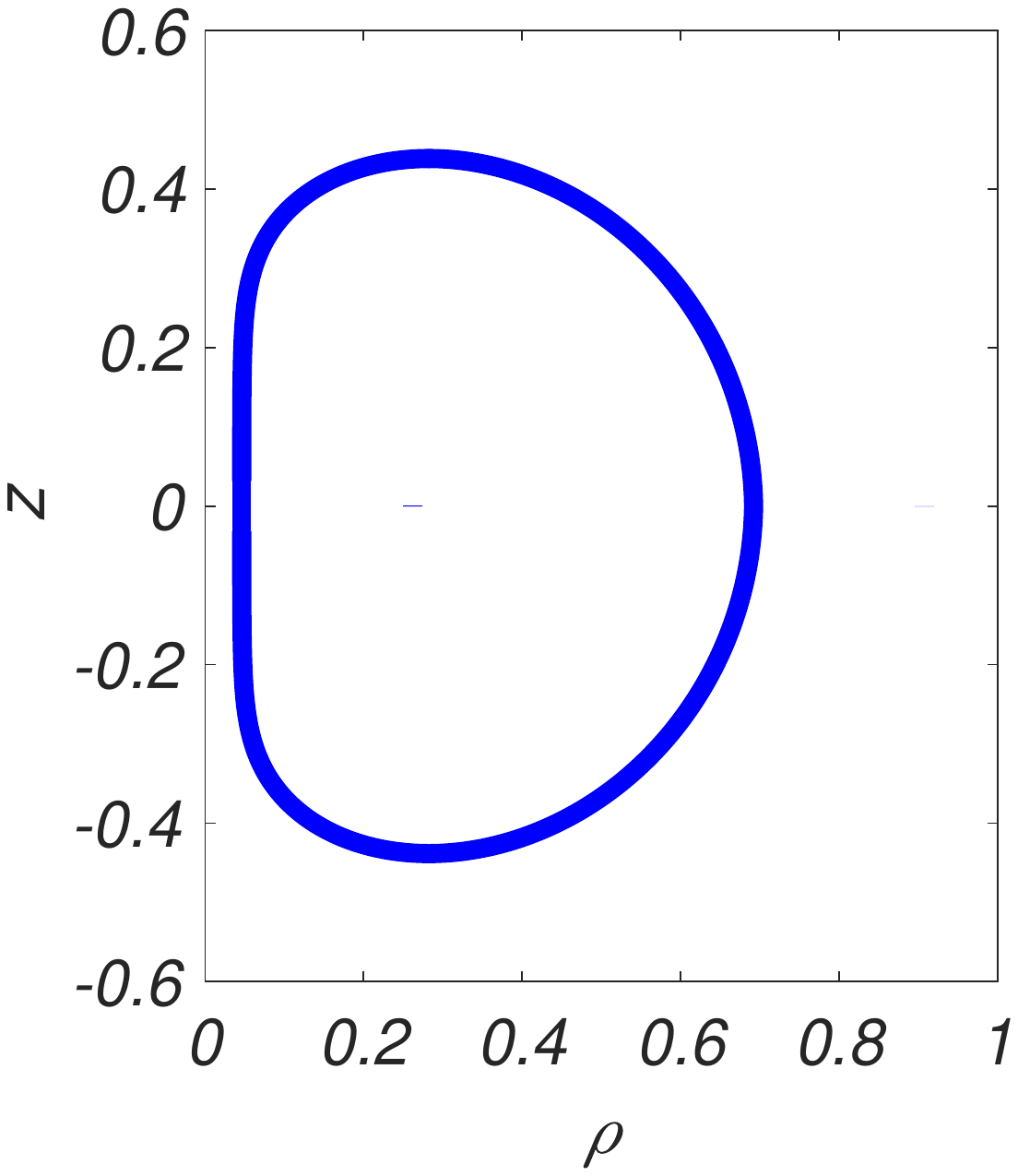}    \includegraphics[width=2.in]
                    {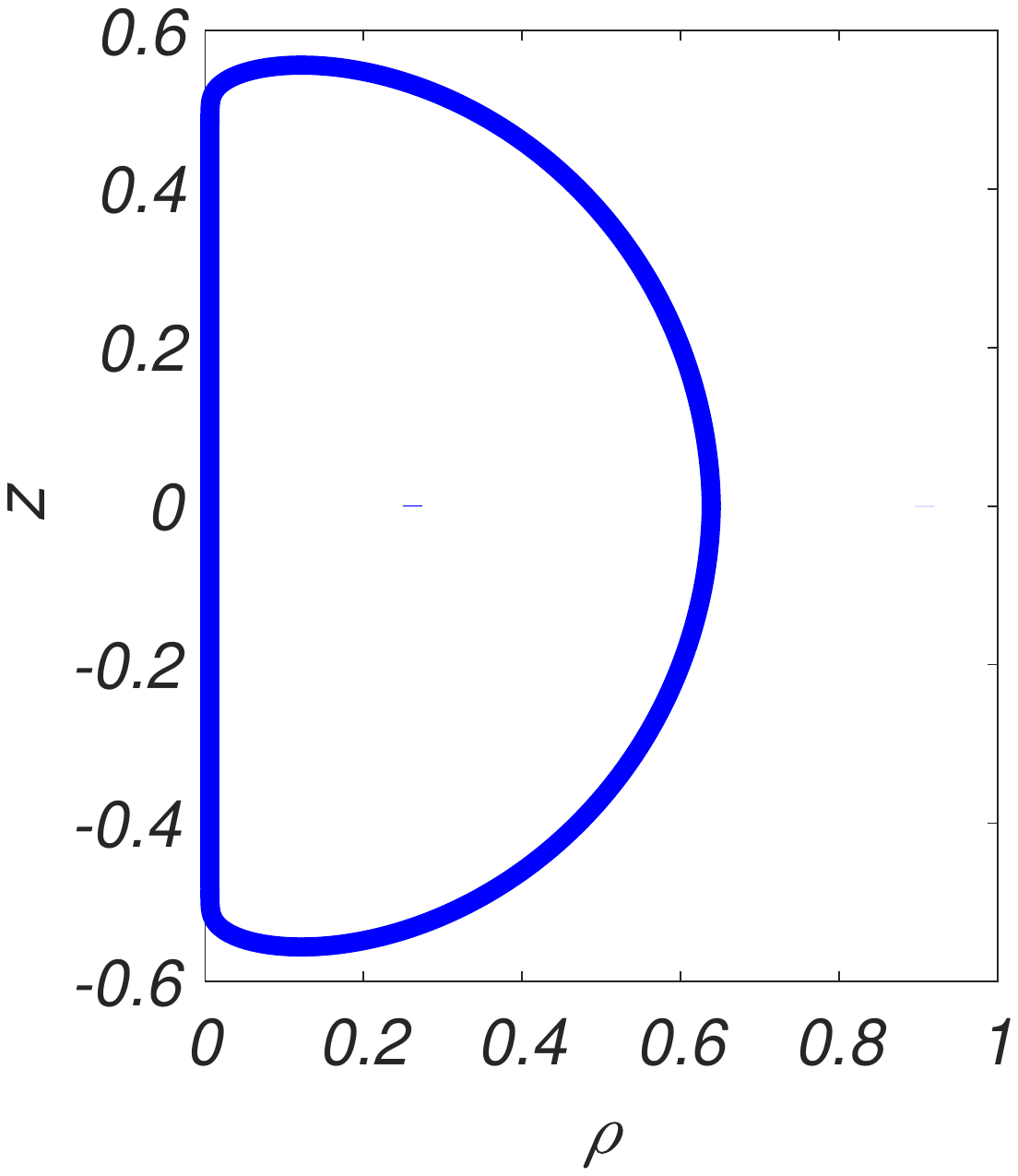}
    \caption{Cross-sections of the energy-minimizing toroids for the isotropic surface energy: $\beta=0.006$ (right), $\beta=0.054$ (middle), $\beta=0.41$ (left).}
  \label{fig1}
\end{figure}

When $\beta$ is small, we observed that there is a critical value of $\beta$, below which $\rho_{min}=t_*$ defined in Claim \ref{c:2}. As it was discussed at the end of Section \ref{san}, this indicates that a vertical facet forms on the side of the torus that faces the $z$-axis; this facet increases in length when $\beta\to0$ as shown in Fig. \ref{fig2}.

\begin{figure}[H]
\centering
    \includegraphics[width=2in]
                    {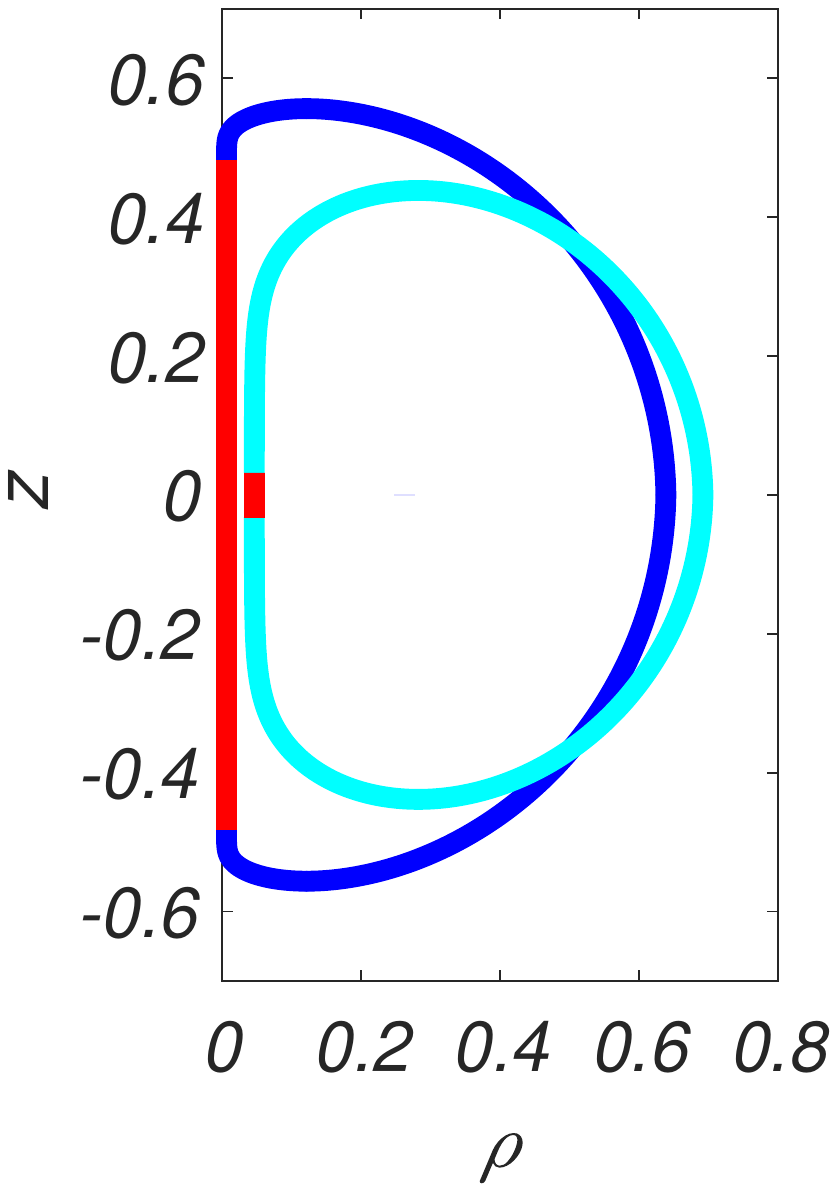}
    \caption{Cross-sections of toroids with vertical facets for the isotropic surface energy (facets are indicated in red). From left to right: $\beta=0.006$ (blue), $\beta=0.054$ (cyan).}
  \label{fig2}
\end{figure}

\begin{figure}[H]
\centering
    \includegraphics[width=2.5in]
                    {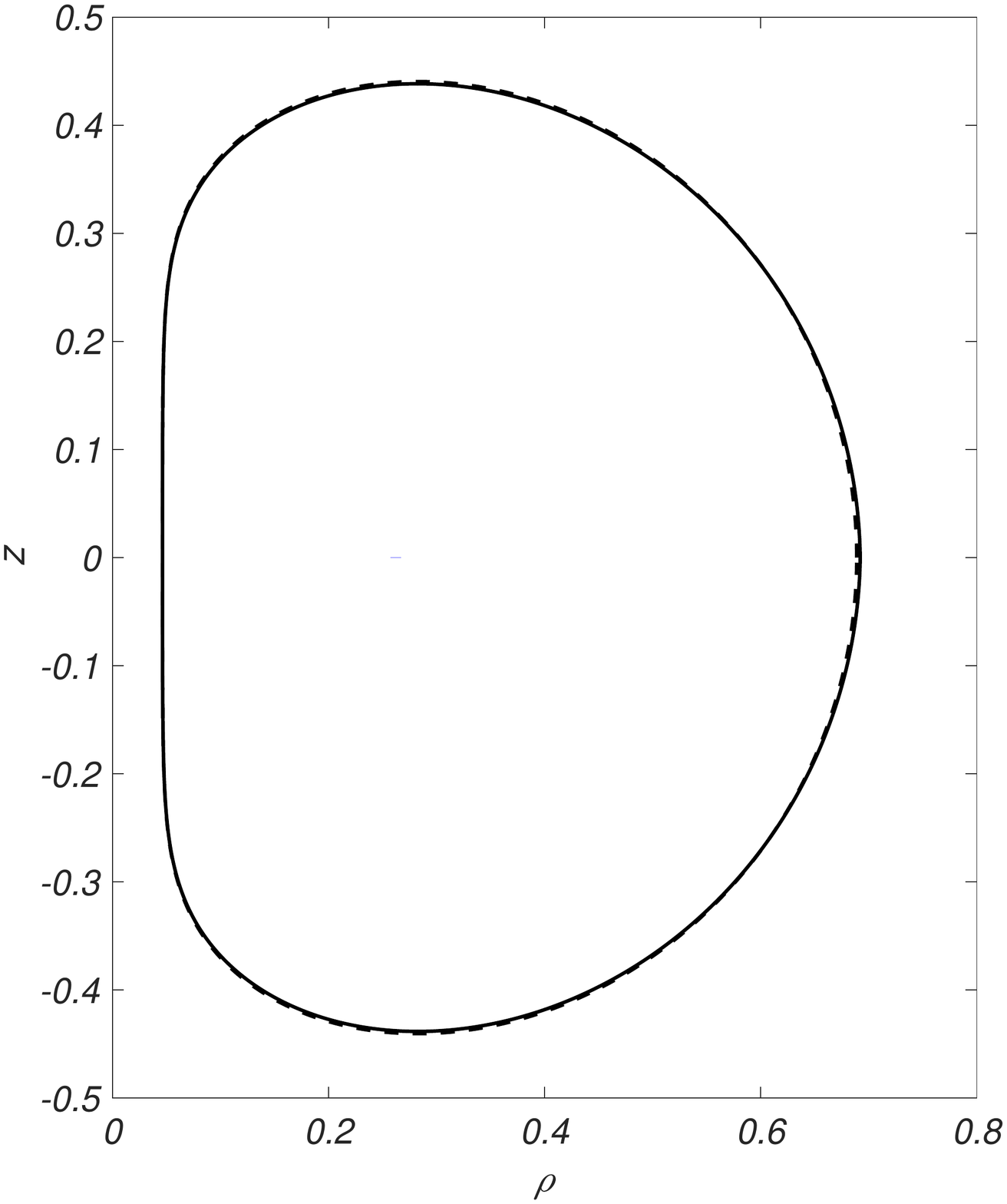}
    \caption{Comparison between the solutions of the system \eqref{eq:simsys} (solid line) vs the minimizer found via gradient flow with $\varepsilon=0$ (dashed line). Here $\beta=0.054$ and we assume that the surface energy is isotropic.}
  \label{fig5}
\end{figure}

Finally, we compared the solutions obtained by using the iterative procedure for solving \eqref{eq:simsys} with the solutions obtained via the gradient flow. The comparison for $\beta=0.054$ and $\gamma\equiv1$ is shown in Fig. \ref{fig5}. Here the gradient flow simulations were done assuming that the regularization parameter $\varepsilon=0$ because the surface energy is isotropic. Fig. \ref{fig5} shows a good match between the predictions of the two methods.

\subsection{Numerical results for weakly anisotropic surface energies.}
\label{sec:low}

\begin{figure}[H]
\centering
    \includegraphics[width=2in]
                    {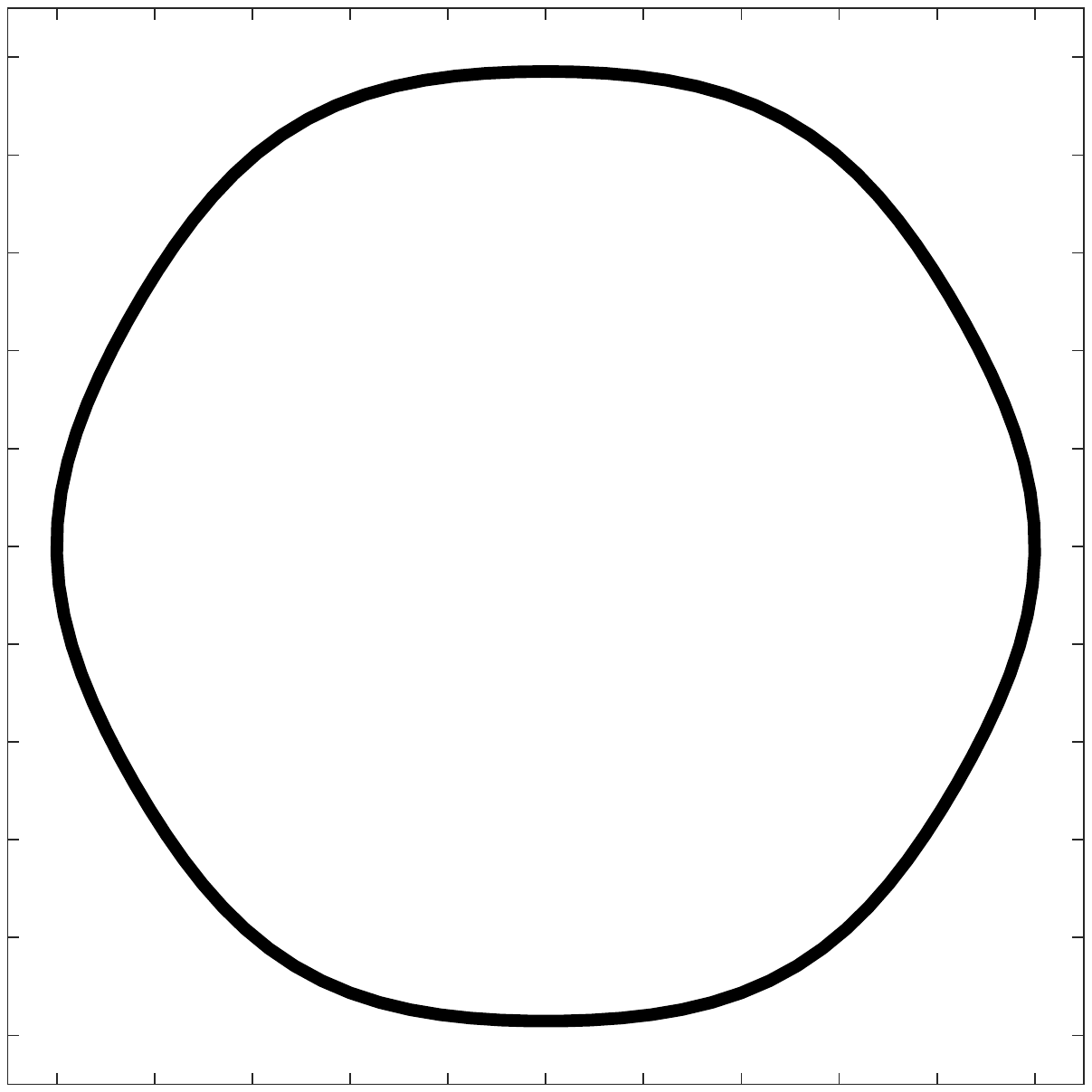} \includegraphics[width=2in]
                    {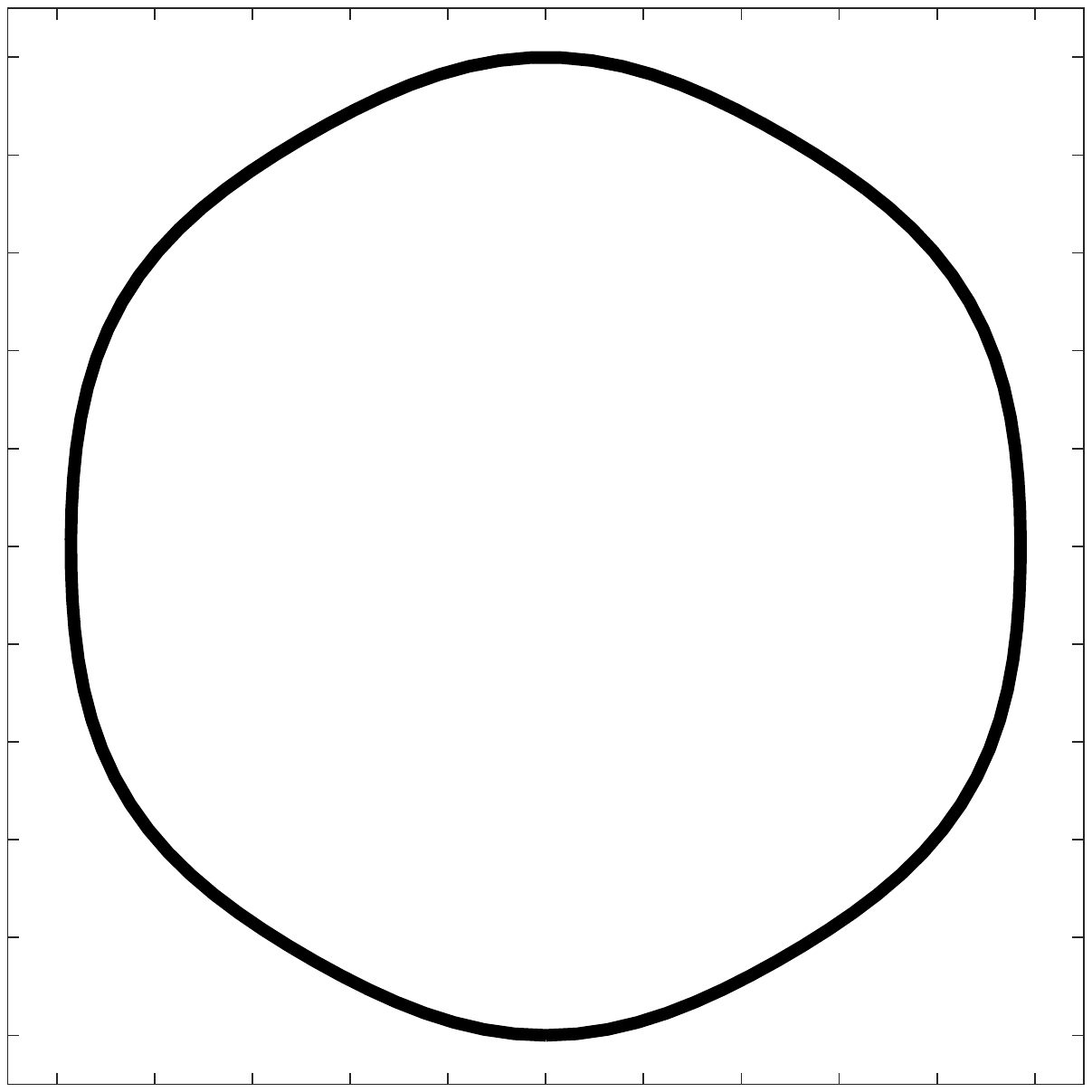}\includegraphics[width=2in]
                    {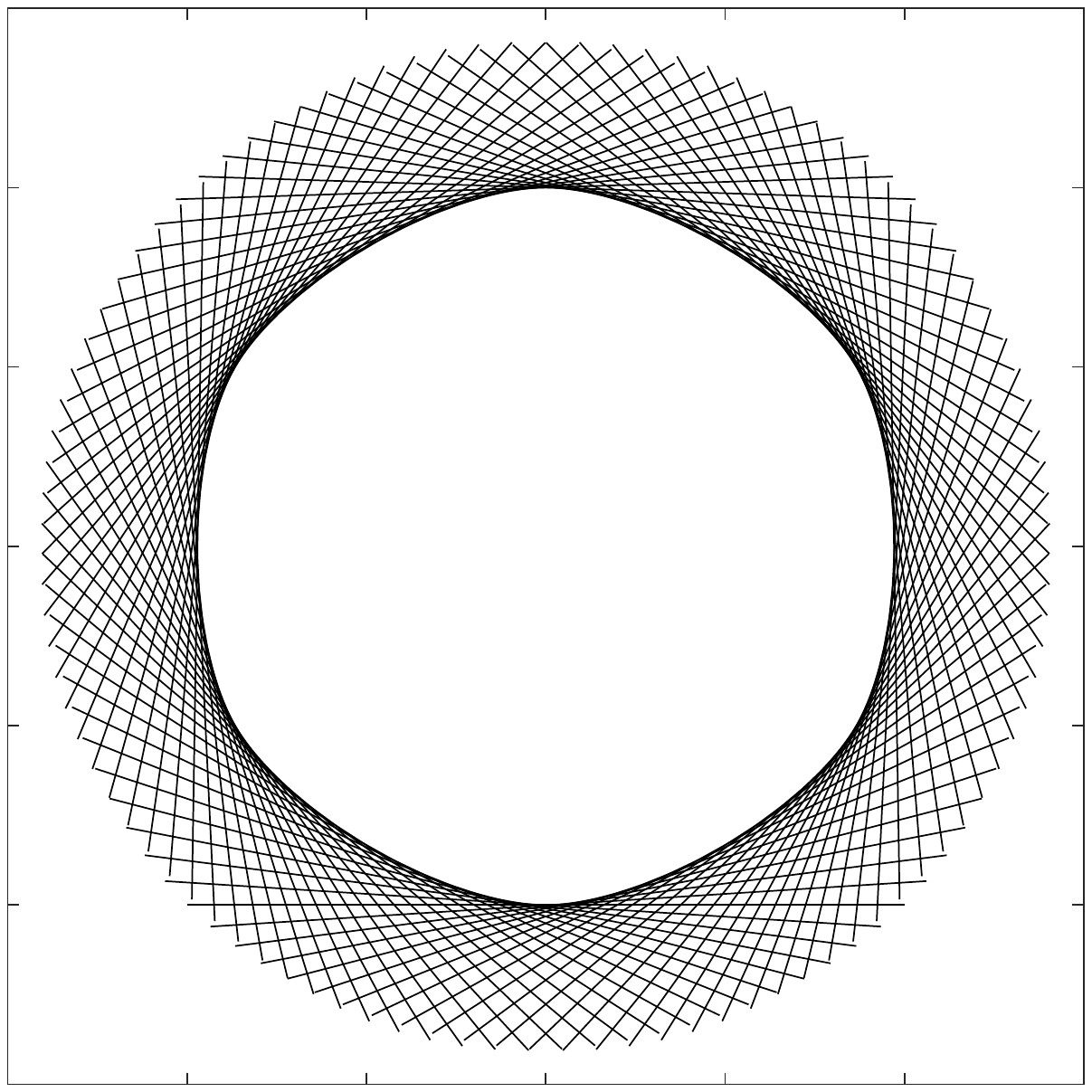}
    \caption{Frank diagram (left), Wulff plot (middle), and Wulff construction for the anisotropic surface energy $\gamma(\theta)=1+\gamma_1\sin^2(3\theta)$ with $\gamma_1=-0.03$.}
  \label{fig6.1}
\end{figure}

\begin{figure}[H]
\centering
    \includegraphics[width=2in]
                    {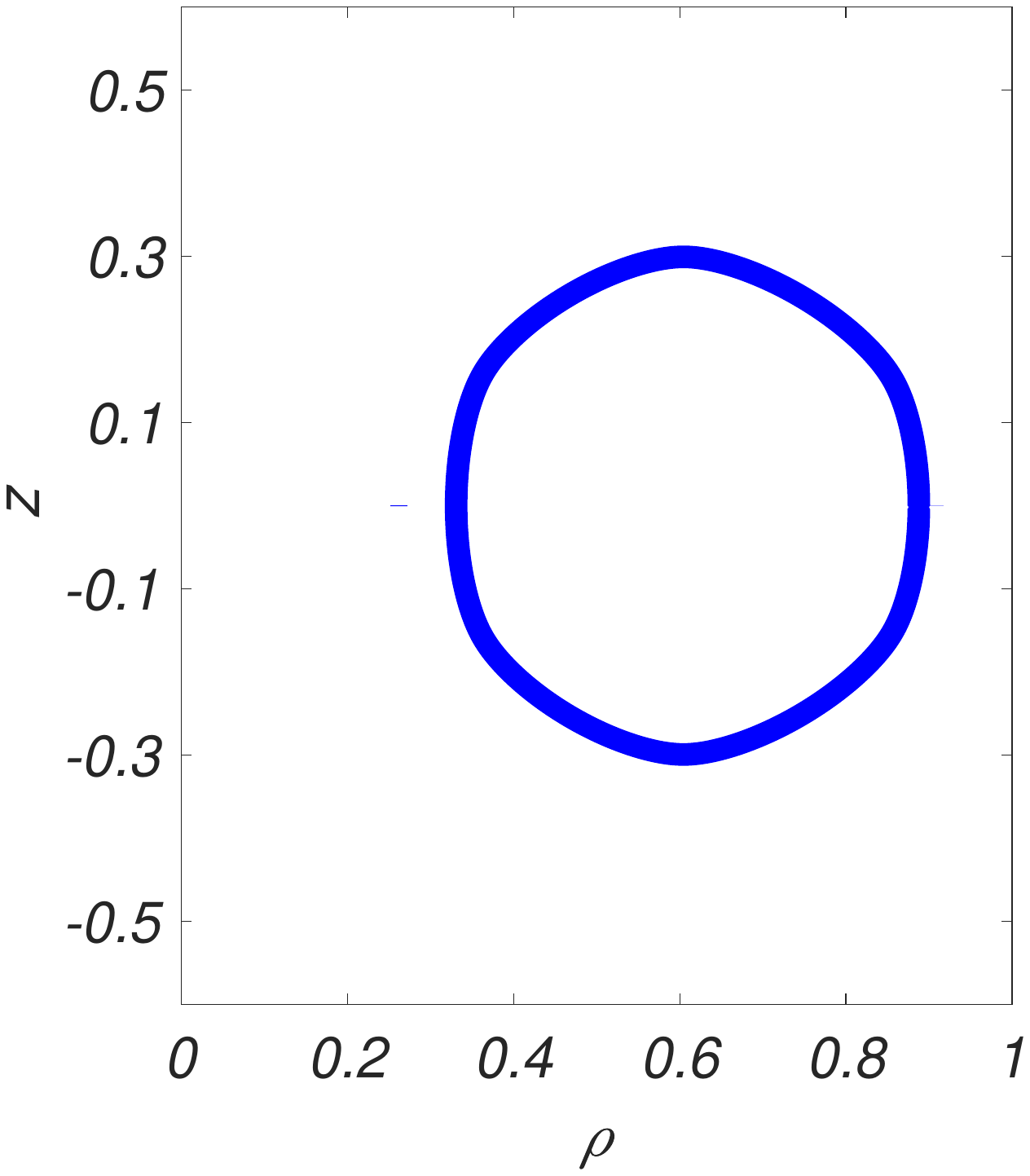} \includegraphics[width=2in]
                    {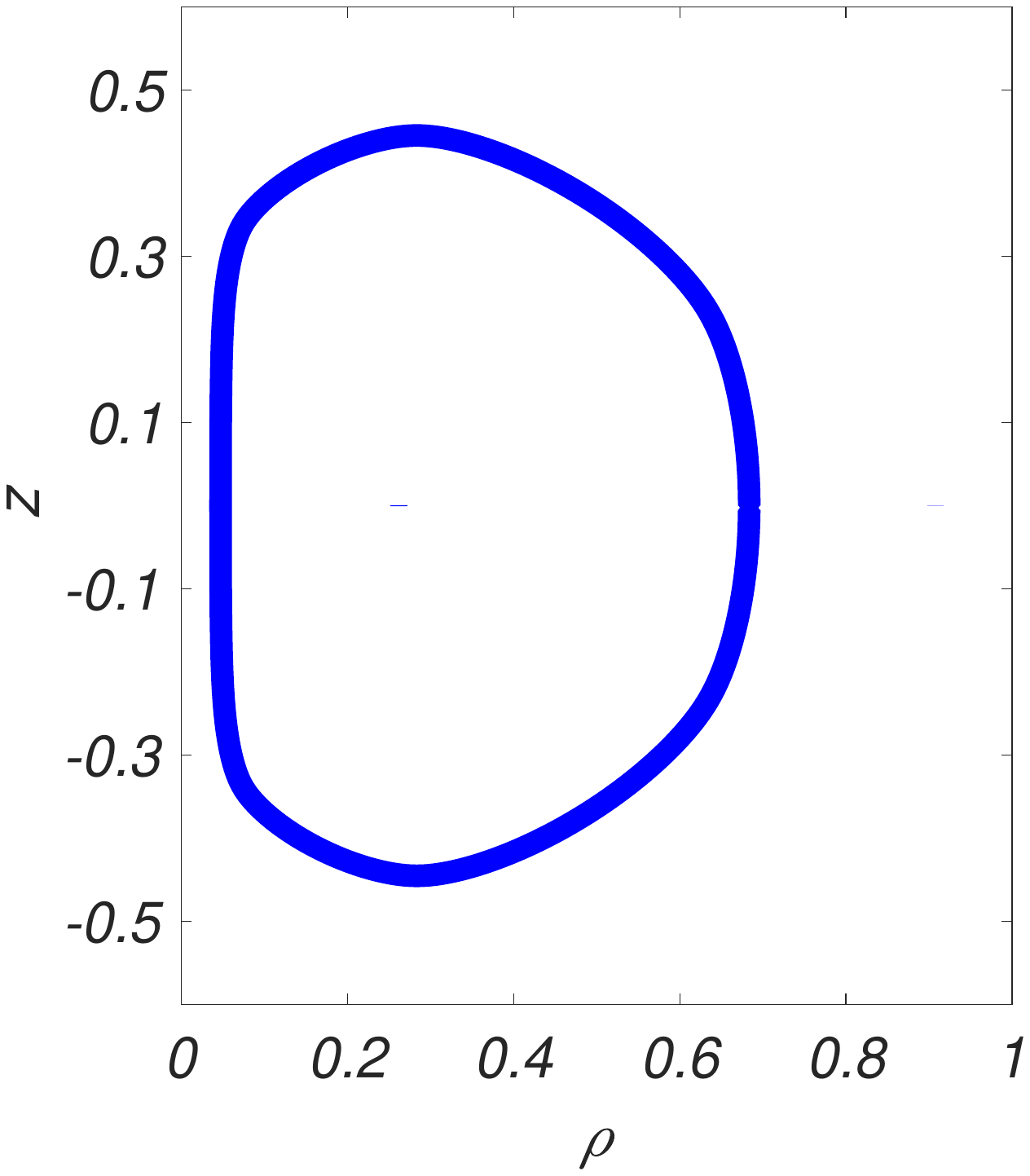}\includegraphics[width=2in]
                    {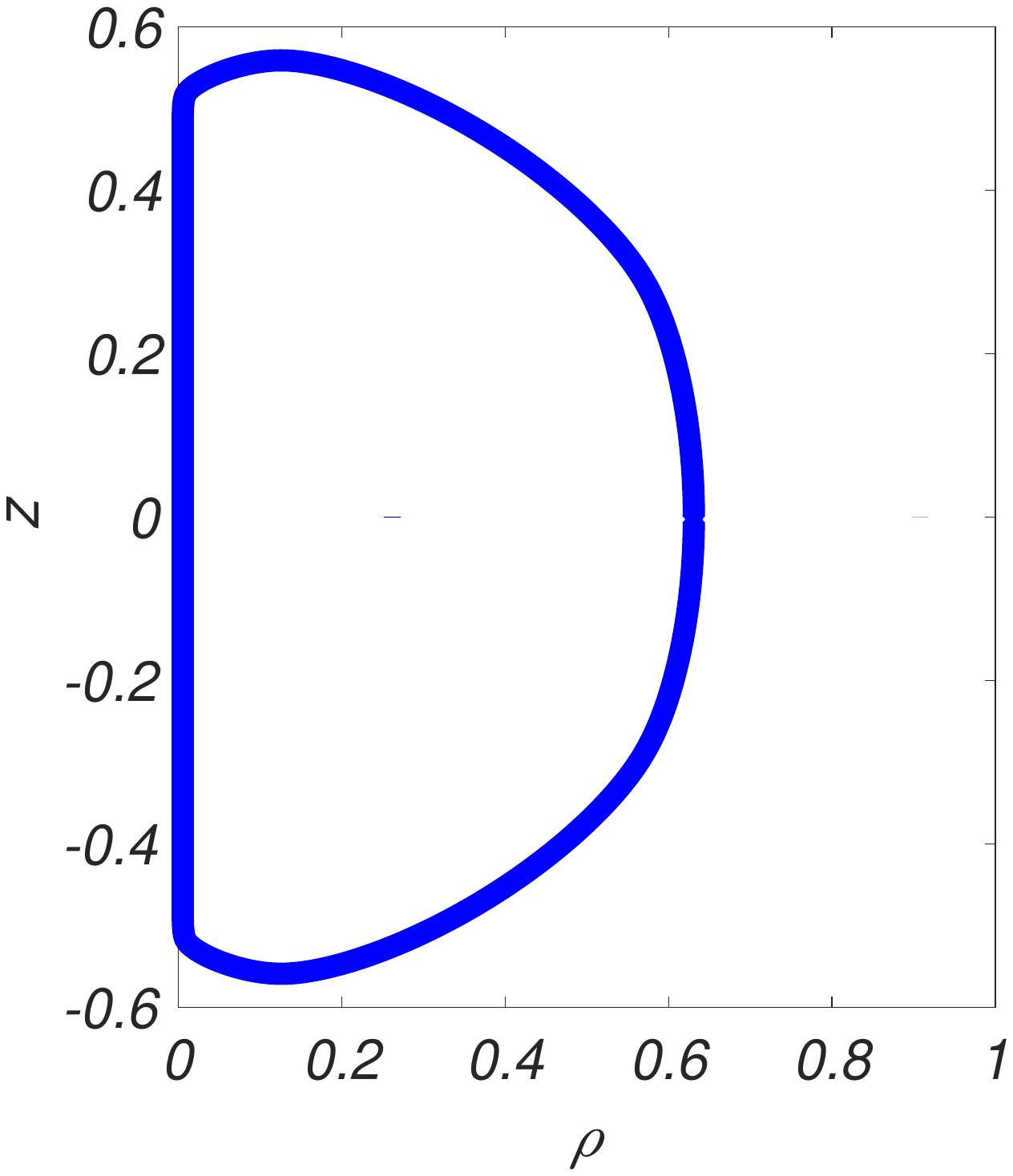} 
    \caption{Cross-sections of the energy-minimizing toroids for the anisotropic surface energy $\gamma(\theta)=1+\gamma_1\sin^2(3\theta)$ with $\gamma_1=-0.03$ and $\varepsilon=0$. Here $\beta=0.54$ (left), $\beta=0.054$ (middle), and $\beta=0.0054$ (right).}
  \label{fig7.1}
\end{figure}

We simulate curves with anisotropic surface energy by setting
\begin{equation}
\label{eq:anis.1}
\gamma(\theta)=1+\gamma_1\sin^2(3\theta)
\end{equation}
First, we consider the case when $\gamma_1=-0.03$. The Frank diagram, Wulff plot, and Wulff construction corresponding to this choice of surface energy are shown in Fig. \ref{fig6.1}. Observe that the Frank diagram in this case is strictly convex and the Wulff construction predicts a configuration with six rounded facets and no sharp corners. This is indeed a shape of the cross-section of the torus that we obtained in simulations in Fig. \ref{fig7.1} (right) for the same $\gamma_1$ and a relatively large value of $\beta$. For such beta, similar to the results in the isotropic case, the major radius $r$ of the torus is large while its minor radius $a$ is small in order to accommodate significant bending energy and the volume constraint. Because $a/r\ll1$, the variation of $\rho$ across the cross-section is smaller than $\rho$ itself and the weight $\rho$ in the integral corresponding to the capillary force can be considered to be essentially constant. As the result, the shape of the cross-section is close to that obtained via the Wulff construction. On the other hand, increasing $K_3$ and $\beta$ or decreasing $\sigma_{||}$ and volume, expands the central narrow core into a wider  "donut hole"  (Fig. \ref{fig7.1}). We also observe that the corners and facets become more rounded with decreasing $\beta$ and the parts of the curve closest to the $z$-axis transform into a facet.

\begin{figure}[H]
\centering
    \includegraphics[width=2in]
                    {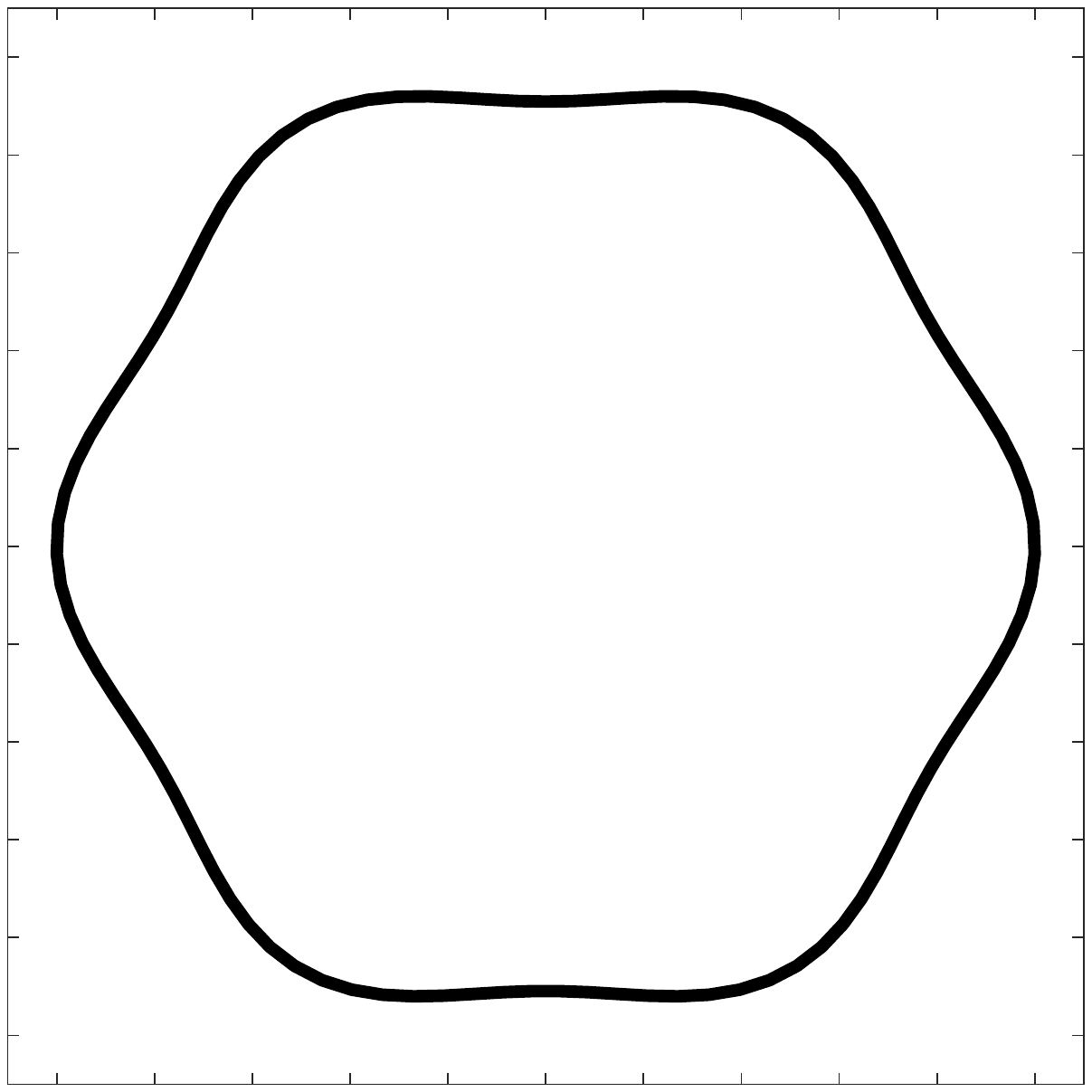} \includegraphics[width=2in]
                    {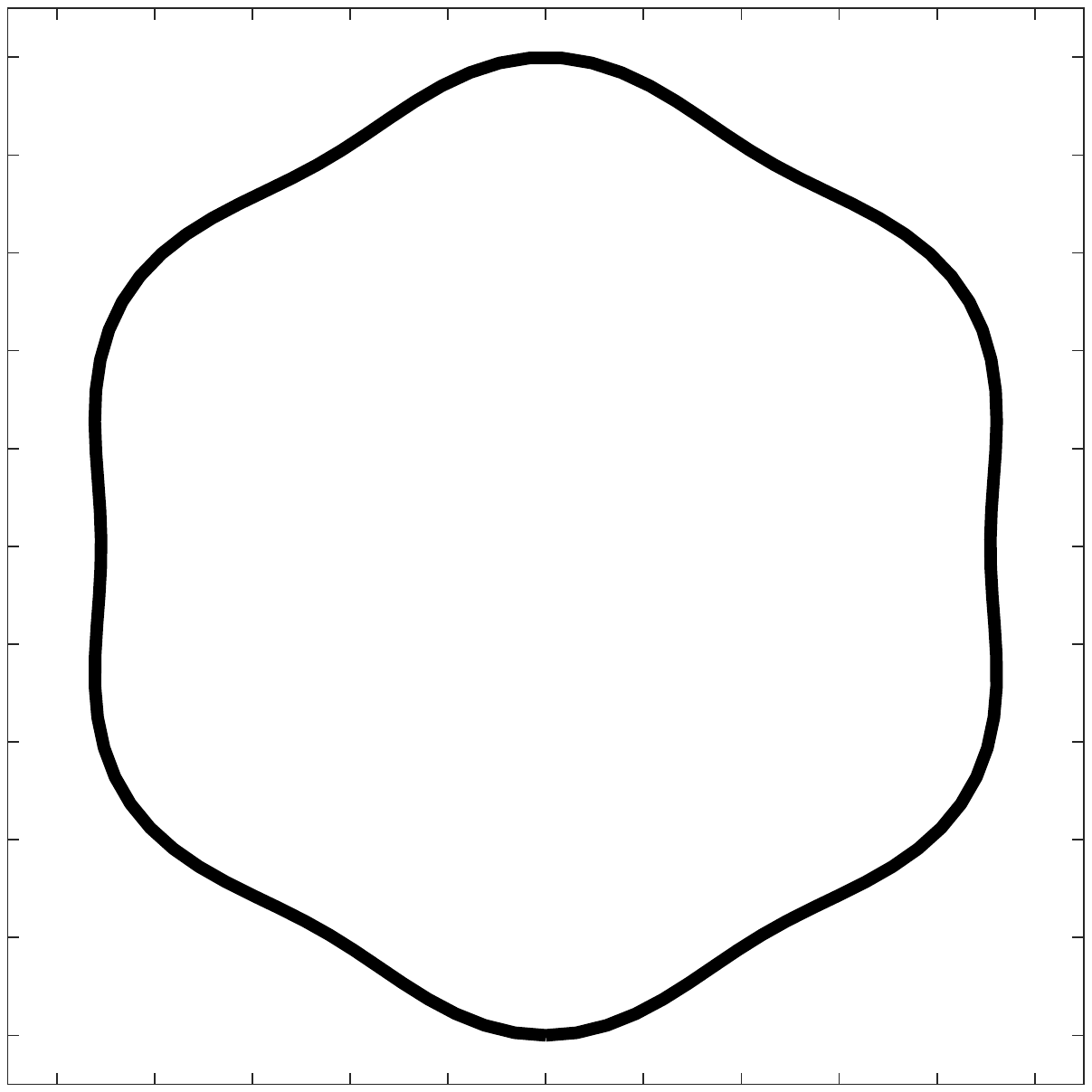}\includegraphics[width=2in]
                    {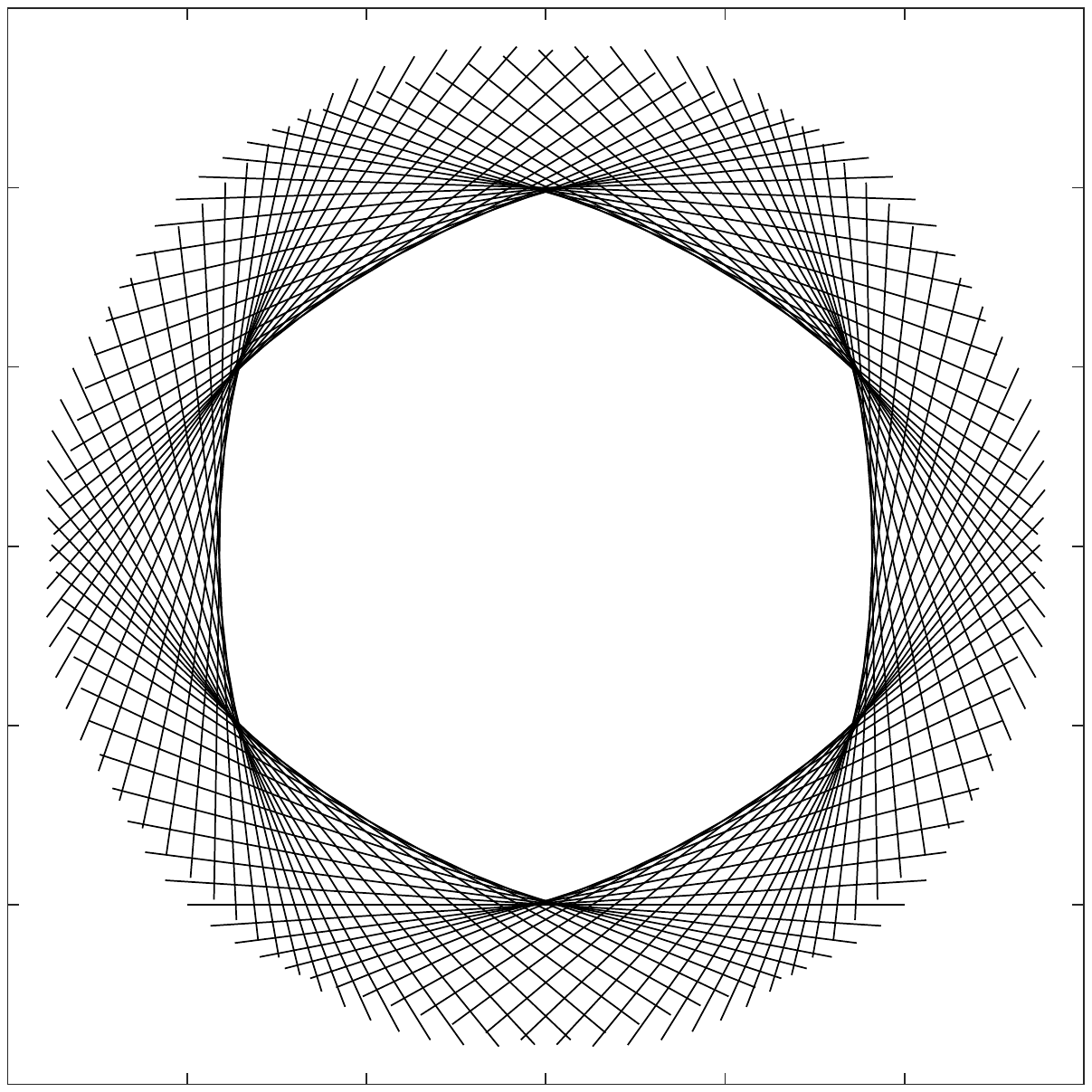}
    \caption{Frank diagram (left), Wulff plot (middle), and Wulff construction for the anisotropic surface energy $\gamma(\theta)=1+\gamma_1\sin^2(3\theta)$ with $\gamma_1=-0.1$.}
  \label{fig8.1}
\end{figure}

\begin{figure}[H]
\centering
    \includegraphics[width=2.5in]
                    {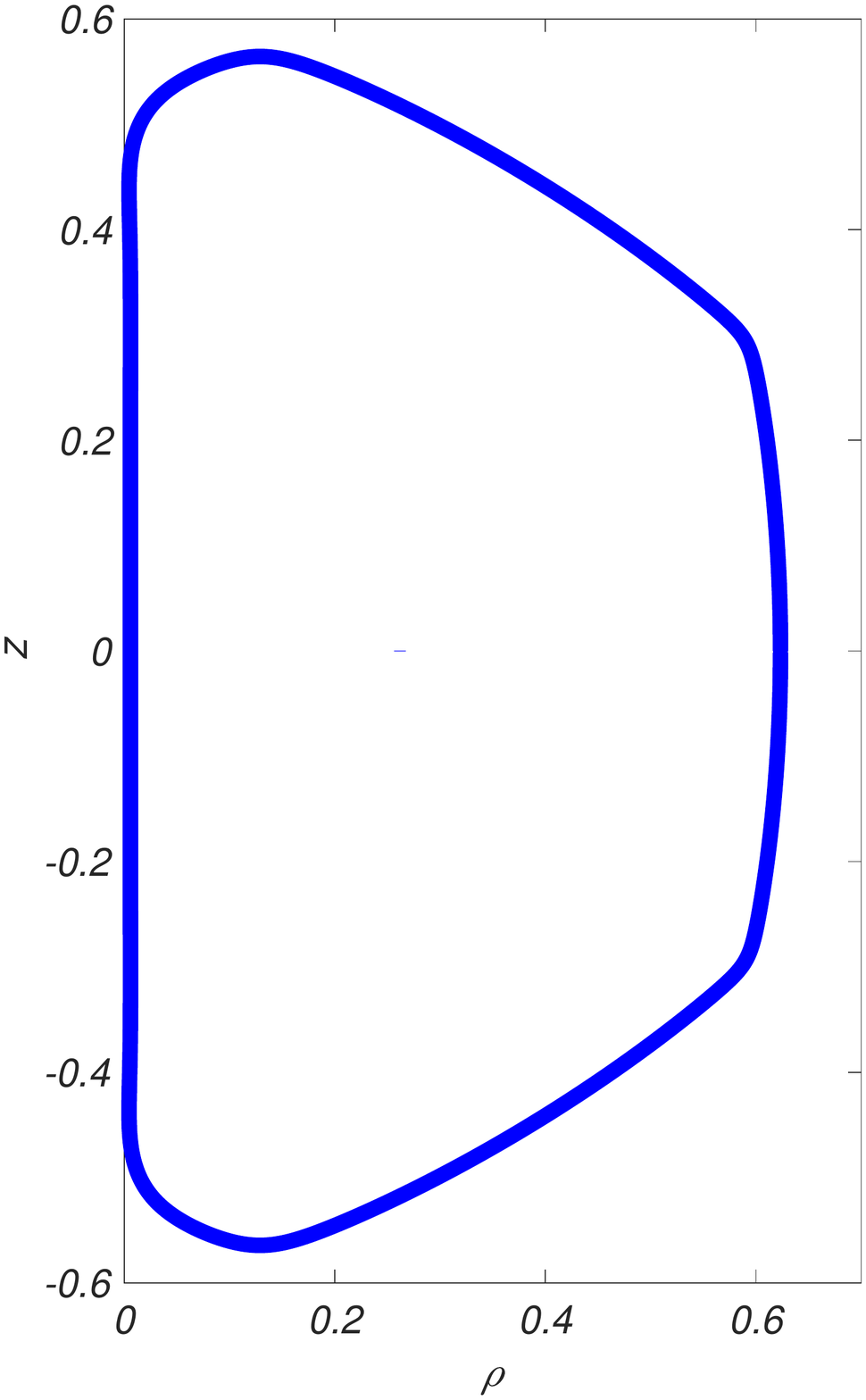}
    \caption{Cross-sections of the energy-minimizing toroid for the anisotropic surface energy $\gamma(\theta)=1+\gamma_1\sin^2(3\theta)$ with $\gamma_1=-0.1$, $\varepsilon=10^{-4}$, and $\beta=0.0054$.}
  \label{fig9.1}
\end{figure}

Enchancing the surface anchoring to $\gamma_1=-0.1$ (See Fig. \ref{fig8.1} for the corresponding Frank diagram, Wulff plot, and Wulff construction) does not change the situation significantly (Fig. \ref{fig9.1}): for the same small $\beta=0.0054$, the part of the crystal closest to the $z$-axis is essentially similar to that for $\gamma=-0.03$. However, away from the axis, faceting becomes more pronounced and is similar to the shape observed for the corresponding Wulff construction Fig. \ref{fig8.1}. Note that, in this case, the Frank diagram is no longer convex and there are six Maxwell lines that indicate that the equilibrium shape must have six corners. The corners in Fig. \ref{fig9.1} are rounded due to regularization employed in the gradient flow simulations (this regularization is not needed when the Wulff construction has a smooth shape).

\bibliographystyleS{ieeetr}
\bibliographyS{chrom,dmref17final,Oleg_Refs,MasterBibTeX.bib}
\articleend
\end{document}